\documentclass[preprint,12pt]{elsarticle}

\usepackage{amssymb}
\usepackage{amsmath}
\journal{Journal of Supercritical Fluids}

\begin{document}

%\begin{frontmatter}

%% Title, authors and addresses

%% use the tnoteref command within \title for footnotes;
%% use the tnotetext command for theassociated footnote;
%% use the fnref command within \author or \affiliation for footnotes;
%% use the fntext command for theassociated footnote;
%% use the corref command within \author for corresponding author footnotes;
%% use the cortext command for theassociated footnote;
%% use the ead command for the email address,
%% and the form \ead[url] for the home page:
%% \title{Title\tnoteref{label1}}
%% \tnotetext[label1]{}
%% \author{Name\corref{cor1}\fnref{label2}}
%% \ead{email address}
%% \ead[url]{home page}
%% \fntext[label2]{}
%% \cortext[cor1]{}
%% \affiliation{organization={},
%%             addressline={},
%%             city={},
%%             postcode={},
%%             state={},
%%             country={}}
%% \fntext[label3]{}

% TITLE
\title{Quantifying the importance of equation of state and transport property compatibility in supercritical fluids}

%% use optional labels to link authors explicitly to addresses:
%% \author[label1,label2]{}
%% \affiliation[label1]{organization={},
%%             addressline={},
%%             city={},
%%             postcode={},
%%             state={},
%%             country={}}
%%
%% \affiliation[label2]{organization={},
%%             addressline={},
%%             city={},
%%             postcode={},
%%             state={},
%%             country={}}

% AUTHORS
\author{Tyler Fara, 
Shashank Yellapantula, 
Bruce Perry, 
Hari Sitaraman, 
Marc Day,
and 
Michael James Martin}

\affiliation{organization={National Laboratory of the Rockies},%Department and Organization
            addressline={15013 Denver Parkway West}, 
            city={Golden},
            postcode={80401}, 
            state={CO},
            country={USA}}

%% Abstract
\begin{abstract}
%% Text of abstract
High-fidelity simulation of supercritical flows requires accurate values of thermodynamic properties including the density $\rho$, specific heat $c_p$, enthalpy $h$, and entropy $s$ as well as transport properties such as the viscosity $\eta$ and thermal conductivity $\lambda$ as functions of the temperature $T$ and pressure $P$. In simulations of mixtures, such as supercritical carbon dioxide ($sCO_{2}$) with contaminants, a generalized equation of state (EoS) is often used, leading to errors in $\rho(T,P)$ when compared to results from correlations developed specifically for individual species. Similarly, generalized correlations for $\eta$ and $\lambda$ are often used, with a corresponding decrease in accuracy. However, a second source of error is introduced in calculations of transport properties.  These correlations are created and computed as functions of $T$ and $\rho$ instead of as functions of $T$ and $P$.  Because of this, when $\eta(T,\rho)$  and $\lambda(T,\rho)$ are calculated, errors in $\rho(T,P)$ propagate through the calculation. This work creates a procedure for tracking both errors caused by using a generalized correlation for $\eta(T,\rho)$ and $\lambda(T,\rho)$, and for tracking the errors caused by using a generalized correlation for $\rho(T,P)$ in these calculations.  The viscosity and thermal conductivity of supercritical $CO_{2}$, $N_{2}$, and $CH_{4}$ are used as examples to demonstrate how to separate the errors due to the generalized property correlations and the errors due to the generalized EoS. Both sources of error are largest near the critical temperature $T_c$. These errors are large enough to impact both the numeric stability and the accuracy of simulations.

\end{abstract}

%%Graphical abstract
%\begin{graphicalabstract}

%\includegraphics[scale=0.2]{visual-abstract-1.png}

%\end{graphicalabstract}

%%Research highlights
%\begin{highlights}
%\item Generalized equations of state used to find the density $\rho(P,T)$ as a function of temperature and pressure contain errors that propogate when used to find transport properties such as viscosity $\eta(\rho,T)$   and thermal conductivity $\lambda(\rho,T)$ using generalized expressions based on density and temperature instead of pressure and temperature
 %\item Comparison with results for $\rho(P,T)$, $\eta(P,T)$, and $\lambda(P,T)$ using species-specific models allows these errors to be bounded, and their impact quantified. In some cases, these errors are large enough to impact the physical accuracy of results, and the numeric stability of simulations.
 %\item These errors can be minimized by developing coupled models for density, viscosity, and thermal conductivity that are calibrated based on equation of state results for density instead of the actual density of the mixture.
%\end{highlights}

%% Keywords

%\end{frontmatter}

% MACROS
% editing commands (DELETE FROM FINAL DRAFT)
%\usepackage{soul}
\newcommand{\tfadd}[1]{\textcolor{red}{#1}}
\newcommand{\tfdel}[2]{\textcolor{blue}{\st{#1}} \textcolor{red}{#2}}
\newcommand{\tfnote}[1]{\textcolor{purple}{[[NOTE: #1]]}}

% variables
\newcommand{\dens}{\rho}
\newcommand{\visc}{\eta}
\newcommand{\cond}{\lambda}

% models
\newcommand{\viscref}{\visc_\text{ref}}
\newcommand{\condref}{\cond_\text{ref}}
\newcommand{\viscgen}{\visc_\text{gen}}
\newcommand{\condgen}{\cond_\text{gen}}
\newcommand{\vischyb}{\visc_\text{test}}
\newcommand{\condhyb}{\cond_\text{test}}
\newcommand{\densn}{\rho_\text{n}}
\newcommand{\viscn}{\visc_\text{n}}
\newcommand{\condn}{\cond_\text{n}}

% eos or correlation
\newcommand{\densref}{\dens_\text{ref}}
\newcommand{\densgen}{\dens_\text{gen}}
\newcommand{\viscrefCorr}{\hat{\visc}_\text{ref}}
\newcommand{\viscgenCorr}{\hat{\visc}_\text{gen}}
\newcommand{\condrefCorr}{\hat{\cond}_\text{ref}}
\newcommand{\condgenCorr}{\hat{\cond}_\text{gen}}

% errors
\newcommand{\densERR}{\Delta \rho}
\newcommand{\denserr}{\delta \rho}
\newcommand{\densERRtot}{\Delta \rho_\text{tot}}
\newcommand{\denserrtot}{\delta \rho_\text{tot}}
\newcommand{\viscERRtot}{\Delta \eta_\text{tot}}
\newcommand{\viscERRcorr}{\Delta \eta_\text{corr}}
\newcommand{\viscERRref}{\Delta \eta_\text{ref}}
\newcommand{\viscERReos}{\Delta \eta_\text{eos}}
\newcommand{\viscerrtot}{\delta \eta_\text{tot}}
\newcommand{\viscerrcorr}{\delta \eta_\text{corr}}
\newcommand{\viscerreos}{\delta \eta_\text{eos}}
\newcommand{\viscERRsecondorder}{O^{(\eta)}((\densERR)^2)}
\newcommand{\viscerrsecondorder}{O^{(\eta)}((\denserr)^2)}
\newcommand{\condERRtot}{\Delta \lambda_\text{tot}}
\newcommand{\condERRcorr}{\Delta \lambda_\text{corr}}
\newcommand{\condERReos}{\Delta \lambda_\text{eos}}
\newcommand{\conderrtot}{\delta \lambda_\text{tot}}
\newcommand{\conderrcorr}{\delta \lambda_\text{corr}}
\newcommand{\conderreos}{\delta \lambda_\text{eos}}
\newcommand{\condERRref}{\Delta \lambda_\text{ref}}
\newcommand{\condERRsecondorder}{O^{(\lambda)}((\densERR)^2)}
\newcommand{\conderrsecondorder}{O^{(\lambda)}((\denserr)^2)}

% additional commands
\newcommand{\dviscgenCorr}{\frac{\partial \viscgenCorr}{\partial \rho}}
\newcommand{\dviscgenCorratdensref}{\left. \frac{\partial \viscgenCorr}{\partial \rho}\right|_{(T,\densref)}}

% updated eos terminology
\newcommand{\viscERRfirst}{\viscERReos^{(1)}}
\newcommand{\viscerrfirst}{\viscerreos^{(1)}}
\newcommand{\viscERRsecond}{\viscERReos^{(2)}}
\newcommand{\viscerrsecond}{\viscerreos^{(2)}}
\newcommand{\condERRfirst}{\condERReos^{(1)}}
\newcommand{\conderrfirst}{\conderreos^{(1)}}
\newcommand{\condERRsecond}{\condERReos^{(2)}}
\newcommand{\conderrsecond}{\conderreos^{(2)}}

% caption labels
\newcommand{\cotwo}{{CO$_2$}}
\newcommand{\ntwo}{{N$_2$}}
\newcommand{\chfour}{{CH$_4$}}

% command to include labels like (a), (b), (c) in figures
\newcommand{\panel}[3][.3\linewidth]{%
  \begin{minipage}{#1}\centering
    \includegraphics[width=\linewidth]{#2}\\[-2pt]
    \hspace{-2em}{\small #3}
  \end{minipage}%
}

% command to include labels like (a), (b), (c) in zoomed-in plot
\newcommand{\panelTwo}[3][.25\linewidth]{%
  \begin{minipage}{#1}\centering
    \includegraphics[width=\linewidth]{#2}\\[-2pt]
    \hspace{0em}{\small #3}
  \end{minipage}%
}

% command to include labels like (a), (b), (c) in cross-species plots
\newcommand{\panelThree}[3][.23\linewidth]{%
  \begin{minipage}{#1}\centering
    \includegraphics[width=\linewidth]{#2}\\[-2pt]
    \hspace{-2em}{\scriptsize #3}
  \end{minipage}%
}

\maketitle

% section: begin
%%%%%%%%%%%%%%%%%%%%%%%%%%%%%%%%%%%%%%%%%%%%%%%
\section{Introduction}
Accurate computation of properties for supercritical fluids often requires complex \textit{equations of state}, or EoSs, to accurately compute thermodynamic properties such as the density $\rho$, specific heat $c_p$, enthalpy $h$, and entropy $s$ as functions of the temperature $T$ and pressure $P$. These are complemented by correlations to calculate transport properties such as viscosity $\eta$ and thermal conductivity $\lambda$. A major challenge for both EoSs and transport property correlations is balancing computational cost, accuracy, and generalizability.  For example, for supercritical carbon dioxide, or s\cotwo{}, the Span--Wagner EoS \cite{SW} or related correlations \cite{SW1} will provide the most accurate values of $\rho$, $c_p$, and $s$. The complexity of these correlations is large enough that machine-learning models have been proposed to accelerate computation of these properties \cite{Ni:2023}. However, when substantial contaminants appear, such as in Venus's atmosphere \cite{Rivera:2023}, oxycombustion in s\cotwo{} ambient systems \cite{DeFrahan:2023, Zheng:2026, Sun:2025}, hydrogen hydrothermal combustion \cite{Fan:2026} or green chemistry processes \cite{Zhang:2014,Aymonier:2018,Rasmussen:2020}, it is necessary to use a \textit{generalized} EoS, such as the Soave--Redlich--Kwong (SRK) \cite{SRK} or Peng--Robinson \cite{PR} EoS.  For a given mixture, these EoSs allow calculation of $\rho(T,P)$, $c_p(T,P)$, and $s(T,P)$.

A similar set of challenges exists for calculation of transport properties such as viscosity $\eta$ and thermal conductivity $\lambda$. The same distinction exists: high-fidelity correlations are available for pure species. For example, for s\cotwo{} the Laesecke--Muzny and Huber correlations for $\eta$ and $\lambda$ are available \cite{laesecke-muzny,huber}. However, once other species are present, generalized models such as the Chung correlation for $\eta$ and $\lambda$ become necessary \cite{chung}.  

A major concern when using these correlations in engineering calculations, including computational fluid dynamics, is how errors in the values in both thermodynamics variables such as $\rho$, $c_p$, and $s$, and transport properties such as $\eta$ and $\lambda$ impact results. Because the variation in properties makes even conduction a highly non-linear process near the critical point \cite{Martin:2019}, small errors in $\lambda$ can lead to large differences in predicted behavior.  Similarly, even when the transport properties are calculated accurately, differences between different equations of state can lead to large differences in flow structure and heat transfer for engineering systems \cite{Rasmussen:2021}. Variations in thermodynamic properties will also impact reacting flow calculations, leading to differences in observed peak temperatures and pollutant formation \cite{DeFrahan:2023}.

\begin{figure}[hbt!]
   \centering
    \includegraphics[width=.5\textwidth]{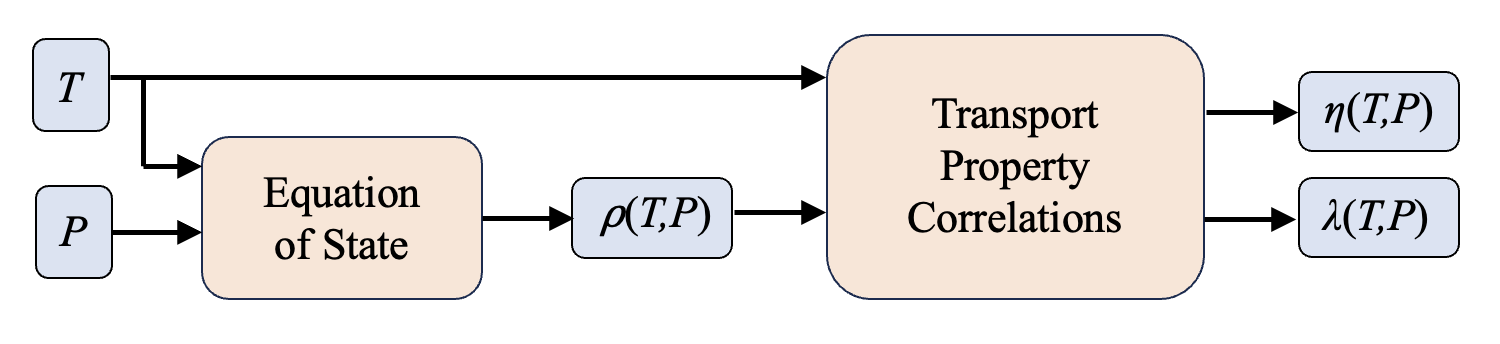}
    \caption{Process flow for calculation of transport properties.}
    \label{fig:EoS}
\end{figure}

An additional potential source of error for calculation of transport properties for supercritical fluids is that they are not derived or calculated as functions of $T$ and $P$, i.e. $\eta = \eta(T,P)$ and $\lambda = \lambda(T,P)$. Instead they are written as functions of temperature and density: $\eta = \eta(T,\rho)$ and $\lambda = \lambda(T,\rho)$. This convention allows the use of a generalized correlations at phase transitions. But in practice, it means that for any calculation of transport properties using these correlations, the calculated values of $\eta$ and $\lambda$ will be influenced by the need to first calculate the density from $T$ and $P$, i.e. $\rho = \rho(T,P)$, using an EoS, as shown in Figure~\ref{fig:EoS}.

The net result is that the error in density created by the EoS propagates into the final values of the transport properties. This has several potential consequences:
\begin{itemize}
    \item {\it Physical errors} in values of $\eta$ and $\lambda$ may lead to inaccurate numeric results. Because flow structure and turbulent flow regimes are often a function of Reynolds number, which is inversely proportional to $\eta$, errors in transport properties may substantially change overall results.
    \item {\it Numeric stability and efficiency} of calculations may be impacted by variations in $\eta$, since CFL conditions for computational fluid dynamics are based on $\eta$.  This is a particular concern in simulation of s\cotwo{} systems, where the combination of stability criteria and the computational costs of complex equations of state can lead to wall times 3 times as long as ideal gas calculations. \cite{DeFrahan:2023,ameli:2018}
    \item {\it Artificial zero or negative values} of $\eta$ and $\lambda$ may be theoretically possible if the combination of the error in $\rho$ and the sensitivity to such error of correlations for $\eta$ and $\lambda$ is large enough.
\end{itemize}

This paper will focus on understanding the relationship between EoS selection and accurate prediction of transport properties. Section~\ref{sec:math_background} introduces terminology and describes an approach for decomposing the total error in viscosity or thermal conductivity into the error due to the choice of correlation and the error due to the choice of EoS. Section~\ref{sec:viscosity} uses the computation of the viscosity of s\cotwo{} as a test case to demonstrate this approach. Section~\ref{sec:viscosity} also describes the impact of viscosity and density error on the Reynolds number and on numeric stability. Section~\ref{sec:conductivity} presents similar results for the decomposition of thermal conductivity errors.  Section~\ref{sec:cross_species} compares the results for supercritical \cotwo{} to similar results for supercritical \ntwo{} and supercritical \chfour{}.

For all species and properties, the generalized EoS will be the SRK EoS and the generalized transport property correlation will be the Chung correlation. However, the approach developed in this paper is intended to be applicable to a broader set of correlations and equations of state.
%%%%%%%%%%%%%%%%%%%%%%%%%%%%%%%%%%%%%%%%%%%%%%%
% section: end

% section: begin
%%%%%%%%%%%%%%%%%%%%%%%%%%%%%%%%%%%%%%%%%%%%%%%
\section{Quantifying sensitivities of transport equations to EoS errors}
\label{sec:math_background}

\subsection{Introducing models for computing density, viscosity and thermal conductivity}

Simulations of supercritical fluids can take multiple approaches to calculating density, viscosity, and thermal conductivity. The highest fidelity can be expected for simulations of a pure species when the equation of state used is designed specifically for that species.  For convenience, this density will be described as a \textit{reference density}, or $\densref(P,T)$. The viscosity and thermal conductivity will be most accurately represented if they are computed using species-specific correlations based on $\densref(P,T)$:

\begin{equation}
    \viscref = \viscref(T,\densref(T,P)),
\end{equation}
\begin{equation}
    \condref = \condref(T,\densref(T,P)),
\end{equation}

Commonly used models for $\text{CO}_2$, $\text{N}_2$, and $\text{CH}_4$ are listed in Table 1.

\begin{table}[hbt!]

\caption{\label{tab:table1} Species-specific property models for representative supercritical fluids}

\centering
\begin{tabular}{lccc}
\hline
Species & $\densref$ model & $\viscref$ model & $\condref$ model\\
    \hline
    $\text{CO}_2$ 
        & Span-Wagner (SW) \cite{span-wagner} 
        & Laesecke-Muzny\cite{laesecke-muzny} 
        & Huber \cite{huber} \\
    $\text{N}_2$  
        & Span \cite{span2000} 
        & Lemmon \cite{lemmon2004} 
        & Lemmon \cite{lemmon2004} \\
    $\text{CH}_4$ 
        & Setzmann \cite{setzmann1991} 
        & Quinones \cite{quinones2006} 
        & Friend \cite{friend1989} \\
    \hline
\end{tabular}
\end{table}

However, as mentioned earlier, there are many situations where it is necessary to calculate the properties of mixtures. For these systems, the density must be calculated using a generalized equation of state, such as the SRK equation, to obtain the \textit{generalized density}, or $\densgen(P,T)$. The viscosity and thermal conductivity are then calculated based on the temperature and $\densgen(P,T)$ using correlations such as the Chung correlation.

\begin{equation}
    \viscgen = \viscgen(T,\densgen(T,P)),
\end{equation}
\begin{equation}
    \condgen = \condgen(T,\densgen(T,P)),
\end{equation}

Finally, a \textit{test} model, where a generalized correlation is used to calculate the viscosity and thermal conductivity, but the reference density is used in the calculation, is useful in understanding potential sources of error in these calculations:
\begin{equation}
    \vischyb = \viscgen(T,\densref(T,P)),
\end{equation}
\begin{equation}
    \condhyb = \condgen(T,\densref(T,P)).
\end{equation}

\subsection{Error and Error Decomposition}
Because generalized models for density, viscosity, and thermal conductivity are critical to a range of engineering calculations, it is useful to quantify both the magnitude and sources of error in their use. Quantifying errors within mixtures is challenging. However, a first estimate of errors in a mixture can be obtained by comparing the densities, viscosities, and thermal conductivities computed from the generalized property correlations with those computed from the reference correlations for each species in the mixture. For a given species, these errors can be written as:

\begin{equation}
    \densERRtot(T,P) = \densgen(T,P) - \densref(T,P),
\end{equation}
\begin{equation}
    \label{eq:total_error_viscosity}
    \viscERRtot(T,P) = \viscgen(T,\densgen(T,P)) - \viscref(T,\densref(T,P)),
\end{equation}
\begin{equation}
    \label{eq:total_error_conductivity}
    \condERRtot(T,P) = \condgen(T,\densgen(T,P)) - \condref(T,\densref(T,P)).
\end{equation}

For both viscosity and thermal conductivity, there are two sources of error. The first source of error comes from using a generalized correlation to calculate transport properties instead of a reference correlation. These errors can be written as:
\begin{equation}
    \viscERRcorr(T,P) = \viscgen(T,\densref(T,P)) - \viscref(T,\densref(T,P)),
\end{equation}
\begin{equation}
    \condERRcorr(T,P) = \condgen(T,\densref(T,P)) - \condref(T,\densref(T,P)).
\end{equation}

A second error is introduced by the fact that the density calculated using a generalized EoS is not the same as that calculated by the reference EoS:
\begin{equation}
    \viscERReos(T,P)
        = \viscgen(T,\densgen(T,P)) 
        - \viscgen(T,\densref(T,P)),
\end{equation}
\begin{equation}
    \condERReos(T,P)
        = \condgen(T,\densgen(T,P)) 
        - \condgen(T,\densref(T,P)).
\end{equation}
This error can be rewritten using Taylor series expansion of $\viscgen(T,\densgen(T,P))$ and $\condgen(T,\densgen(T,P))$ about $\densref$:
\begin{equation}
    \viscERReos(T,P) 
        = \densERRtot(T,P)
            \left.
            \frac{\partial \viscgen}{\partial \rho} 
            \right|_{(T,\densref)}
        + \frac{(\densERRtot(T,P))^2}{2} 
            \left.
            \frac{\partial^2\viscgen}{\partial\rho^2} 
            \right|_{(T,\densref)}
        + \dots,
\end{equation}
\begin{equation}
    \condERReos(T,P)  
        = \densERRtot(T,P) 
            \left.
            \frac{\partial \condgen}{\partial \rho} 
            \right|_{(T,\densref)}
        + \frac{(\densERRtot(T,P))^2}{2}
            \left.
            \frac{\partial^2\condgen}{\partial\rho^2} 
            \right|_{(T,\densref)}
        + \dots.
\end{equation}
A first-order approximation of this error can be written in terms of sensitivity coefficients $S_1$ and $S_2$, where $S_1$ is the partial derivative of the generalized viscosity and $S_2$ is the partial derivative of the generalized thermal conductivity, each with respect to density:
\begin{equation}
    \viscERRfirst(T,P) = \densERRtot(T,P) \; S_1(T,P),
\end{equation}
\begin{equation}
    \condERRfirst(T,P) = \densERRtot(T,P) \; S_2(T,P).
\end{equation}

These two distinct sources of error can be summarized by the equations:
\begin{equation}
    \viscERRtot(T,P) 
        = \viscERRcorr(T,P) 
        + \viscERRfirst(T,P)
        + \viscERRsecond(T,P),
\end{equation}
\begin{equation}
    \condERRtot(T,P) 
        = \condERRcorr(T,P) 
        + \condERRfirst(T,P)
        + \condERRsecond(T,P),
\end{equation}
where $\viscERRsecond$ and $\condERRsecond$ represent the higher order terms of the respective Taylor series expansions. It can be shown that $\viscERRsecond$ and $\condERRsecond$ are small relative to the other sources of error, so that $\viscERRtot$ is well-approximated by the sum of $\viscERRcorr$ and $\viscERRfirst$, and similar for $\condERRtot$. 

Finally, these error decompositions can be nondimensionalized by dividing each term by the appropriate reference value:
\begin{equation}
    \denserrtot = \frac{\densERRtot}{\densref},
\end{equation}
\begin{equation}
    \viscerrtot 
        = \frac{\viscERRtot}{\viscref}
        = \viscerrcorr
        + \viscerrfirst 
        + \viscerrsecond,
\end{equation}
\begin{equation}
    \conderrtot 
        = \frac{\condERRtot}{\condref}
        = \conderrcorr
        + \conderrfirst 
        + \conderrsecond.
\end{equation}

%%%%%%%%%%%%%%%%%%%%%%%%%%%%%%%%%%%%%%%%%%%%%%%
% section: end

% section: begin
%%%%%%%%%%%%%%%%%%%%%%%%%%%%%%%%%%%%%%%%%%%%%%%
\section{Supercritical \cotwo{} Viscosity Results: Accuracy, Kinematic Viscosity, and Stability}
\label{sec:viscosity}
\subsection{Viscosity Error Analysis}
In many cases of engineering interest, including oxycombustion in the Allam-Fetvedt cycle and the Venusian atmosphere, s\cotwo{} is the dominant component in a mixture, with relatively small amounts of trace gases. Therefore, carrying out an error analysis for s\cotwo{} is likely to give an approximation of the error for the mixture. In these cases, reference viscosity values computed by $\viscref(T,\densref(T,P))$ using the Span--Wagner EoS and the Laesecke--Muzny correlation can be compared to those values computed by $\viscgen(T,\densgen(T,P))$ using the generalized SRK EoS and the Chung viscosity correlation. 

The supercritical region, where both pressure and temperature are greater than the critical point $T_c = 304.128$ K and $P_c = 72.808$ atmospheres, is of particular interest. All results are show in the region $\Omega = [T^*, 3T^*] \times [P^*, 3P^*]$ where $T^* = T / T_c$ and $P^* = P / P_c$. 

\begin{figure}[!hbt]
    \centering
    \panel{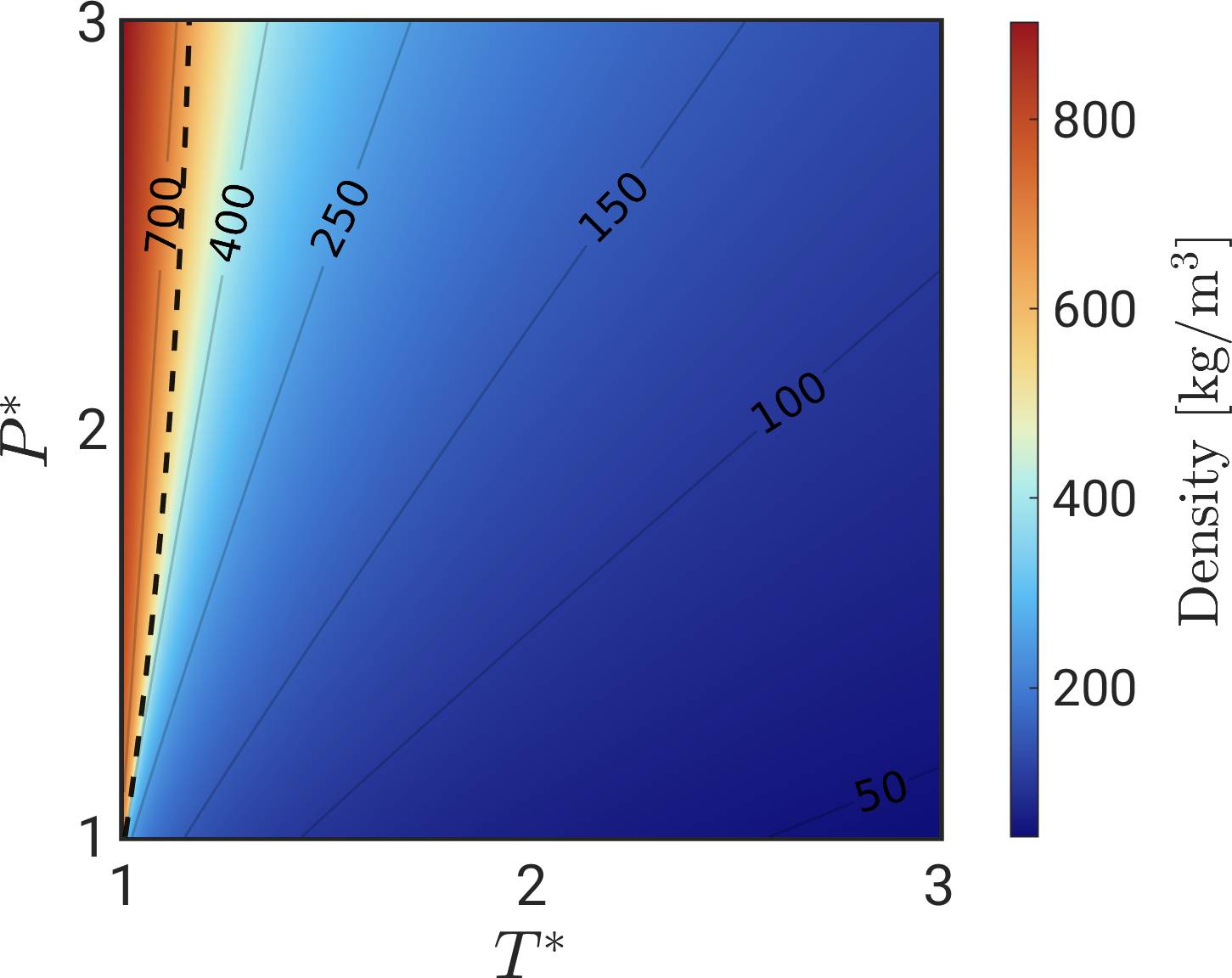}{(a) \cotwo{}, $\densref$}
    \hfill
    \panel{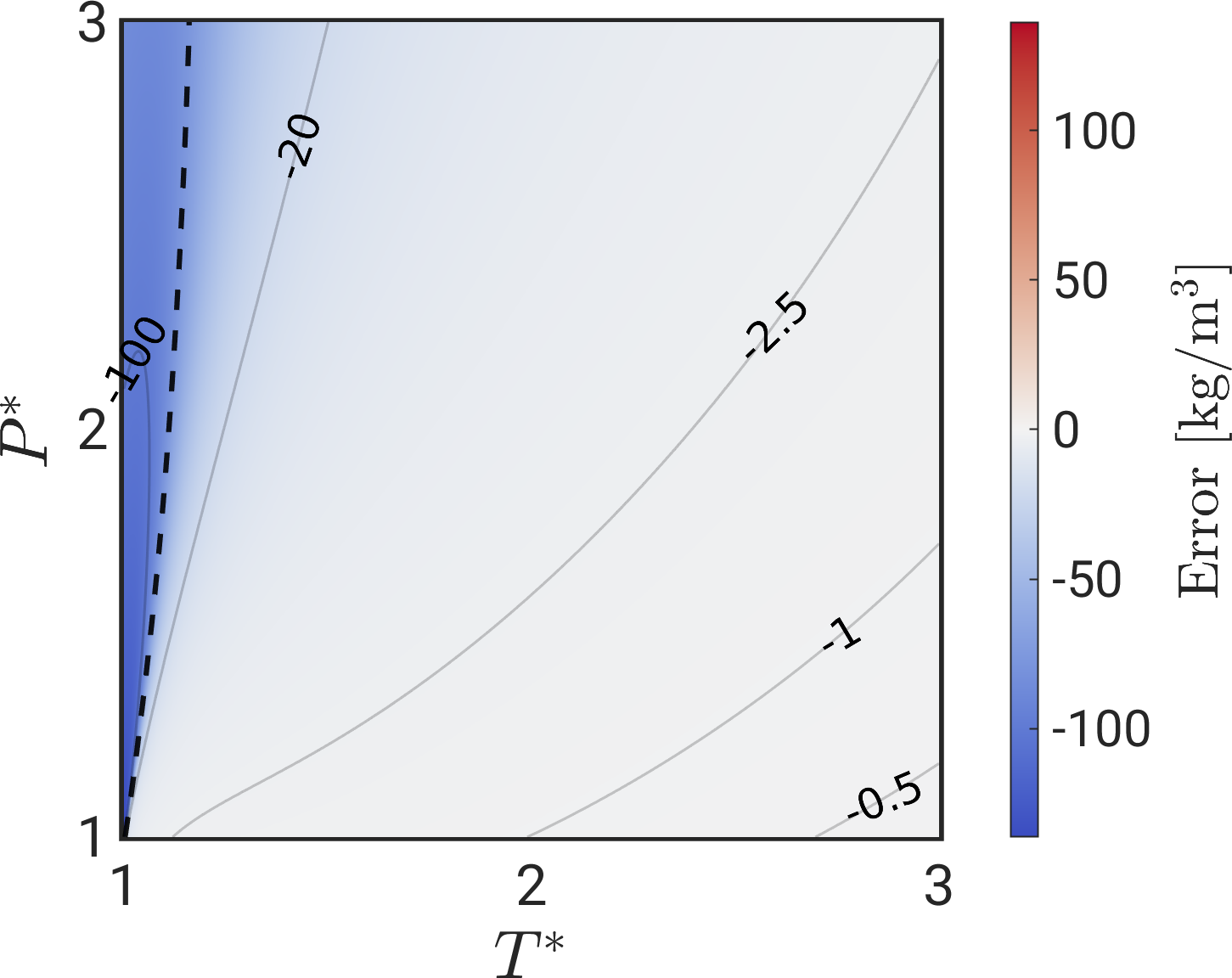}{(b) \cotwo{}, $\densERRtot$}
    \hfill
    \panel{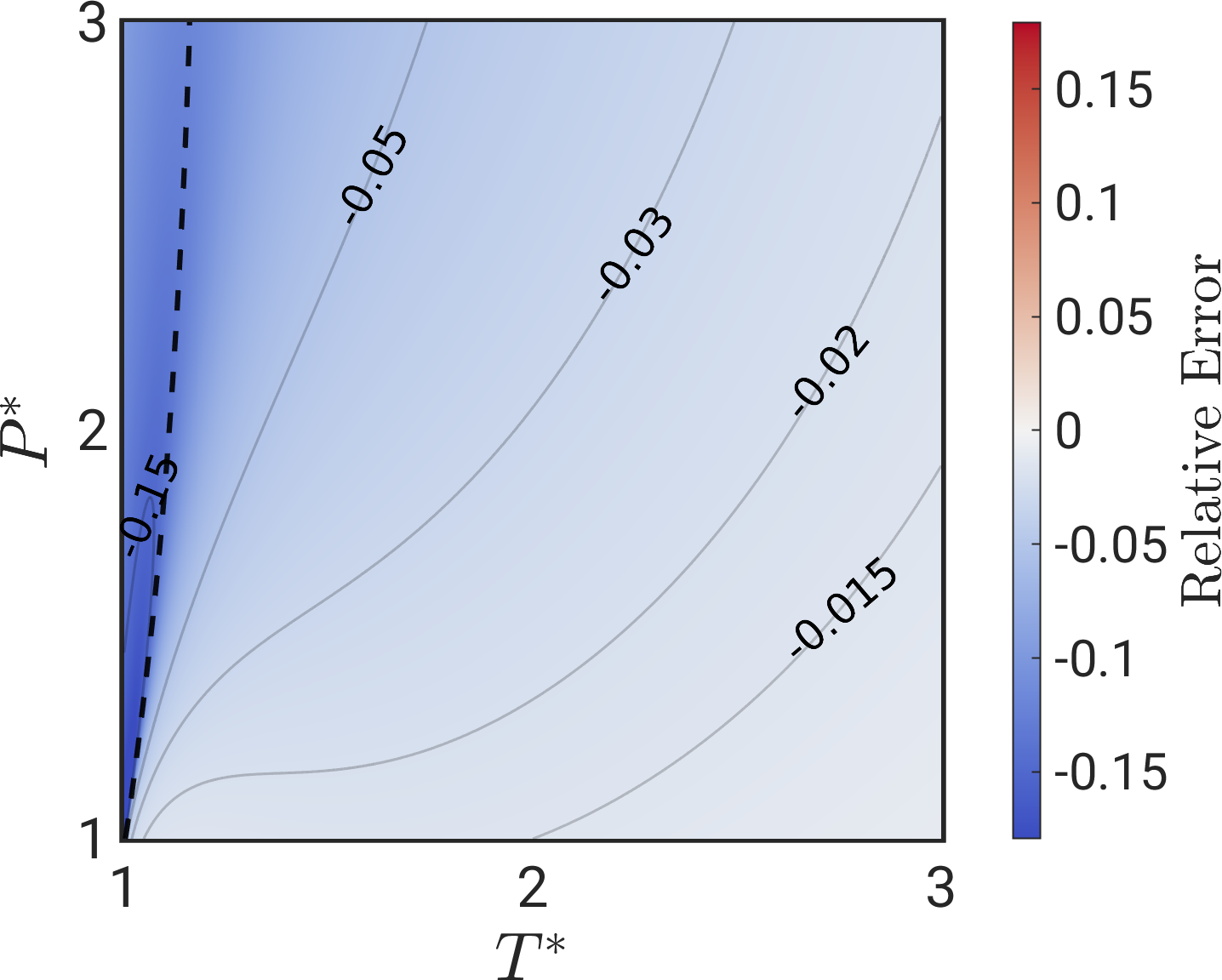}{(c) \cotwo{}, $\denserrtot$}
    \caption{\cotwo{} density results. The figures shows (a) reference values for $\densref$ as computed by the SW EoS, (b) the signed error $\densERRtot$, and (c) the relative error $\denserrtot$. The dashed line is the Widom line.}
    \label{fig:CO2_density}
\end{figure}

To begin, the reference density computed by the SW EoS is compared against the density computed by the SRK EoS. Figure~\ref{fig:CO2_density} shows the reference density $\densref$, the total error $\densERRtot$, and the relative error $\denserrtot$, defined as the local total error divided by the local reference density. Throughout $\Omega$, the SRK EoS underestimates the density by as much as $-22.89\%$ relative to the SW EoS, with the most significant errors occurring in the liquid-like region, i.e. left of the Widom line.

\begin{figure}[!hbt]
    \centering
    \panel{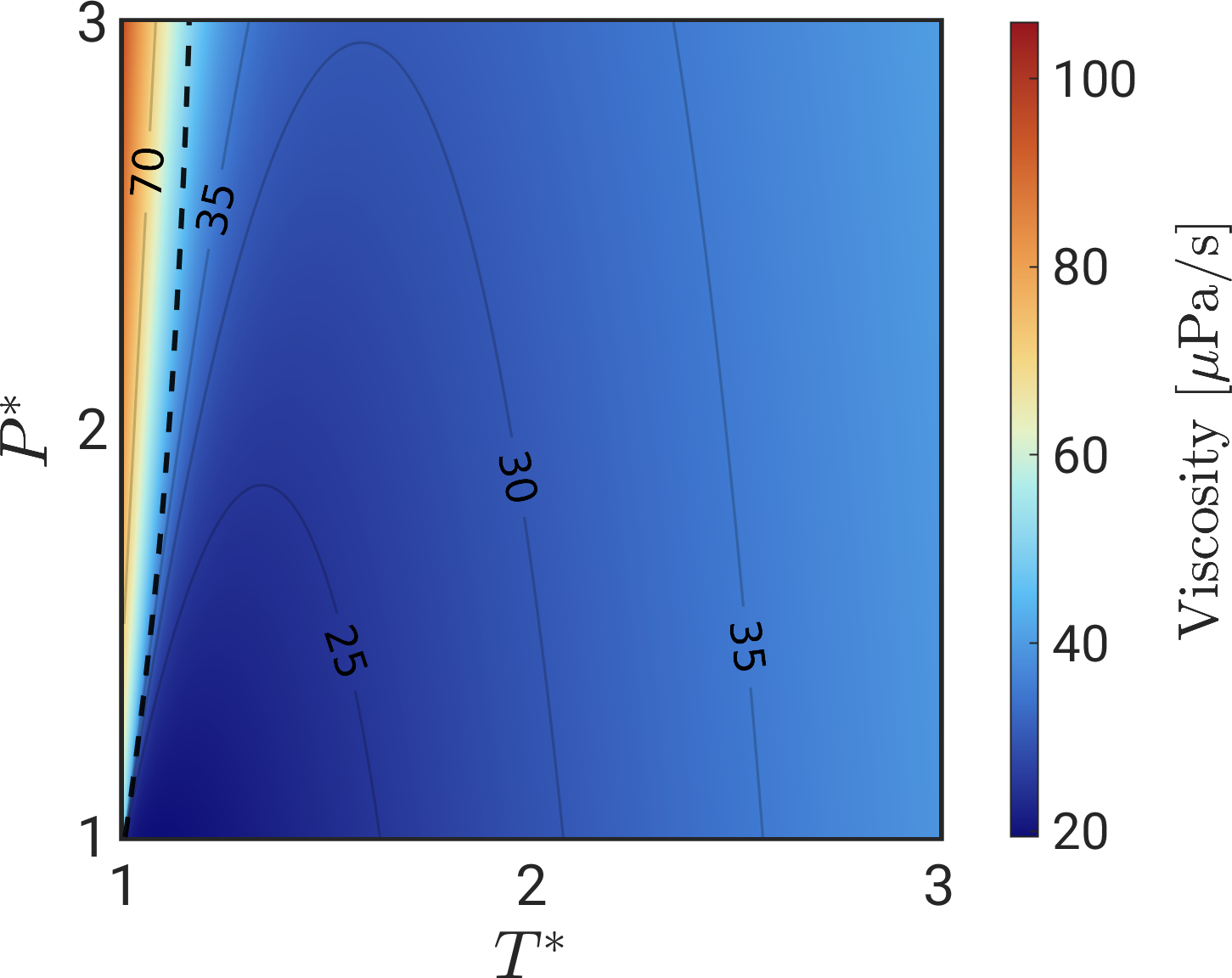}{(a) \cotwo{}, $\viscref$}
    \hfill
    \panel{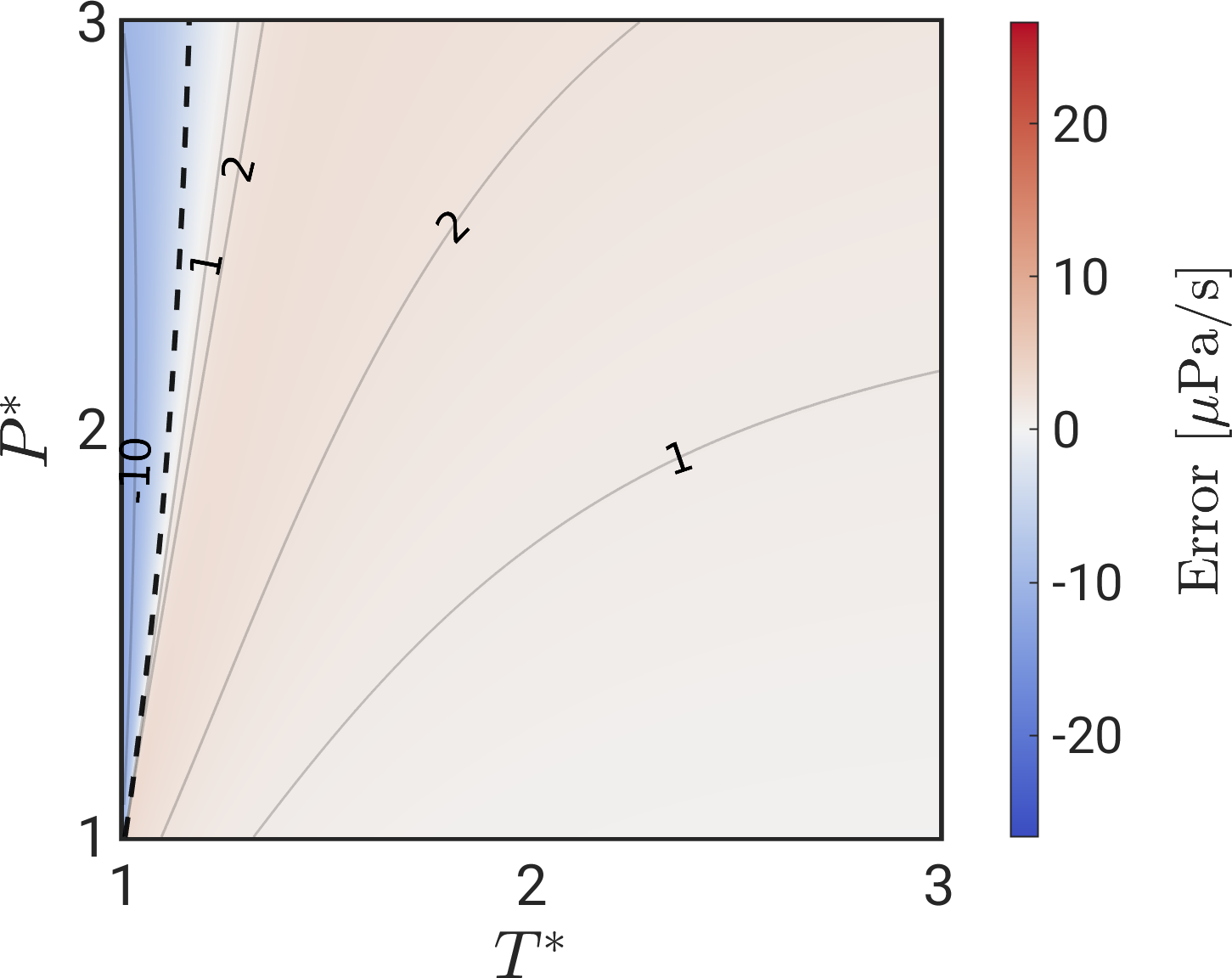}{(b) \cotwo{}, $\viscERRtot$}
    \hfill
    \panel{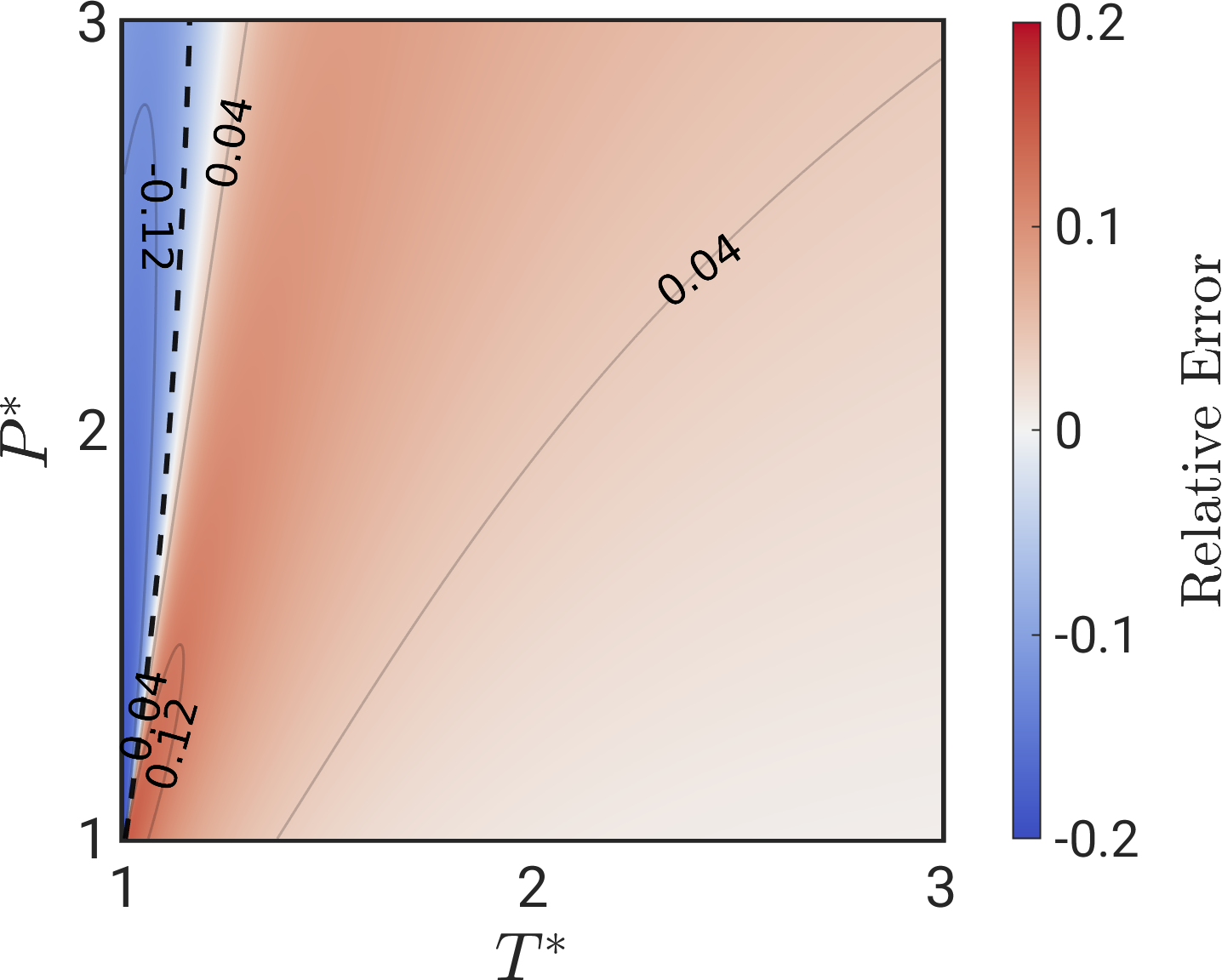}{(c) \cotwo{}, $\viscerrtot$}
    \caption{\cotwo{} viscosity results. The figure shows (a) $\viscref(T,\densref)$ computed by the reference EoS and correlation, (b) the signed error $\viscERRtot$, and (c) the relative error $\viscerrtot$. The dashed line is the Widom line.}
    \label{fig:CO2_viscosity}
\end{figure}

Next, the reference viscosity $\viscref(T,\densref(T,P))$ is compared against the viscosity $\viscgen(T,\densgen(T,P))$ computed by the generalized property correlations. Figure~\ref{fig:CO2_viscosity} presents the reference viscosity $\viscref$ and the errors $\viscERRtot$ and $\viscerrtot$. Viscosity is underestimated by $\viscgen$ by as much as $-18.50\%$ in the liquid-like region of $\Omega$, while $\viscgen$ overestimates viscosity by as much as $14.57\%$ in the gas-like region, with these maximum errors occurring near the critical point. 

Comparing Figures~\ref{fig:CO2_density}(c) and \ref{fig:CO2_viscosity}(c), the region where viscosity is underestimated by the combination of the generalized correlation and the generalized EoS roughly overlaps with the region where density is underestimated by the generalized EoS. By contrast, in the region where viscosity is overestimated, the density errors are small. This pattern reflects the decomposition of the total viscosity error into those errors caused by the use of the generalized correlation itself and those caused by the propagation of density error, as described in Section~\ref{sec:math_background}. In particular, Figures~\ref{fig:CO2_density} and \ref{fig:CO2_viscosity} suggest that the propagation of density error dominates in the liquid-like region and, likewise, that the correlation error itself dominates in the gas-like region. 

\begin{figure}[!hbt]
    \centering
    \panel{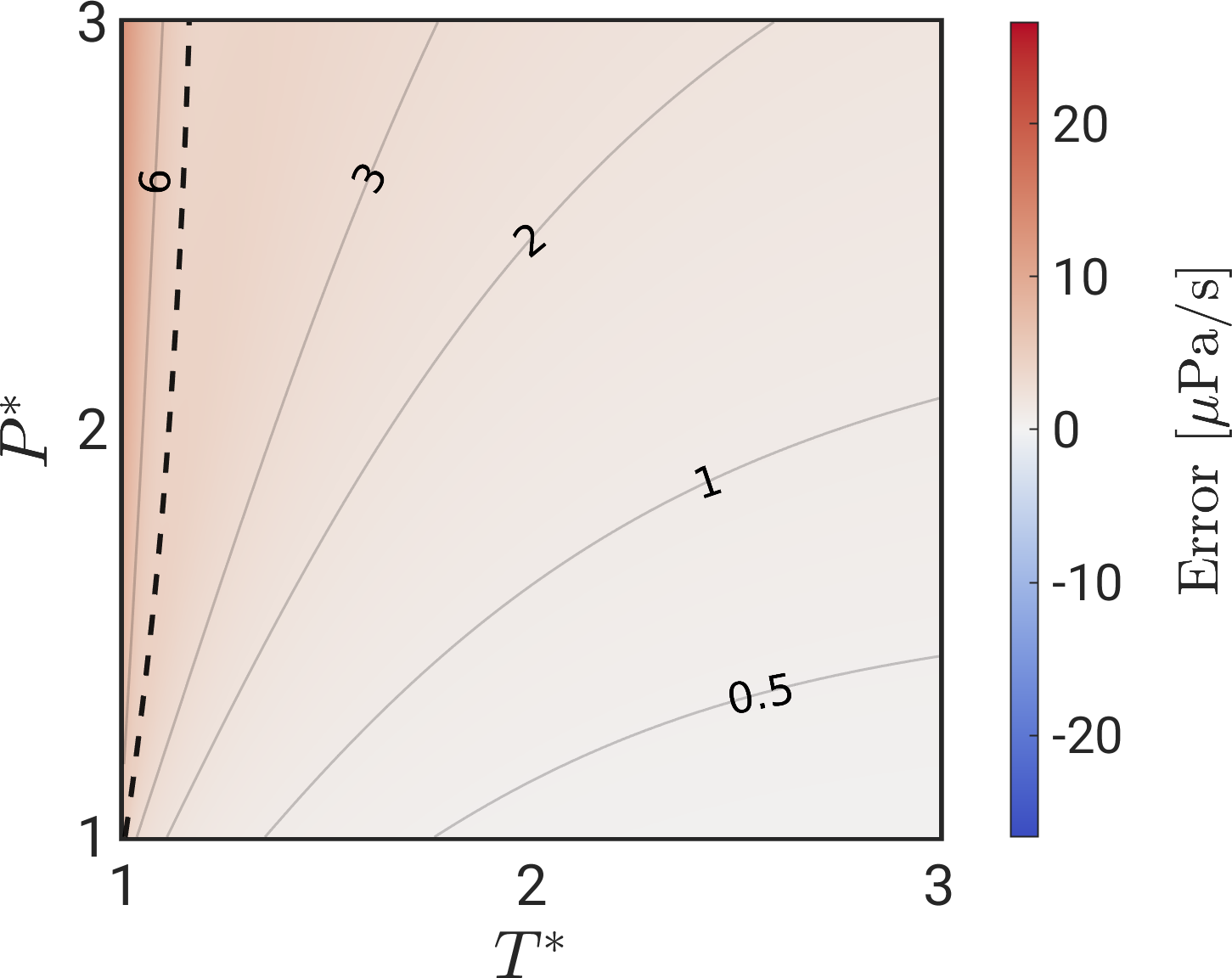}{(a) \cotwo{}, $\viscERRcorr$}
    \hfill
    \panel{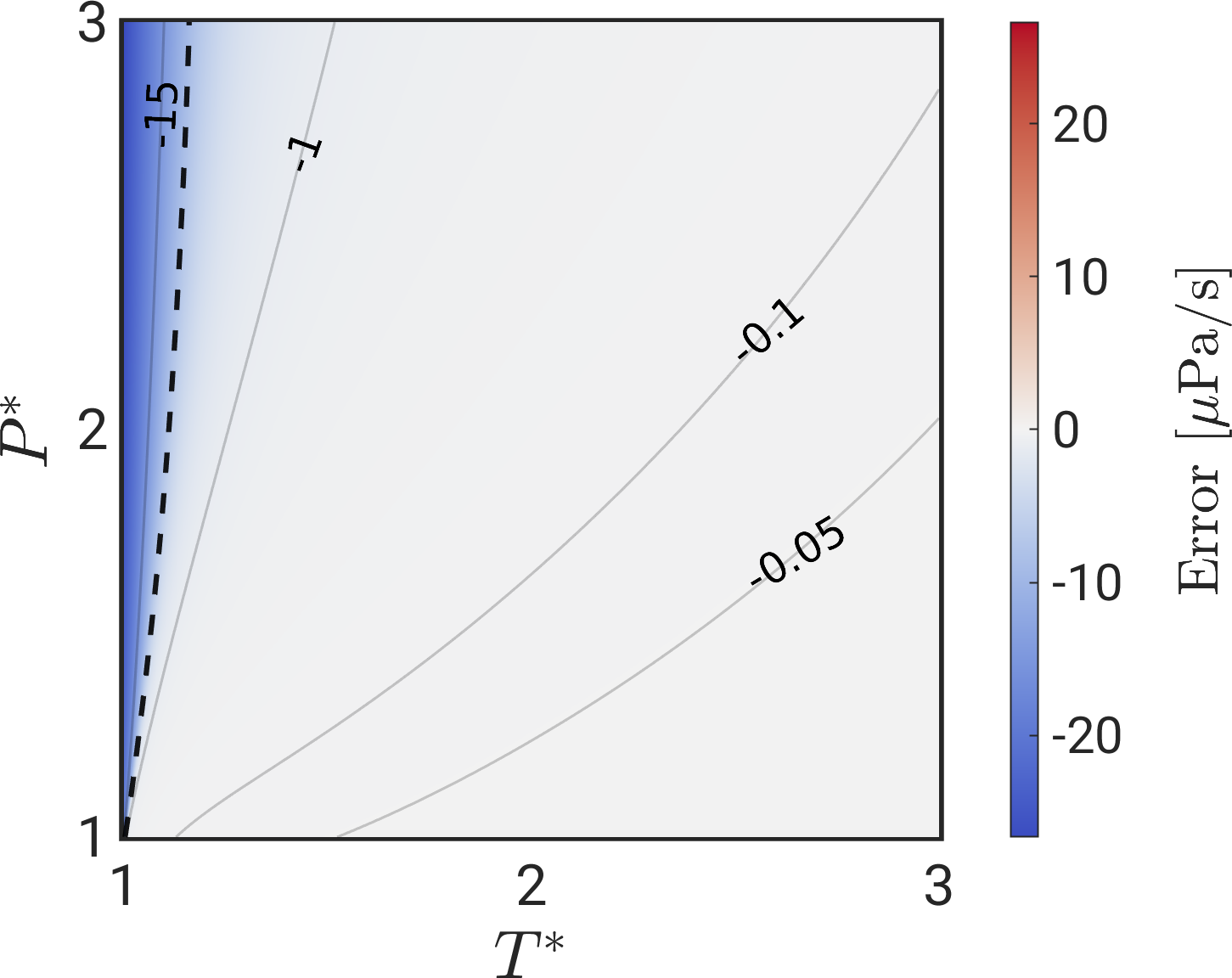}{(b) \cotwo{}, $\viscERRfirst$}
    \hfill
    \panel{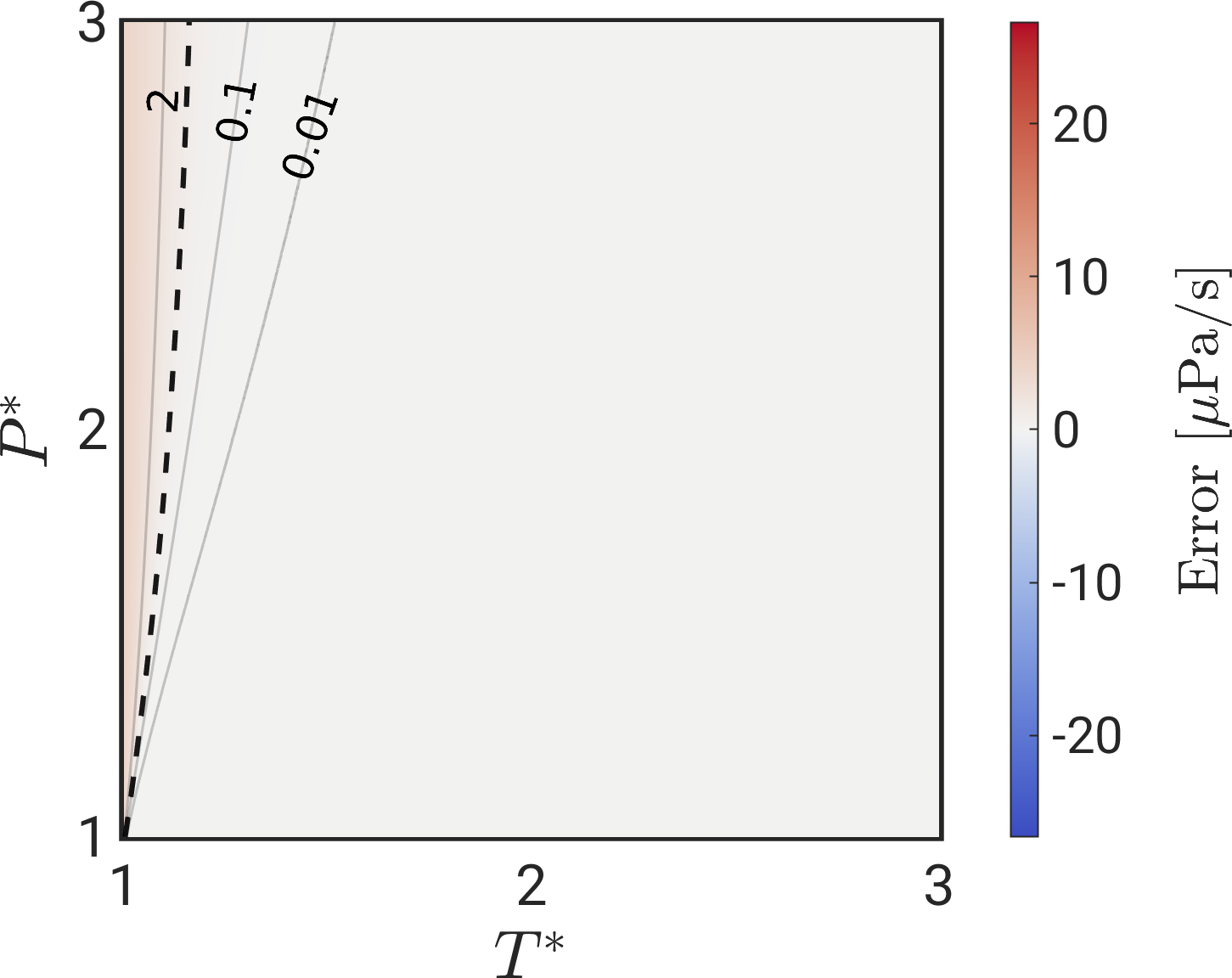}{(c) \cotwo{}, $\viscERRsecond$}
    \vspace{1em} \\
    \hrule \vspace{1em}
    \panel{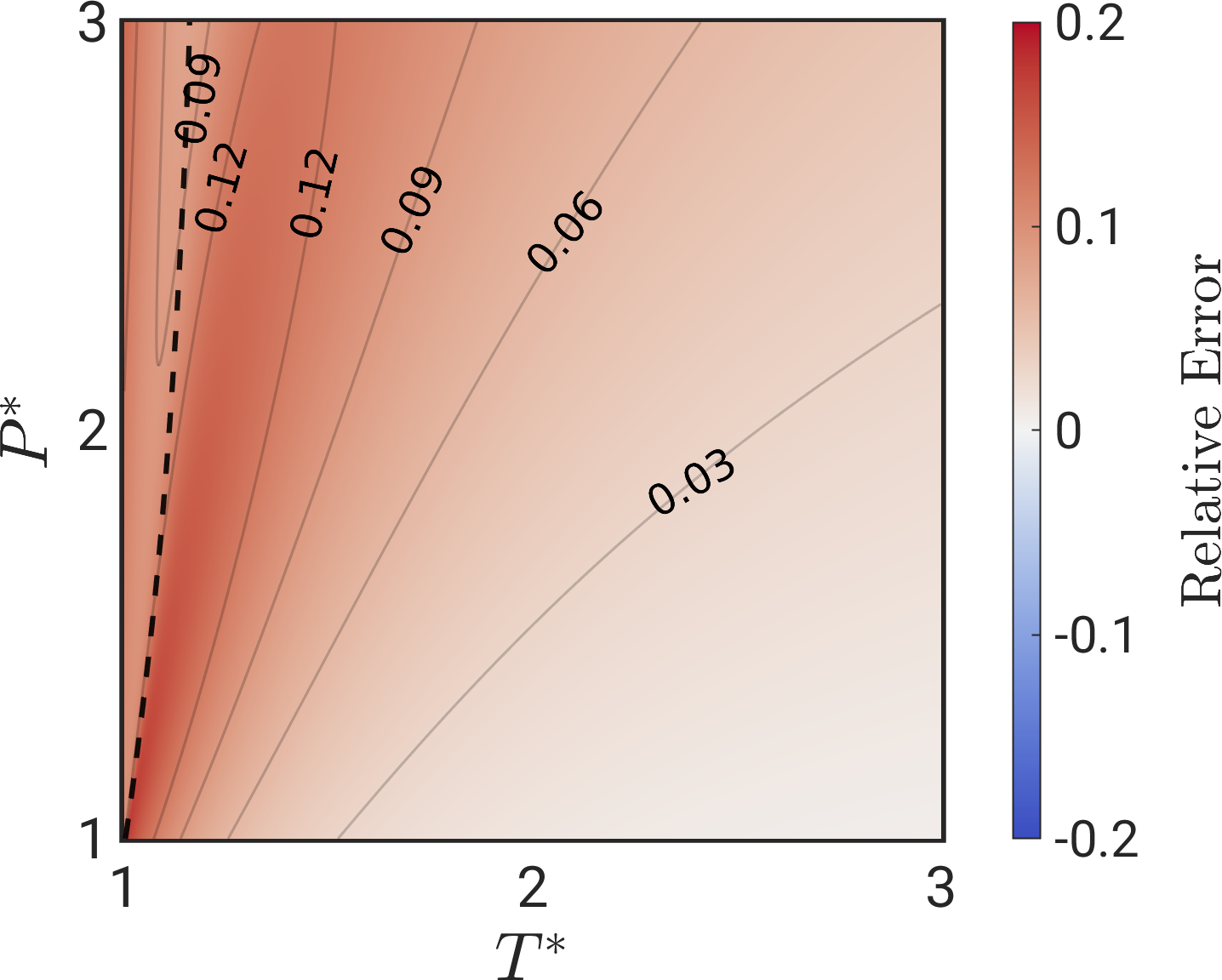}{(d) \cotwo{}, $\viscerrcorr$}
    \hfill
    \panel{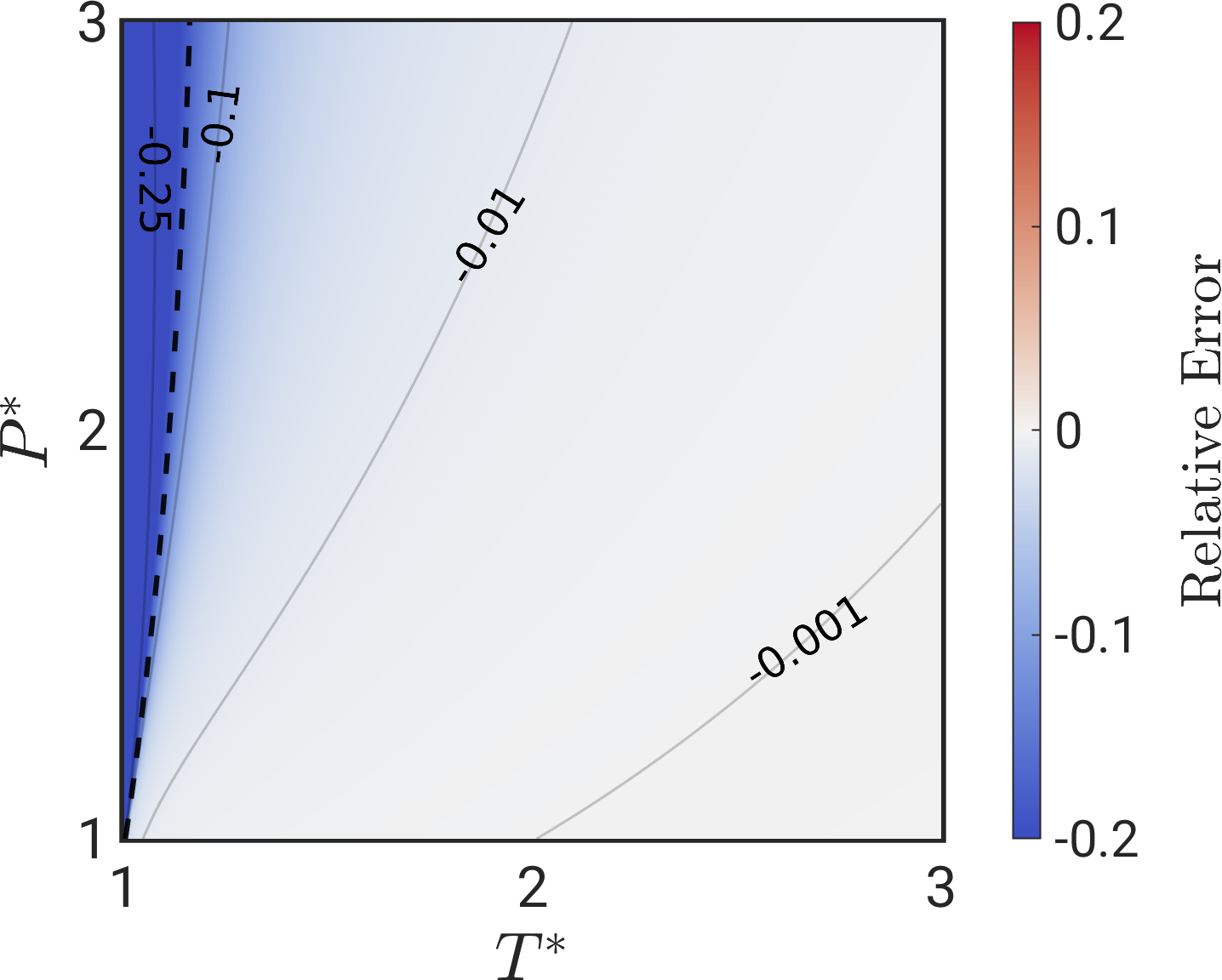}{(e) \cotwo{}, $\viscerrfirst$}
    \hfill
    \panel{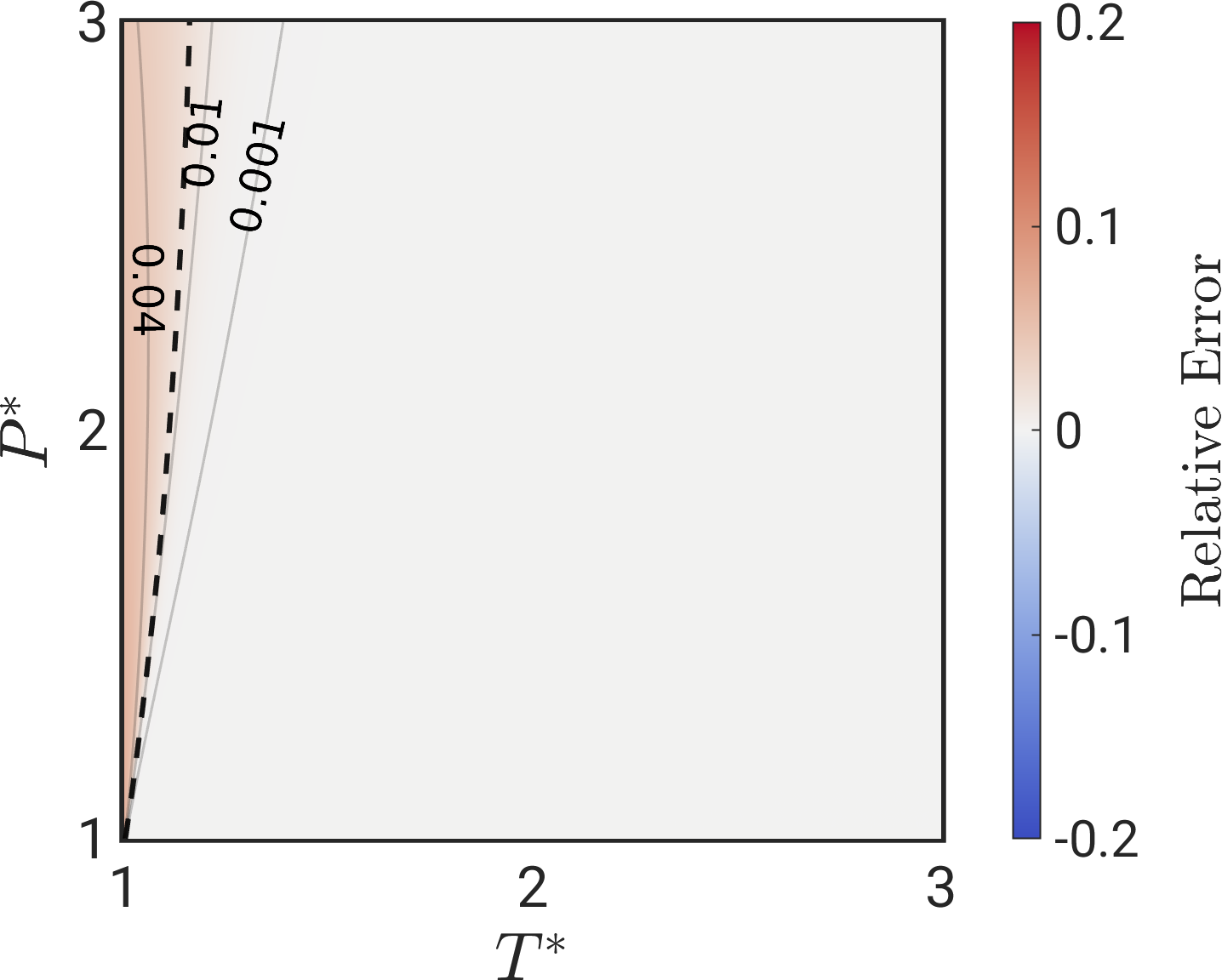}{(f) \cotwo{}, $\viscerrsecond$}
    \caption{\cotwo{} viscosity error decomposition. The top row shows the signed errors (a) $\viscERRcorr = \viscgen(T,\densref) - \viscref(T,\densref)$, (b) $\viscERRfirst =  \densERRtot S_1$, and (c) $\viscERRsecond$. The bottom row shows the corresponding relative errors, i.e. (d) $\viscerrcorr$, (e) $\viscerrfirst$, and (f) $\viscerrsecond$.}
    \label{fig:CO2_viscosity_error_decomposition}
\end{figure}

To demonstrate that this is the case, Figure~\ref{fig:CO2_viscosity_error_decomposition} shows $\viscERRcorr$, $\viscERRfirst$, and $\viscERRsecond$ and $\viscerrcorr$, $\viscerrfirst$, and $\viscerrsecond$. Figure~\ref{fig:CO2_viscosity_error_decomposition}(b) shows that $\viscERRfirst$ contributes strongly to the underestimation of viscosity in the liquid-like region, with relative error as much as $-35.04\%$. Interestingly, Figure~\ref{fig:CO2_viscosity_error_decomposition}(a) shows that the Chung correlation itself tends to overestimate viscosity in this same region, as well as throughout $\Omega$, with the strongest overestimation being $17.62\%$ in the gas-like region near the critical point. However, as expected, the propagation of density error dominates the correlation error in the liquid-like region and this explains the overall underestimation of viscosity there, while in the gas-like region the correlation error dominates leading to the overall overestimation of viscosity there. Lastly, as Figure~\ref{fig:CO2_viscosity_error_decomposition}(b) and (c) shows, the linear propagation term, $\viscERRfirst$, dominates the higher order terms $\viscERRsecond$, though the effect of $\viscERRfirst$ is offset slightly by $\viscERRsecond$, which has maximum relative value of $6.51\%$ occurring in the liquid-like region.

\begin{figure}[!hbt]
    \centering
    \panel{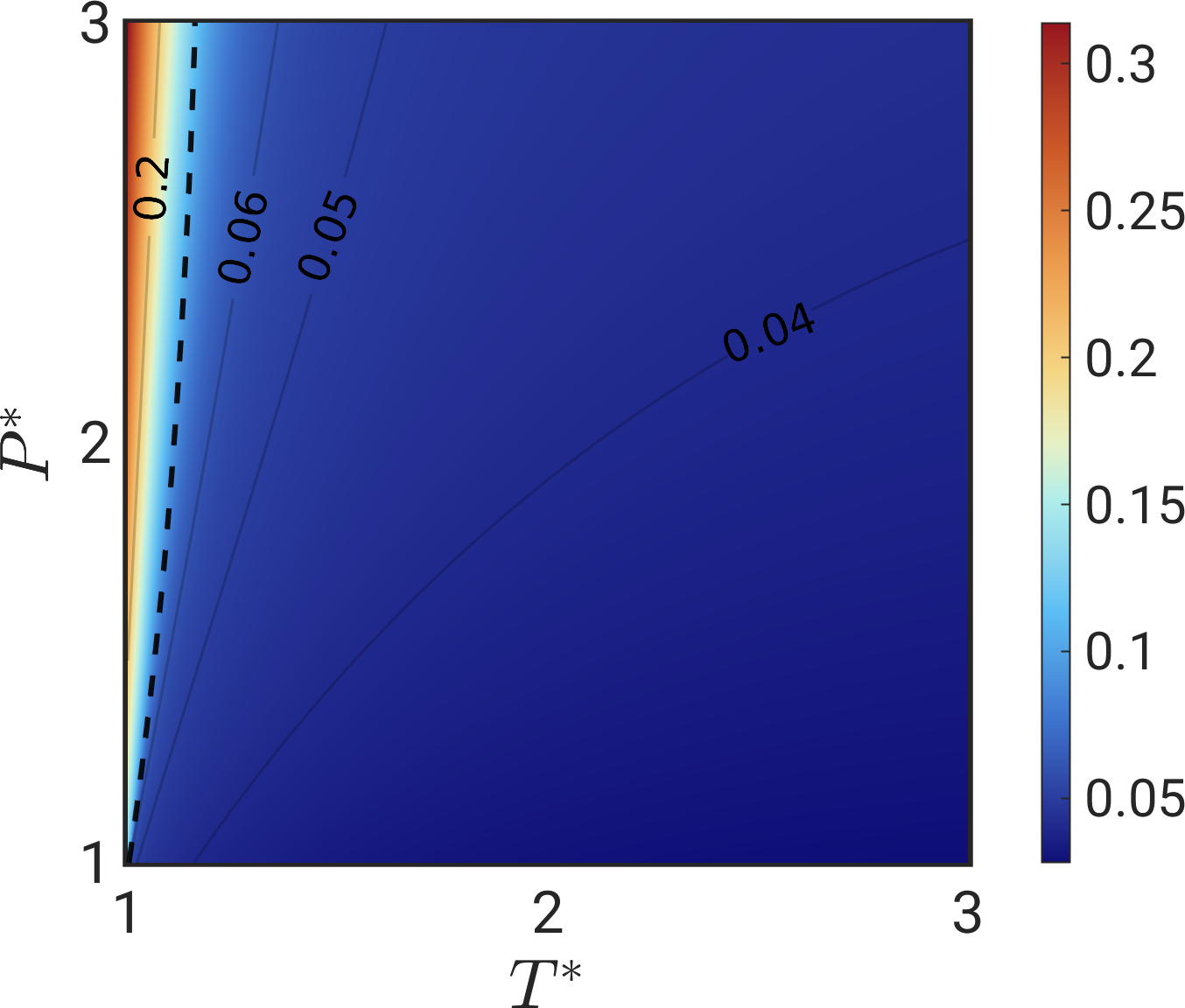}{(a) \cotwo{}, $S_1$}
    \hfill
    \panel{images/2d_results_NEW/CO2_2d_dens_ERR.png}{(b) \cotwo{}, $\densERRtot$}
    \hfill
    \panel{images/2d_results_NEW/CO2_2d_visc_ERR_eos.png}{(c) \cotwo{}, $\viscERRfirst$}
    \caption{\cotwo{} first order error factors for viscosity. Shown are (a) $S_1 = \left. \tfrac{\partial \viscgen}{\partial \rho} \right|_{(T, \densref)}$, (b) $\densERRtot$, and (c) $\viscERRfirst$.}
    \label{fig:CO2_viscosity_first_order}
\end{figure}

The sensitivity term $S_1$ quantifies the propagation of density error. Figure~\ref{fig:CO2_viscosity_first_order} shows the quantity $S_1$. Recalling that $\viscERRfirst = \densERRtot \, S_1$, the plots of $\densERRtot$ and $\viscERRfirst$, which originally appeared in Figures~\ref{fig:CO2_density} and \ref{fig:CO2_viscosity_error_decomposition} respectively, are shown again for ease of comparison. As Figure~\ref{fig:CO2_viscosity_first_order} shows, $S_1$ has the largest magnitude in the liquid-like region, meaning that the density error is largest in the same region where the generalized correlation is most sensitive to these errors. These two factors together help explain the dominance of $\viscERRfirst$ in this region, which leads to the underestimation of $\viscgen$ in this region.

%

%Finally, Figure~\ref{fig:CO2_viscosity_error_summed} shows $\viscERRtot$ against $\viscERRcorr + \viscERRfirst$. This result shows that the decomposition $\viscERRcorr + \viscERRfirst$ provides a reasonable first-order approximation of $\viscERRtot$ and, in particular, provides useful separation between the portion of the error in viscosity attributable to the Chung correlation and the portion arising from propagated density errors caused by the SRK EoS. Figure~\ref{fig:CO2_viscosity_error_summed} further shows that $\viscERRfirst$ provides a useful first-order approximation of the effect of the propagation of density error in the Chung correlation.

To summarize, in the case of s\cotwo{}, the Chung correlation overestimates viscosity and this overestimation is most pronounced in the liquid-like region. However, the SRK EoS underestimates density which contributes to an underestimation of viscosity and this effect is most pronounced in the liquid-like region. The combined effect is an overestimation of viscosity in the gas-like region, where the error from the Chung correlation dominates, and an underestimation of viscosity in the liquid-like region, where the propagation of density error due to the SRK EoS dominates. 

Other EoSs, such as the Peng--Robinson EoS, will have different density errors, which may or may not have the same overall pattern. However, since the sensitivity factor $S_1$ is due to the Chung correlation itself, and is independent of the choice of EoS, this term is a good indicator of the sensitivity of the Chung correlation to errors in density.

%%%
\subsection{Impact on Kinematic Viscosity and Reynolds Number}
The kinematic viscosity $\nu$, defined as the ratio of the viscosity to the density, is critical to determining the Reynolds Number $Re_L$ of any fluid system:
\begin{equation}
    \\Re_L = \frac{\rho  U L}{\mu}= \frac{U L}{\nu},
\end{equation}
where $U$ is the characteristic velocity of the system, and $L$ is the characteristic length.

The kinematic viscosity $\nu_\text{ref}$ based on the reference density and viscosity will be given by:
\begin{equation}
    \nu_\text{ref} = \frac{\viscref(T,\densref(T,P))}{\densref(T,P))},
\end{equation}
while the kinematic viscosity computed from the generalized density and viscosity is given by:
\begin{equation}
    \nu_\text{gen} = \frac{\viscgen(T,\densgen(T,P))}{\densgen(T,P))}.
\end{equation}
The ratio of these two values will determine the impact on the Reynolds number:
\begin{equation}
    \frac{\nu_\text{ref}} {\nu_\text{gen}}= \frac{\viscref(T,\densref(T,P))}{\viscgen(T,\densgen(T,P))}\frac{\densgen(T,P)}{\densref(T,P)} = \frac{Re_{L,\text{gen}}}{Re_{L,\text{ref}}}.
\end{equation}

\begin{figure}[!hbt]
    \centering
    \panel{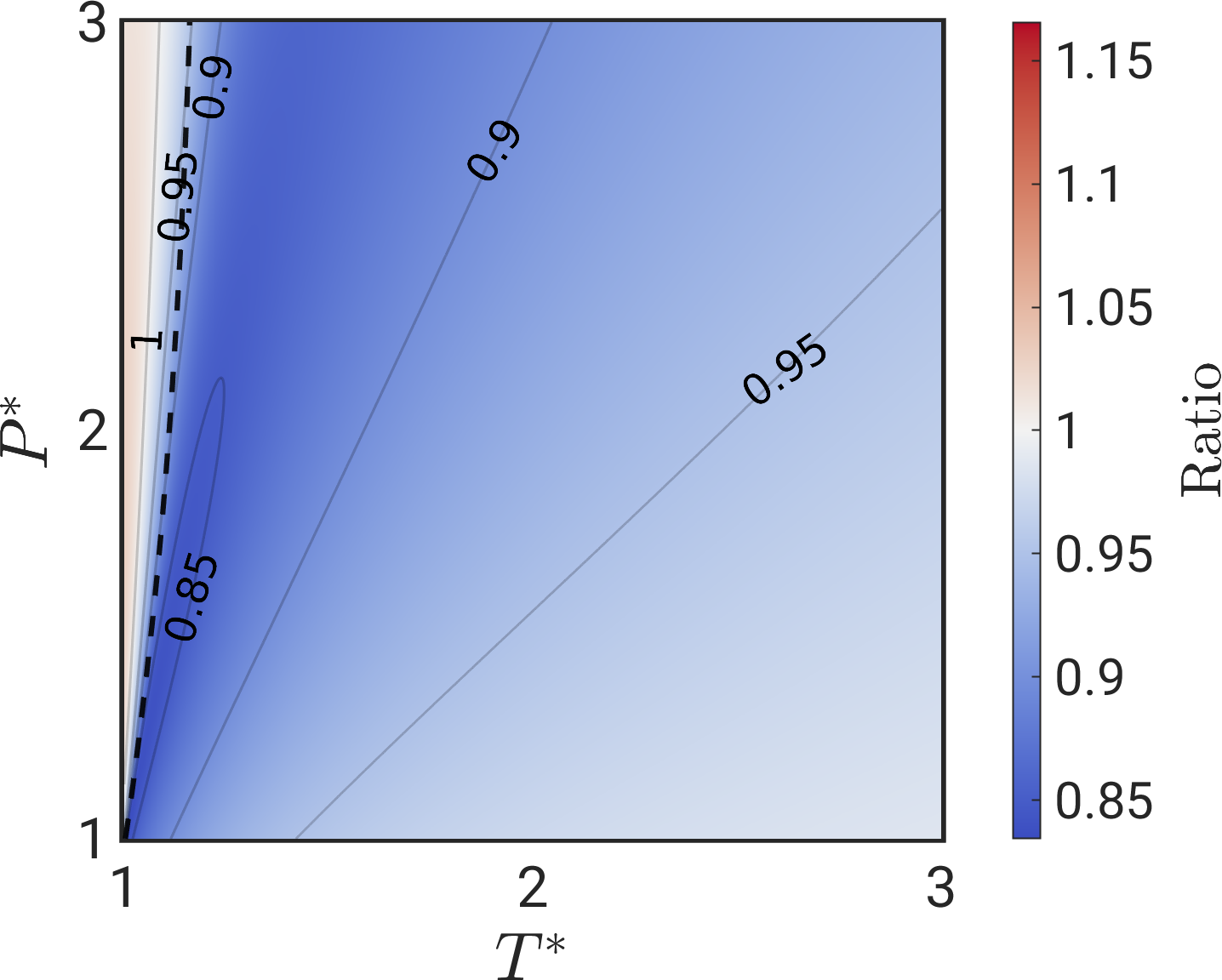}{(a) \cotwo{}, $\nu_\text{ref} / \nu_\text{gen}$}
    \caption{\cotwo{} ratio of kinematic viscosities.}
    \label{fig:CO2_kinematic_viscosity}
\end{figure}
%

%%%
\subsection{Impacts on Numeric Stability}
\label{sec:viscosity_stability}
Numeric simulations of supercritical fluids are computationally more expensive than simulations using the ideal gas equations. This reflects both the increased computational costs of calculating fluid properties for supercritical fluids, and the need to use relatively conservative stability criteria.  Both reacting \cite{DeFrahan:2023} and non-reacting \cite{ameli:2018} simulations of s\cotwo{} flows showed that the wall time needed for the same configuration increased by a factor of 3 when supercritical equations of state and transport property models were used. 

The stability criteria for numeric simulations often depend on fluid properties. For instance, the maximum allowable time step for the Marker-and-Cell (MAC) methods \cite{Harlow:1965}, on which multiple computational fluid dynamics algorithms are built, is given by 
\begin{equation}
    \label{eqn:stability}
    \Delta(t)^\text{max} = \frac{1}{4}\frac{(\Delta (x))^2}{\nu}
\end{equation}
The change in the maximum allowable time step due to errors in the density and viscosity would then be given by:
\begin{equation}
   \label{eqn:stability1}
    \frac{\Delta(t)^\text{max}_\text{gen}}
        {\Delta(t)^\text{max}_\text{ref}} 
    = \frac{\nu_\text{ref}(T,\densref(T,P))}
        {\nu_\text{gen}(T,\densgen(T,P))}.
\end{equation}
The changes in maximum allowable time step will therefore follow the pattern seen in Figure \ref{fig:CO2_kinematic_viscosity} of the previous section, meaning that changes in the necessary time step are likely to be on the order of 10--15 percent. 

The use of generalized EoSs and transport property correlations create a second possible cause of numeric instability. If the error in density is large enough, it might lead to calculation of a negative viscosity. This would lead to both unphysical results and numerical instability. The maximum allowable $\densERRtot$ or $\denserrtot$ to avoid a negative viscosity can be found using the Taylor expansion of $\viscgen(T,\densgen)$ about $\densref$:
\begin{equation}
    \viscgen(T,\densgen(T,P))
        \approx \viscgen(T,\densref(T,P))
        + S_1(T,P) \, \densERRtot(T,P).
\end{equation}
Setting the right hand side equal to zero and solving for the density error gives
\begin{equation}
    \label{eq:density_floor_viscosity}
    \densERR_*^{(\visc)}(T,P) 
        = -\frac{\viscgen(T,\densref(T,P))}{S_1(T,P)},
\end{equation}
where $\densERR_*^{(\visc)}$ is a threshold value of density error near which $\viscgen$ may become negative. The corresponding $\denserr_*^{(\visc)}$ is defined by dividing by the reference density. This threshold depends on the reference density, the chosen generalized viscosity correlation, and its sensitivity coefficient $S_1(T,P)$, but not on the generalized EoS. 

\begin{figure}[!hbt]
    \centering
    \panel{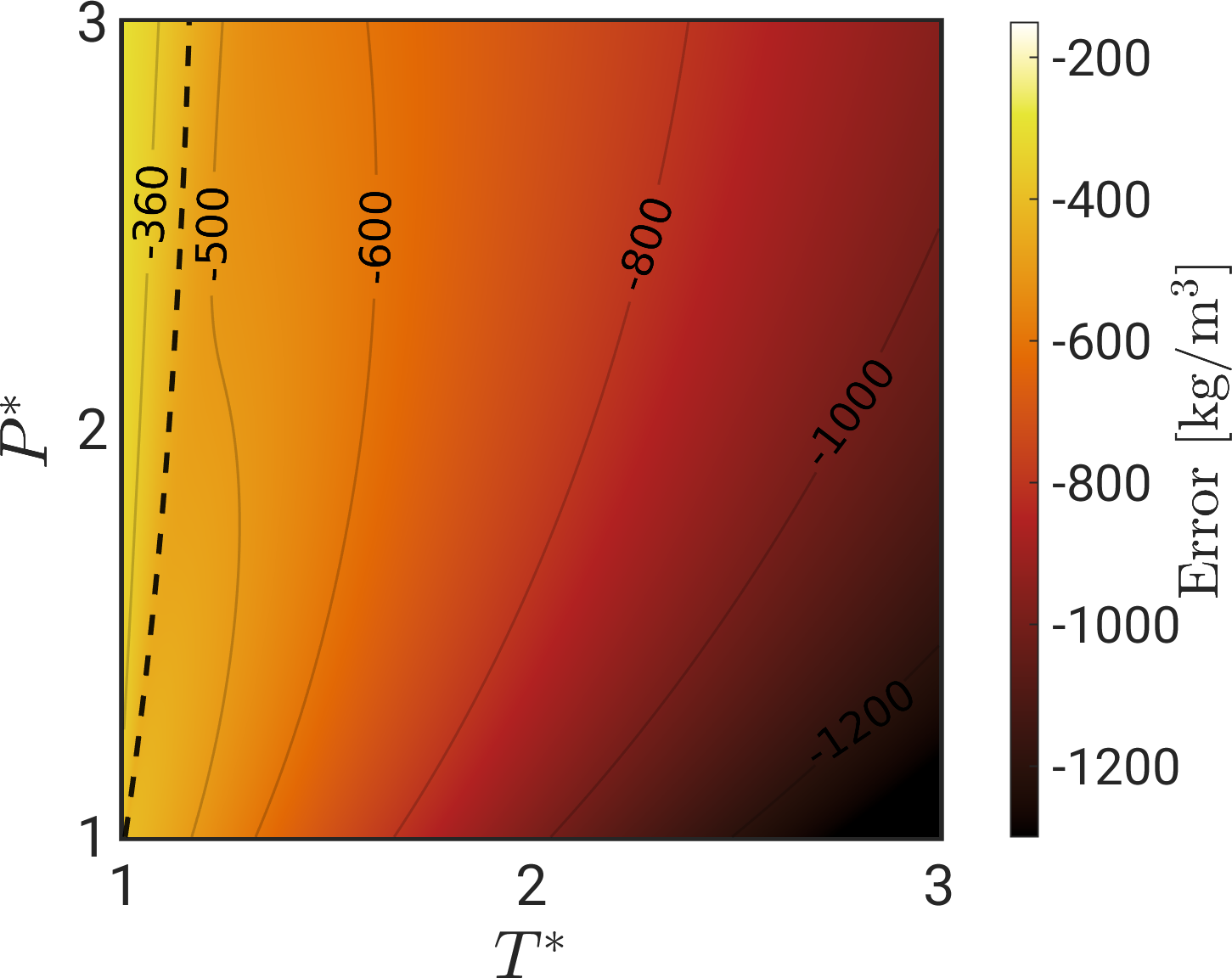}{(a) \cotwo{}, $\densERR_*^{(\visc)}$}
    \hspace{1cm}
    \panel{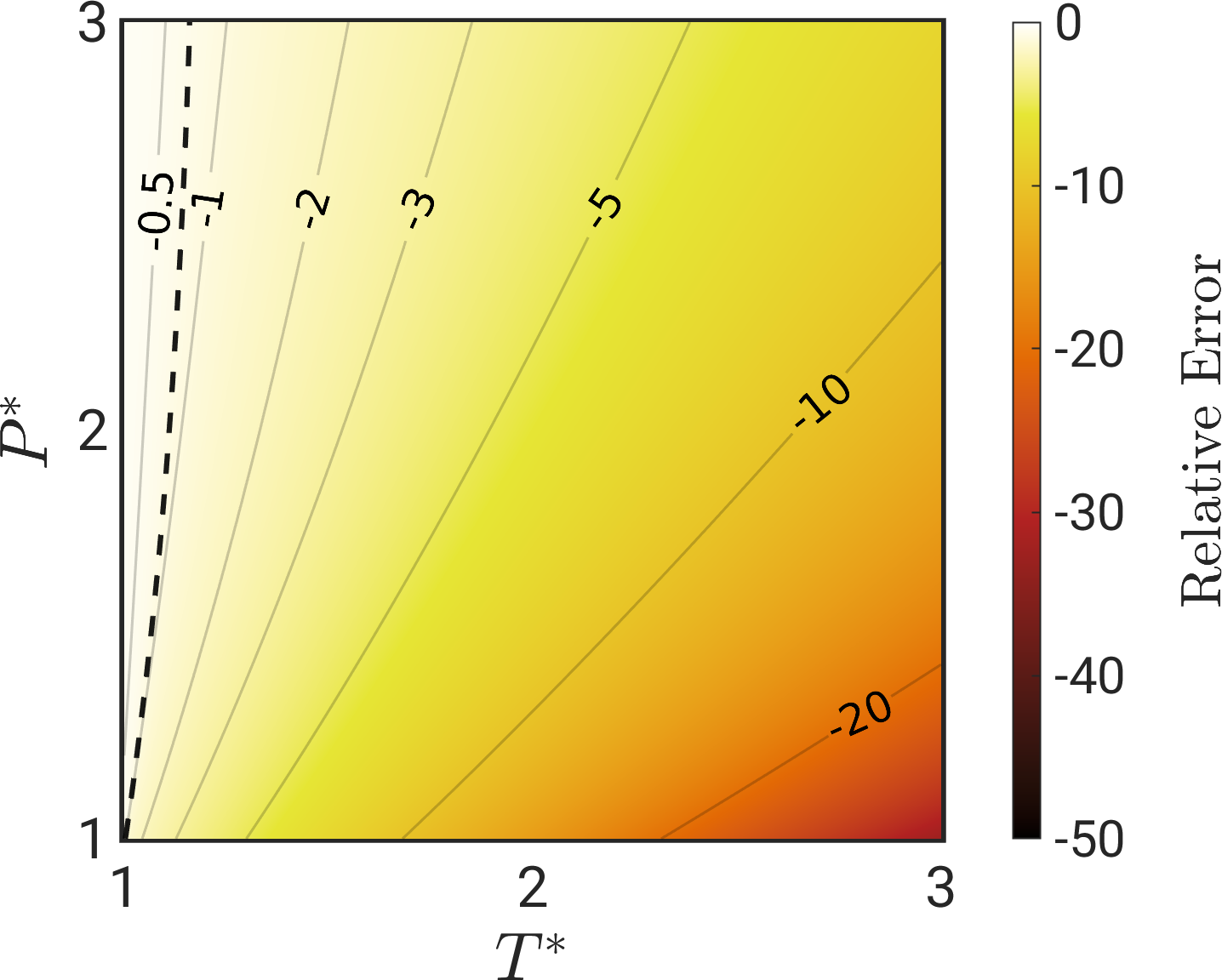}{(b) \cotwo{}, $\denserr_*^{(\visc)}$}
    \caption{\cotwo{} floor for density error in viscosity. Shown are (a) $\densERR_*^{(\visc)}$ and (b) $\denserr_*^{(\visc)}$.}
    \label{fig:CO2_viscosity_density_floor}
\end{figure}

Figure~\ref{fig:CO2_viscosity_density_floor}, illustrates $\densERR_*^{(\visc)}$ and $\denserr_*^{(\visc)}$ for the Chung correlation. For example, in the liquid-like region, an EoS that underestimates density by about 360 kg/m$^3$, i.e. that outputs a density that is about half the reference density, could cause the Chung correlation to produce negative values of viscosity. This scenario is plausible. However, in Figure~\ref{fig:CO2_viscosity_density_floor}(b), regions where $\denserr_*^{(\visc)} < -1$ are also regions where the chosen EoS would need to output negative densities in order to produce negative viscosities in the Chung correlation. If an EoS can be assumed  to produce positive values of density, then density errors could only realistically lead to negative viscosity values in the liquid-like region.

%

%%%%%%%%%%%%%%%%%%%%%%%%%%%%%%%%%%%%%%%%%%%%%%%
% section: end

% section: begin
%%%%%%%%%%%%%%%%%%%%%%%%%%%%%%%%%%%%%%%%%%%%%%%
\section{Supercritical \cotwo{} Conductivity Results}
\label{sec:conductivity}

%%%
\subsection{Thermal Conductivity Error Analysis}
Because thermal conductivity for supercritical fluids is also calculated using correlations based on temperature and thermal conductivity, it is subject to similar errors.  The total error in thermal conductivity  $\condERRtot$ can be decomposed in a manner similar to viscosity:
\begin{equation}
    \label{eq:lambda_decomposition}
    \condERRtot = \condERRcorr + \condERRfirst + \condERRsecond.
\end{equation}
Section~\ref{sec:viscosity}, presents results on the rectangle $\Omega = [T^*, 3T^*] \times [P^*, 3P^*]$.

\begin{figure}[!hbt]
    \centering
    \panel{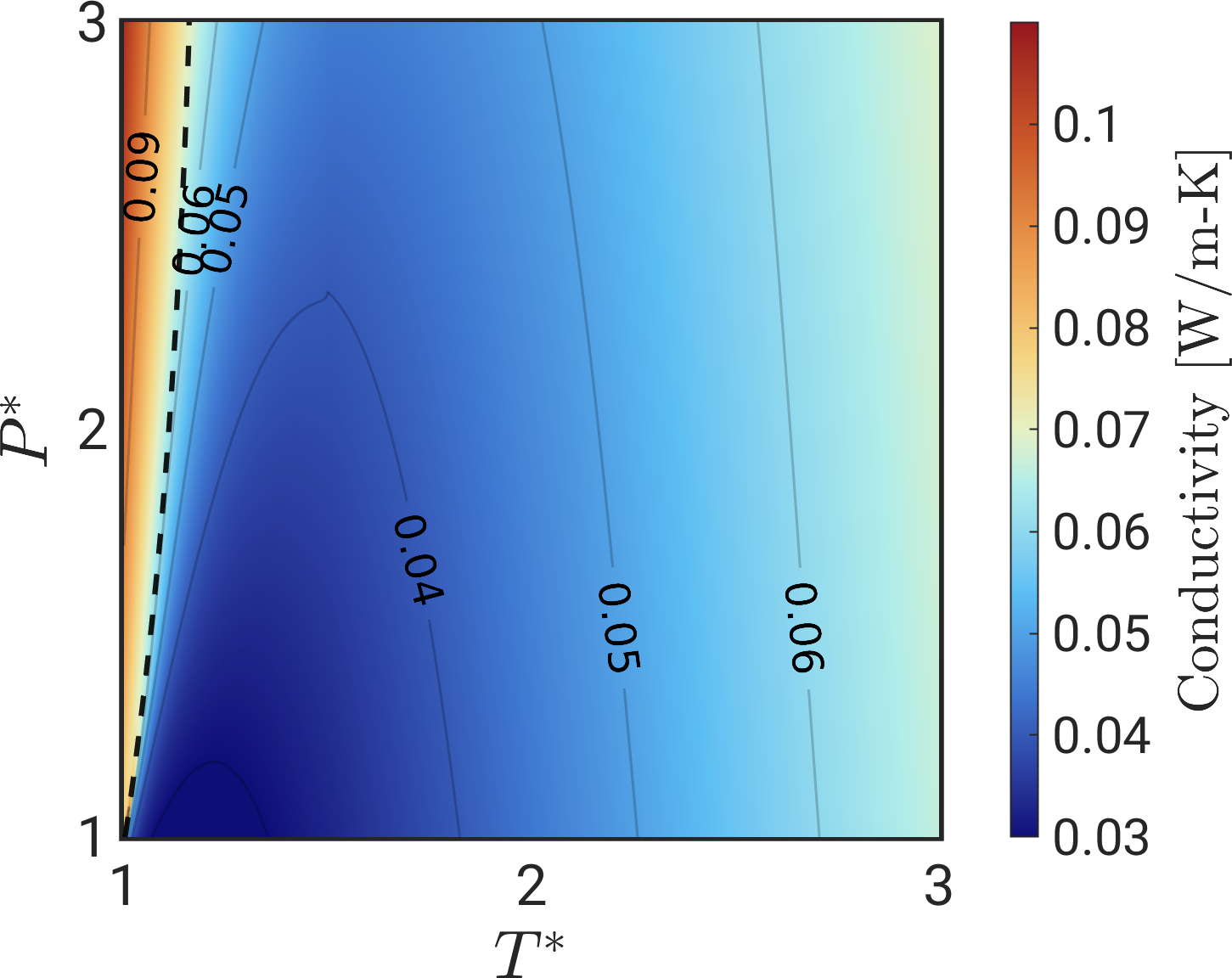}{(a) \cotwo{}, $\condref$}
    \hfill
    \panel{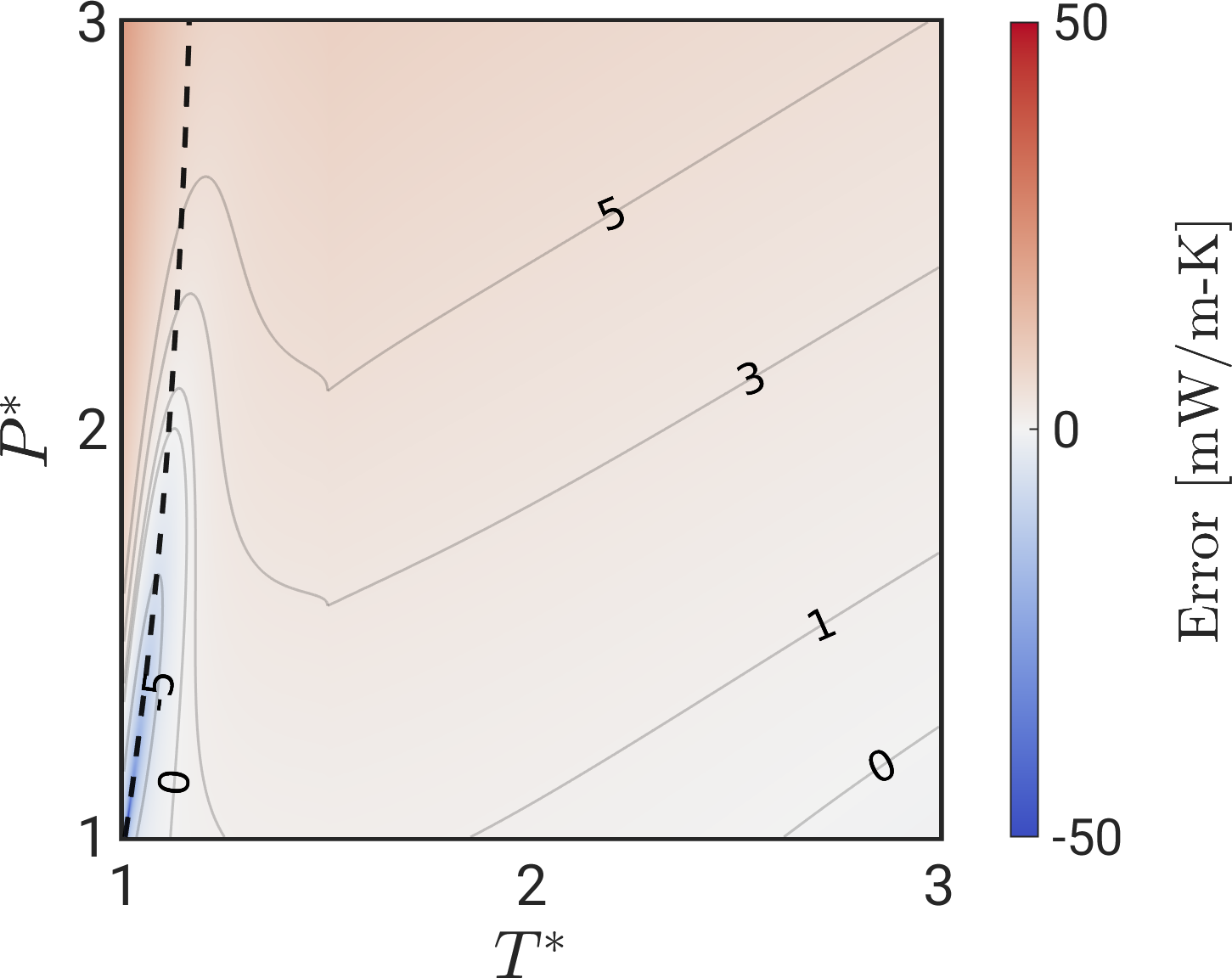}{(b) \cotwo{}, $\condERRtot$}
    \hfill
    \panel{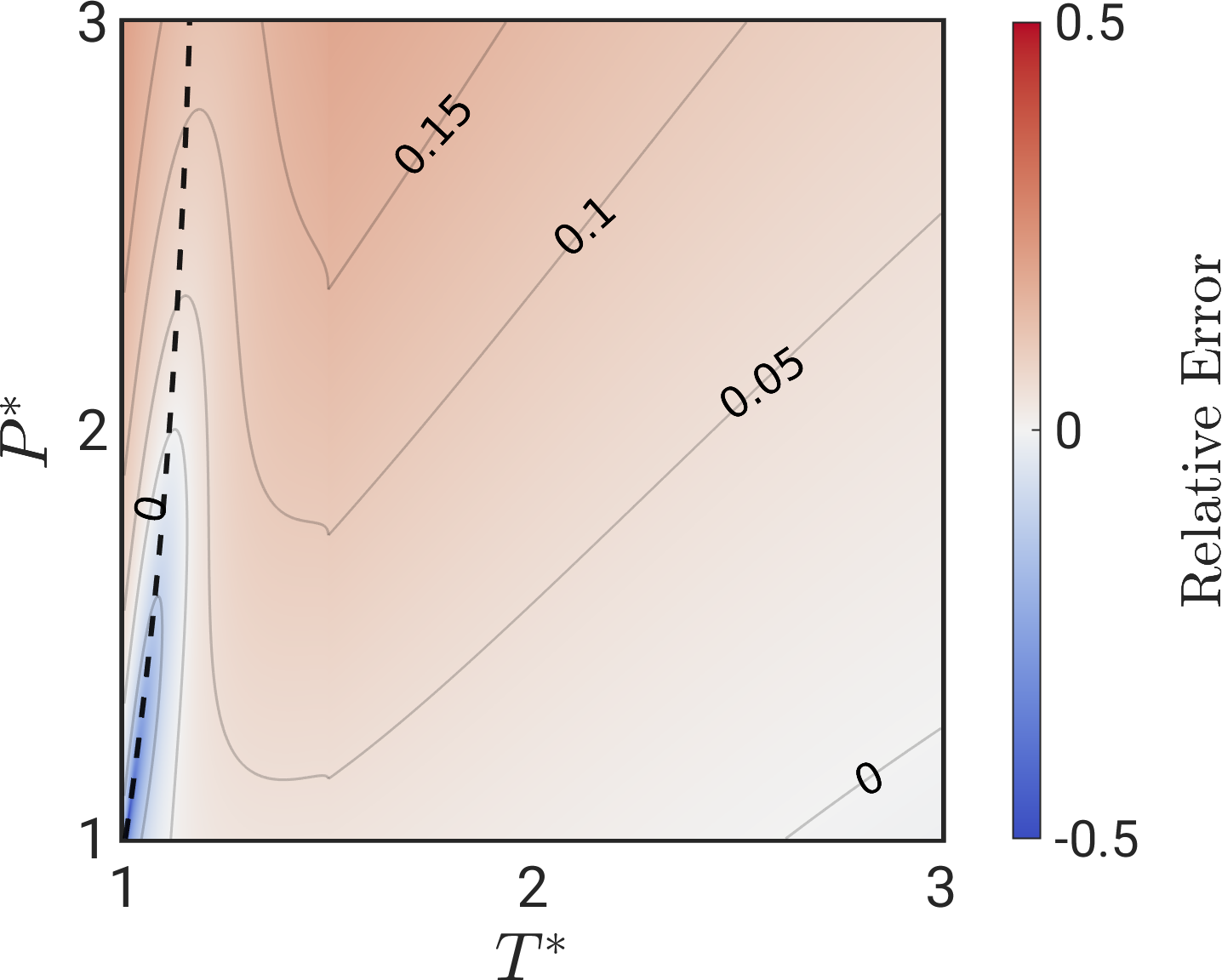}{(c) \cotwo{}, $\conderrtot$}
    \caption{\cotwo{} thermal conductivity results. The figure shows (a) $\condref$ computed by the reference EoS and correlation, (b) the signed error $\condERRtot$, and (c) the relative error $\conderrtot$. The dashed line is the Widom line.}
    \label{fig:CO2_conductivity}
\end{figure}

To begin,  $\condref$, computed by combining the SW reference EoS and the Huber--Sykioti--Assael--Perkins (HSAP) reference correlation \cite{huber} are compared against $\condgen$ values, computed by combining the SRK generalized EoS and the Chung generalized correlation. Figure~\ref{fig:CO2_conductivity}, presents the reference thermal conductivity $\condref$ and the errors $\condERRtot$ and $\conderrtot$. Thermal conductivity is overestimated by $\condgen$ throughout most of the region considered, but is underestimated in a region near the critical point. These over- and underestimations can be quite severe. The maximum overestimation is $20.61\%$. The maximum underestimation is $-93.71\%$, which occurs right at the critical point, though the magnitude of this underestimation rapidly reduces to an underestimation of about $-10\%$ just away from the critical point.

\begin{figure}[!hbt]
    \centering
    \panelTwo{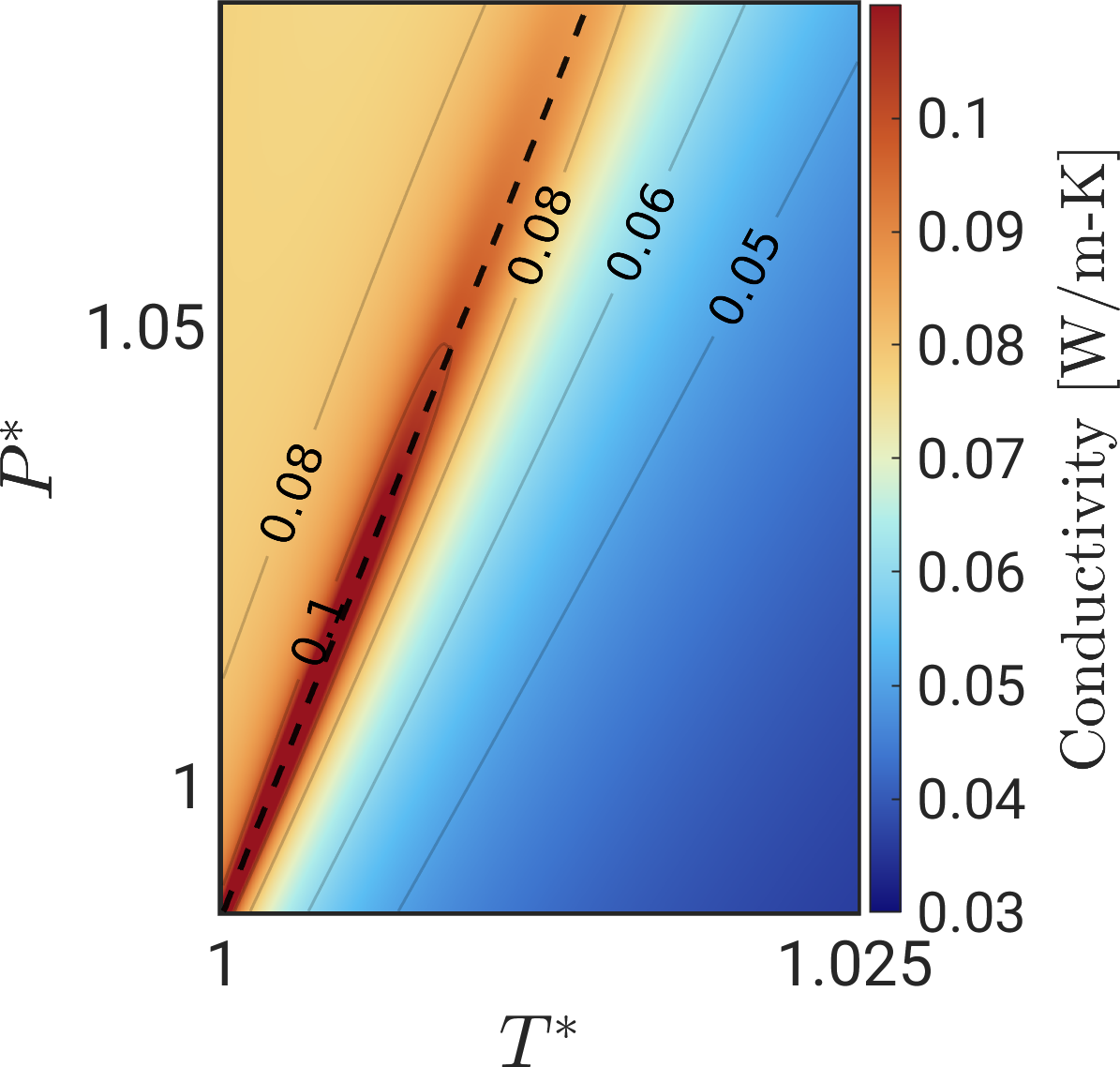}{(a) \cotwo{}, $\condref$}
    \hspace{1.5em}
    \panelTwo{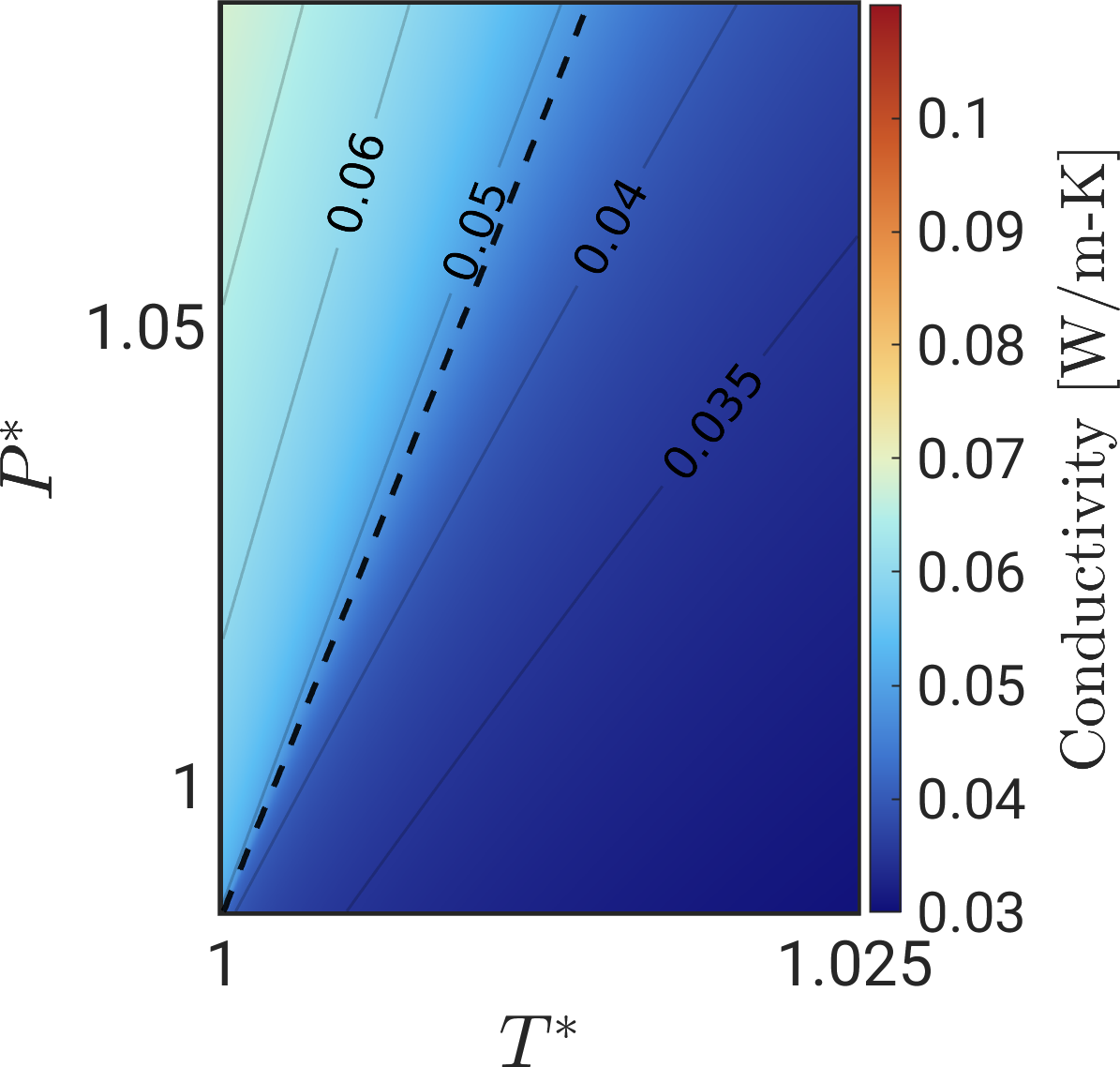}{(b) \cotwo{}, $\condgen$}
    \hspace{1.5em}
    \panelTwo{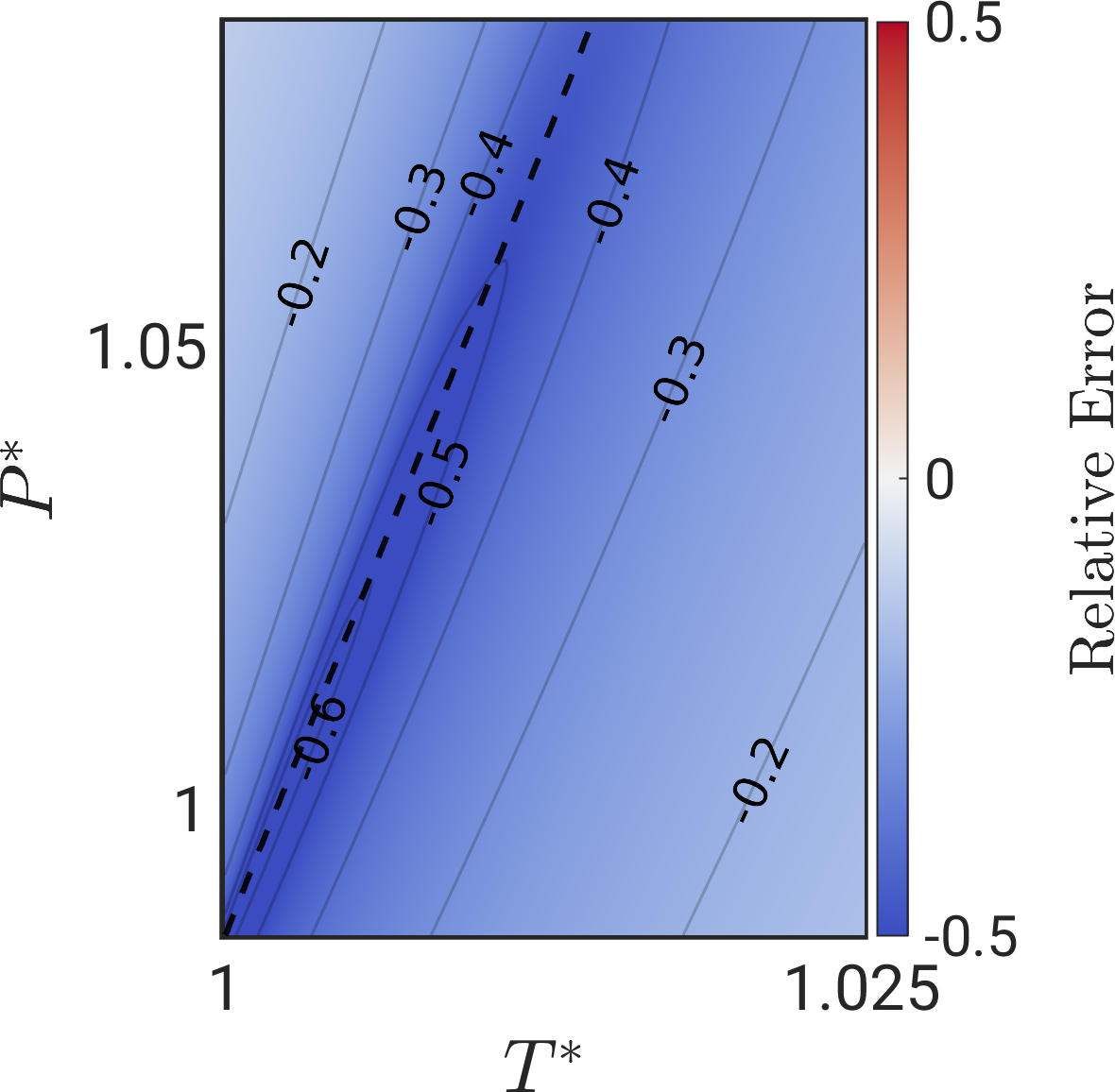}{(c) \cotwo{}, $\conderrtot$}
    \caption{\cotwo{} thermal conductivity, zoom-in near critical point. The figure shows (a) $\condref$ computed by the reference EoS and correlation, (b) $\condgen$ computed by the SRK EoS and Chung correlation, and (c) the relative error $\conderrtot$. The dashed line is the Widom line.}
    \label{fig:CO2_conductivity_zoom}
\end{figure}

To help explain these results, Figure~\ref{fig:CO2_conductivity_zoom} shows $\condref$, $\condgen$, and $\conderrtot$ on a magnified subsection of $\Omega$ near the critical point. This figure illustrates that the HSAP reference correlation includes a critical enhancement term which approximates singular behavior near the critical point, while the Chung correlation does not include such a term. As a result, $\condgen$ drastically underestimates thermal conductivity in a small region that traces the Widom line near the critical point.

\begin{figure}[!hbt]
    \centering
    \panel{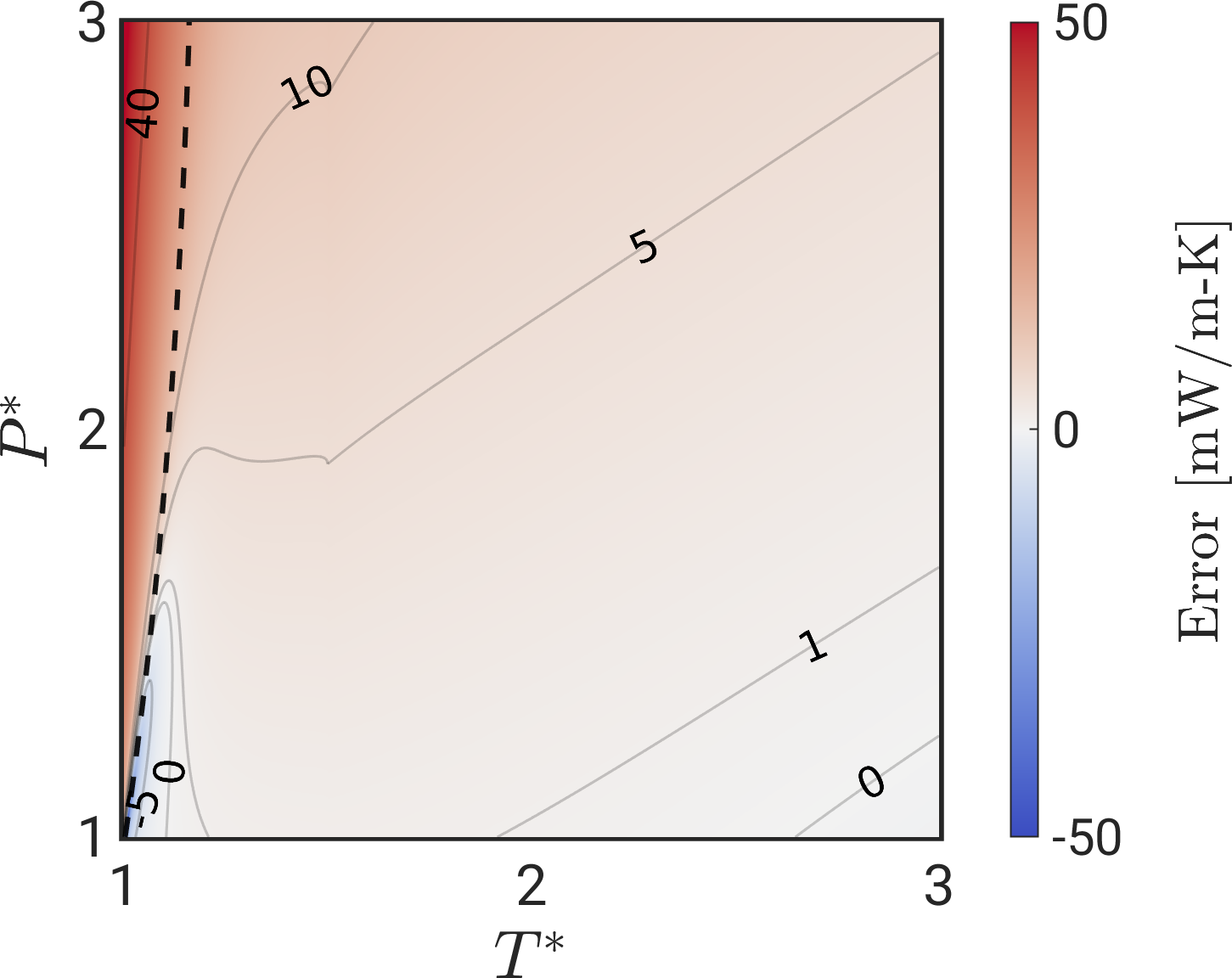}{(a) \cotwo{}, $\condERRcorr$}
    \hfill
    \panel{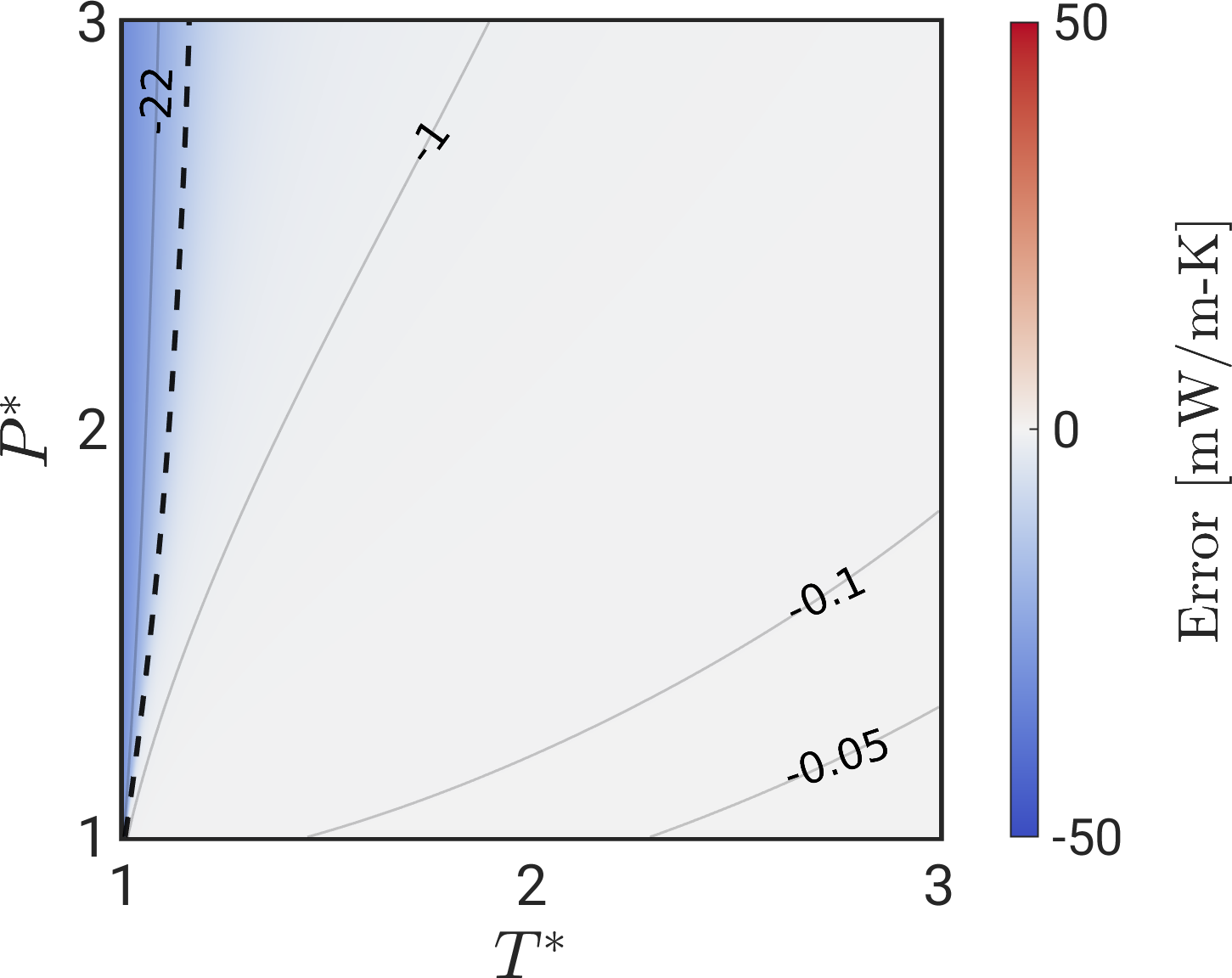}{(b) \cotwo{}, $\condERRfirst$}
    \hfill
    \panel{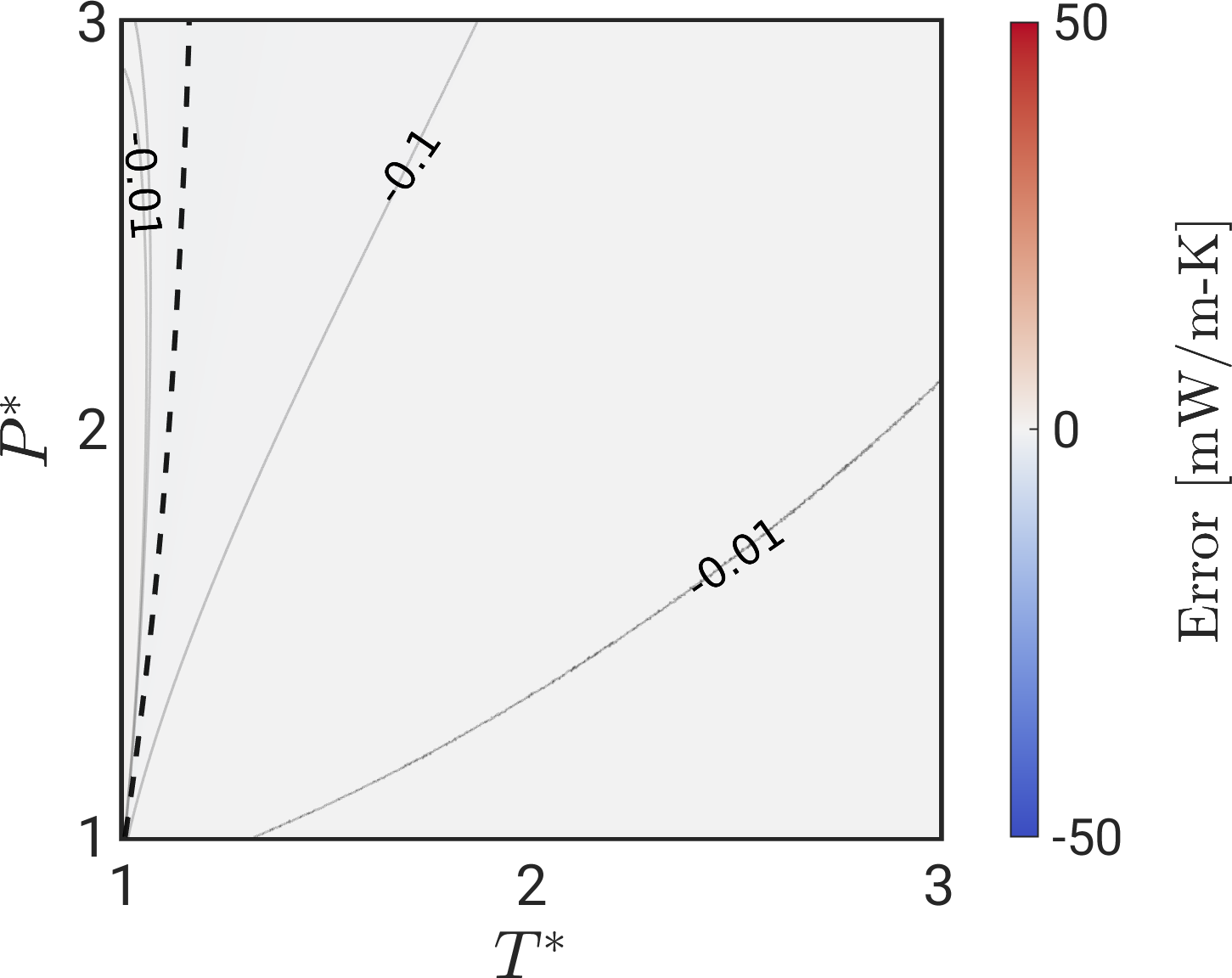}{(c) \cotwo{}, $\condERRsecond$}
    \vspace{1em} \\
    \hrule \vspace{1em}
    \panel{images/2d_results_NEW/CO2_2d_cond_err_corr.png}{(d) \cotwo{}, $\conderrcorr$}
    \hfill
    \panel{images/2d_results_NEW/CO2_2d_cond_err_eos.png}{(e) \cotwo{}, $\conderrfirst$}
    \hfill
    \panel{images/2d_results_NEW/CO2_2d_cond_err_secondorder.png}{(f) \cotwo{}, $\conderrsecond$}
    \caption{\cotwo{} thermal conductivity error decomposition. The top row shows the signed errors (a) $\condERRcorr$, (b) $\condERRfirst$, and (c) $\condERRsecond$. The bottom row shows the corresponding relative errors, i.e. (d) $\conderrcorr$, (e) $\conderrfirst$, and (f) $\conderrsecond$.}
    \label{fig:CO2_conductivity_error_decomposition}
\end{figure}

The decomposition of errors suggested by Equation~\eqref{eq:lambda_decomposition} can be examined in greater detail. Figure~\ref{fig:CO2_conductivity_error_decomposition} presents the errors $\condERRcorr$, $\condERRfirst$ and $\condERRsecond$. Recalling that $\condERRcorr$ is the portion of the error due specifically to the Chung correlation, it is clear the Chung correlation tends to overestimate thermal conductivity throughout most of $\Omega$, with the most extreme overestimation of $49.53\%$ in the liquid-like region. $\condERRcorr$ is negative only in the small region near the critical point described in the previous paragraph, where it takes a minimum value of  $-93.71\%$. This reinforces that the error in this region is due to the Chung correlation, and not due to the propagation of density error.

On the other hand, the values for $\condERRfirst$ shown in Figure~\ref{fig:CO2_conductivity_error_decomposition} demonstrate that the underestimation of density due to the SRK EoS propagates into the Chung correlation and partly offsets the overestimation due to the Chung correlation itself.  $\condERRfirst$ is negative throughout $\Omega$, with a most extreme underestimation of $-34.67\%$ in the liquid-like region. $\condERRsecond$ is responsible for only a very small amount of the error.

\begin{figure}[!hbt]
    \centering
    \panel{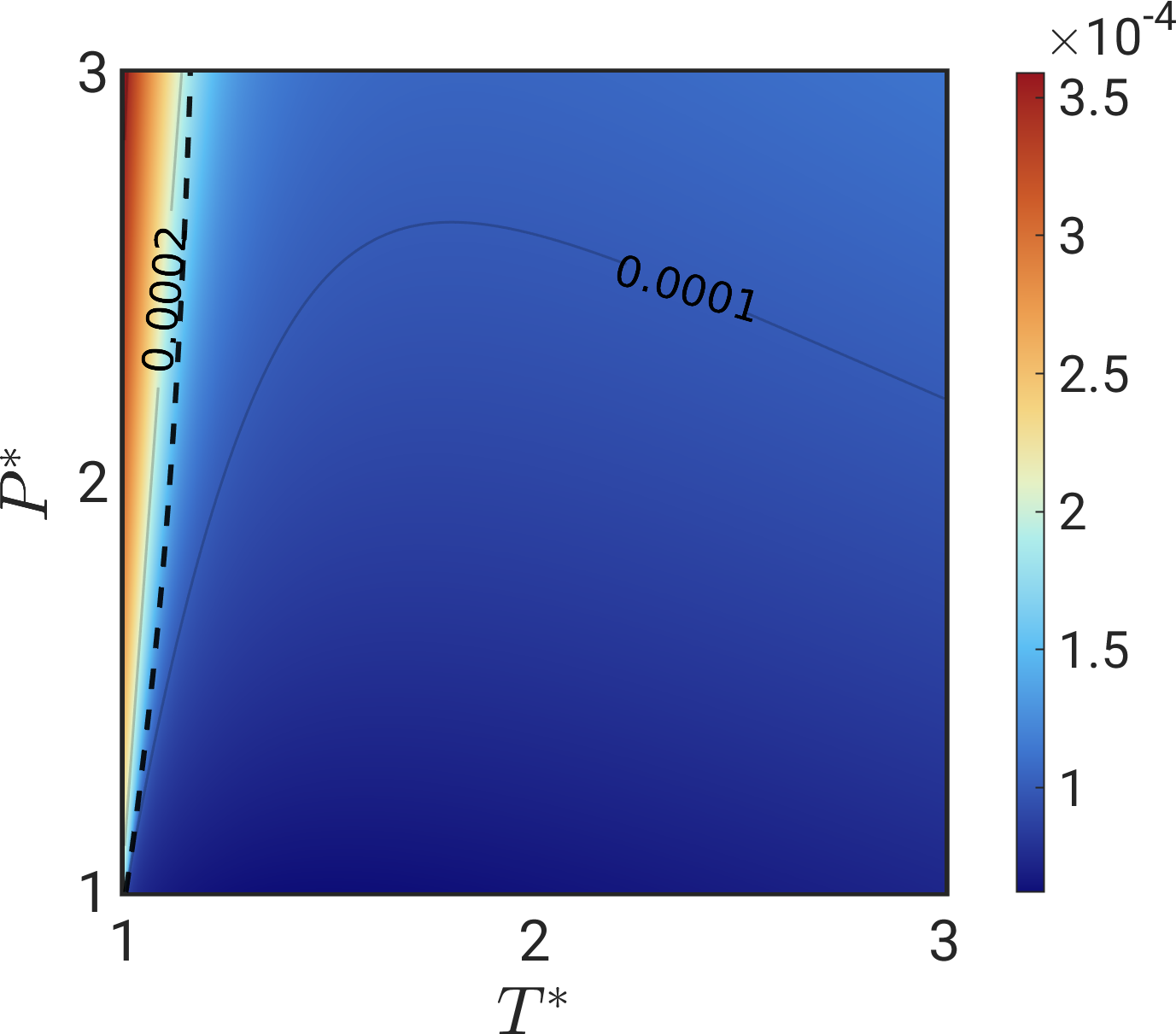}{(a) \cotwo{}, $S_2$}
    \hfill
    \panel{images/2d_results_NEW/CO2_2d_dens_ERR.png}{(b) \cotwo{}, $\densERRtot$}
    \hfill
    \panel{images/2d_results_NEW/CO2_2d_cond_ERR_eos.png}{(c) \cotwo{}, $\condERRfirst$}
    \caption{\cotwo{} first order error factors for thermal conductivity. Shown are (a) $S_2 = \left. \tfrac{\partial \condgen}{\partial \rho} \right|_{(T, \densref)}$, (b) $\densERRtot$, and (c) $\condERRfirst$.}
    \label{fig:CO2_conductivity_first_order}
\end{figure}

Indeed, by considering the sensitivity term $S_2$, which represents the sensitivity of the Chung correlation to errors in density, it is clear that the Chung correlation is most sensitive to errors in density in the liquid-like region. Figure~\ref{fig:CO2_conductivity_first_order} plots  the quantity $S_2$. Recalling that $\condERRfirst = \densERRtot \, S_2$, the plots of $\densERRtot$ and $\condERRfirst$, which originally appeared in Figures~\ref{fig:CO2_density} and \ref{fig:CO2_conductivity_error_decomposition} respectively, are repeated for ease of comparison. 

\begin{figure}[!hbt]
    \centering
    \panel{images/2d_results_NEW/CO2_2d_cond_ERR_tot.png}{(a) \cotwo{}, $\condERRtot$}
    \hspace{1cm}
    \panel{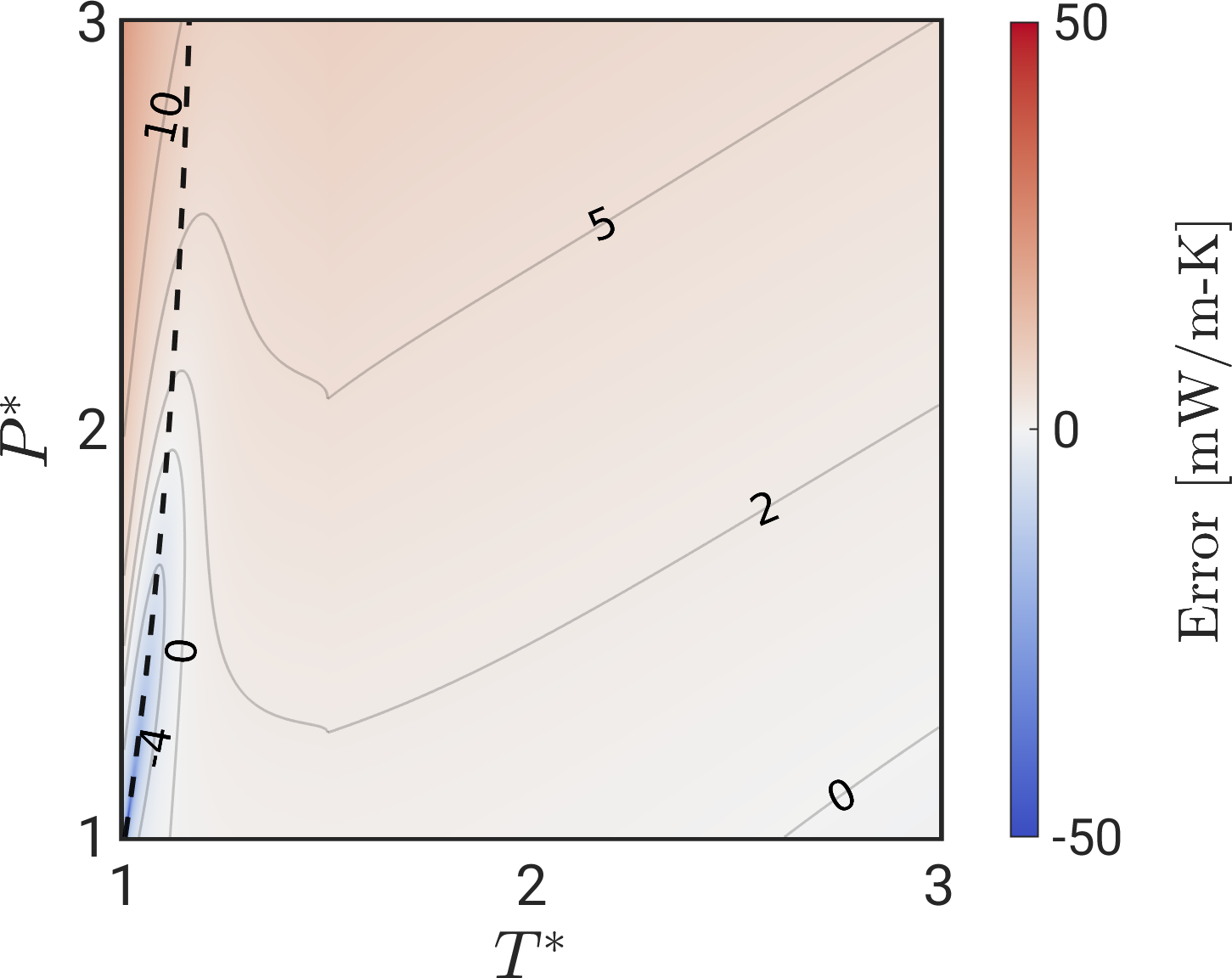}{(b) \cotwo{}, $\condERRcorr + \condERRfirst$}
    \caption{\cotwo{} total error and first order error decomposition for thermal conductivity. Shown are (a) $\condERRtot$ and (b) $\condERRcorr + \condERRfirst$.}
    \label{fig:CO2_conductivity_error_summed}
\end{figure}

Finally, Figure~\ref{fig:CO2_conductivity_error_summed} shows $\condERRtot$ against $\condERRcorr + \condERRfirst$, where $\condERRfirst$ is computed by multiplying $S_2$ and $\densERRtot$. This figure shows that the first order approximation $\condERRcorr + \condERRfirst = \condERRcorr + S_2 \densERRtot$ is a good approximation of the total error. 

%

%%%
\subsection{Impacts on Numeric Stability}
\label{sec:conducivity_stability}
In analogy with equation \eqref{eq:density_floor_viscosity} from Section~\ref{sec:viscosity_stability}, density errors that can potentially create negative thermal conductivities are possible, and can be quantified and avoided. Such errors are computed by:
\begin{equation}
    \label{eq:density_floor_conductivity}
    \densERR_*^{(\cond)}(T,P) 
        = -\frac{\condgen(T,\densref(T,P))}{S_2(T,P)},
\end{equation}
where $\densERR_*^{(\cond)}$ is a threshold value of density error near which $\condgen$ may become negative, and the corresponding $\denserr_*^{(\cond)}$ is defined by dividing by the reference density. 

\begin{figure}[!hbt]
    \centering
    \panel{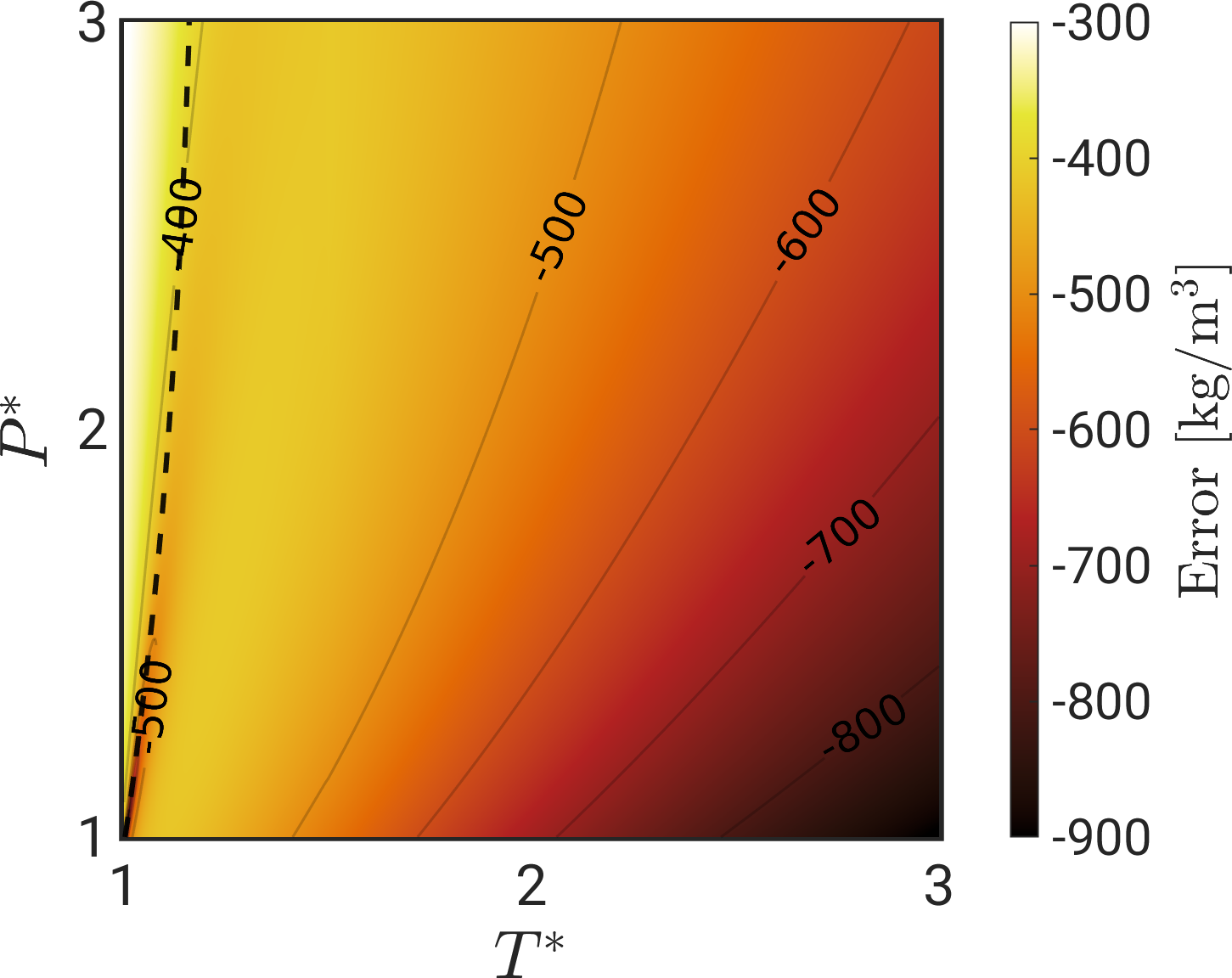}{(a) \cotwo{}, $\densERR_*^{(\cond)}$}
    \hspace{1cm}
    \panel{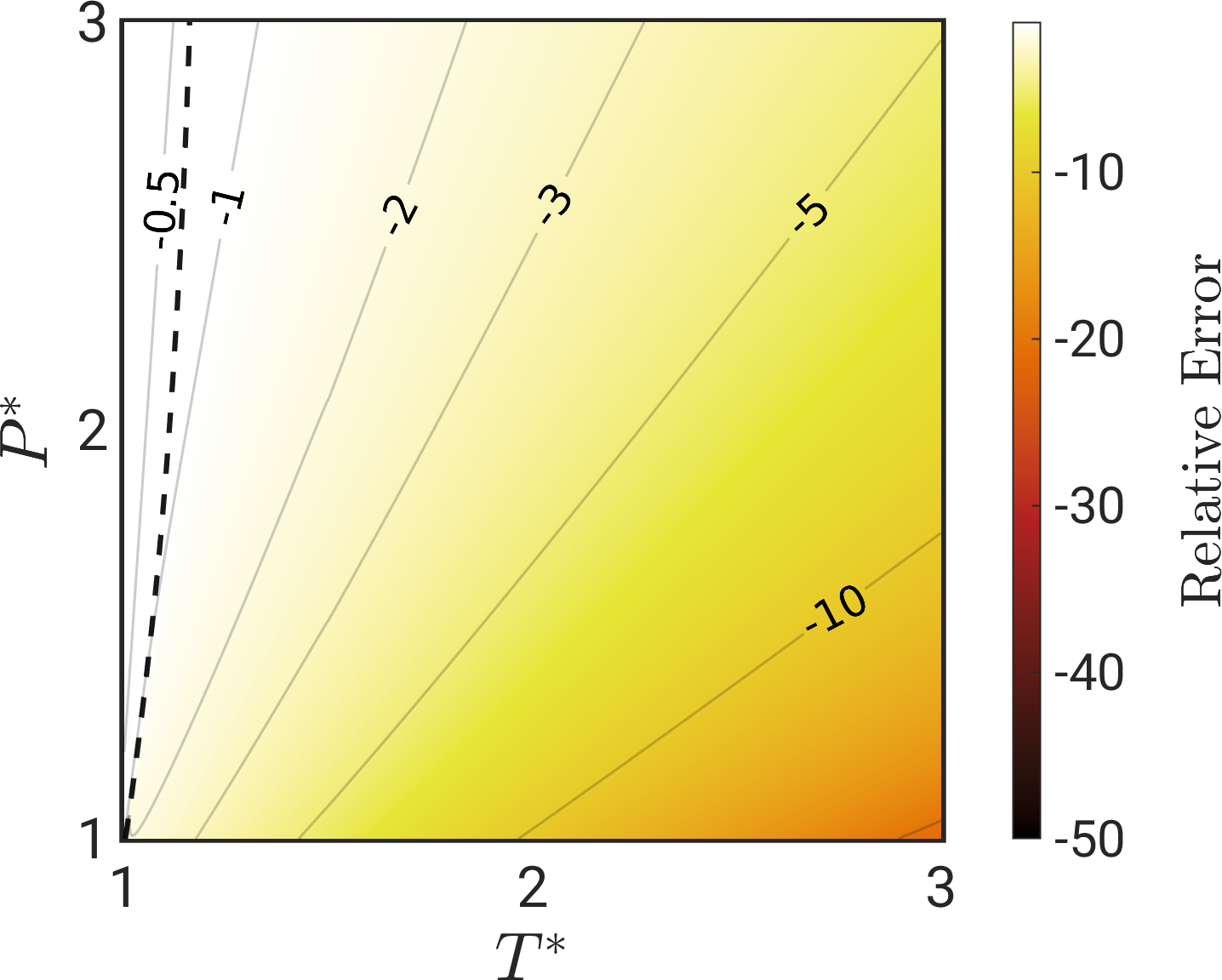}{(b) \cotwo{}, $\denserr_*^{(\cond)}$}
    \caption{\cotwo{} floor for density error in thermal conductivity. Shown are (a) $\densERR_*^{(\cond)}$ and (b) $\denserr_*^{(\cond)}$.}
    \label{fig:CO2_conductivity_density_floor}
\end{figure}

Figure~\ref{fig:CO2_conductivity_density_floor} illustrates $\densERR_*^{(\cond)}$ and $\denserr_*^{(\cond)}$ for the Chung correlation for thermal conductivity. In the liquid-like region, an EoS that underestimates density by about 400 kg/m$^3$, i.e. by about half the reference value, could cause the Chung correlation to produce negative values of viscosity. This scenario is plausible. However, as Figure~\ref{fig:CO2_viscosity_density_floor}(b) shows, just outside the liquid-like region, such an EoS would have to produce negative values of density in order to cause the Chung correlation to produce negative values of thermal conductivity. 

%

%%%
\subsection{Comparison of viscosity and thermal conductivity results}
The decomposition of thermal conductivity error is similar to the decomposition of viscosity error presented in Section~\ref{sec:viscosity}. Specifically, the Chung correlation itself overestimates both thermal conductivity and viscosity throughout most of $\Omega$ and, for both properties, the most severe overestimation occurs in the liquid-like region. At the same time, for both thermal conductivity and viscosity, the errors due to the Chung correlation are partly offset by the fact that the SRK EoS underestimates density in the liquid-like region, which is reflected in the quantities $\condERRfirst$ and $\viscERRfirst$. 

For both thermal conductivity and viscosity, the Chung correlation is most sensitive to errors in density in the liquid-like region, which is precisely where the SRK EoS most severely underestimates the density. For both thermal conductivity and viscosity, in the liquid-like region it is theoretically possible for the Chung correlation to produce negative values of either property; this would require the choice of a hypothetical EoS that underestimates density by about half its reference value in this region. While the SRK EoS does underestimate density in this region, the underestimation is not so severe as to cause negative values of either viscosity or thermal conductivity.  

A final similarity is that, for both thermal conductivity and viscosity, the sums $\condERRcorr + S_2 \densERRtot$ and $\viscERRcorr + S_1 \densERRtot$ provide good approximations of the total errors $\condERRtot$ and $\viscERRtot$ respectively. So, for both thermal conductivity and viscosity, most of the total error can be explained in terms of (1) the error produced by the Chung correlation itself, (2) the sensitivity of the Chung correlation to errors in density, and (3) the actual density error due to the choice of EoS. 

On the other hand, a major difference between the error patterns for thermal conductivity compared to viscosity is that the Chung correlation also radically underestimates thermal conductivity in a small region that follows the Widom line near the critical point. The Chung correlation also underestimates viscosity near the critical point, but the underestimation is about five times less severe.

%%%%%%%%%%%%%%%%%%%%%%%%%%%%%%%%%%%%%%%%%%%%%%%
% section: end

% section: begin
%%%%%%%%%%%%%%%%%%%%%%%%%%%%%%%%%%%%%%%%%%%%%%%
\section{Cross-Species Results}
\label{sec:cross_species}
s\cotwo{} is often considered a particularly challenging fluid to work with. So it is important to establish if the same effects appear in other fluids. This can be done by repeating the analysis for other species, such as  \ntwo{} and \chfour{}. To provide meaningful comparisons between these species, the same domain $\Omega = [T^*, 3T^*] \times [P^*, 3P^*]$ used in Sections~\ref{sec:viscosity} and \ref{sec:conductivity} is used for these species.

%%% 
\subsection{Cross-Species Density Results}
\begin{figure}[!hbt]
    \centering
    \panel{images/2d_results_NEW/CO2_2d_dens_err.png}{(a) \cotwo{}, $\denserrtot$}
    \hfill
    \panel{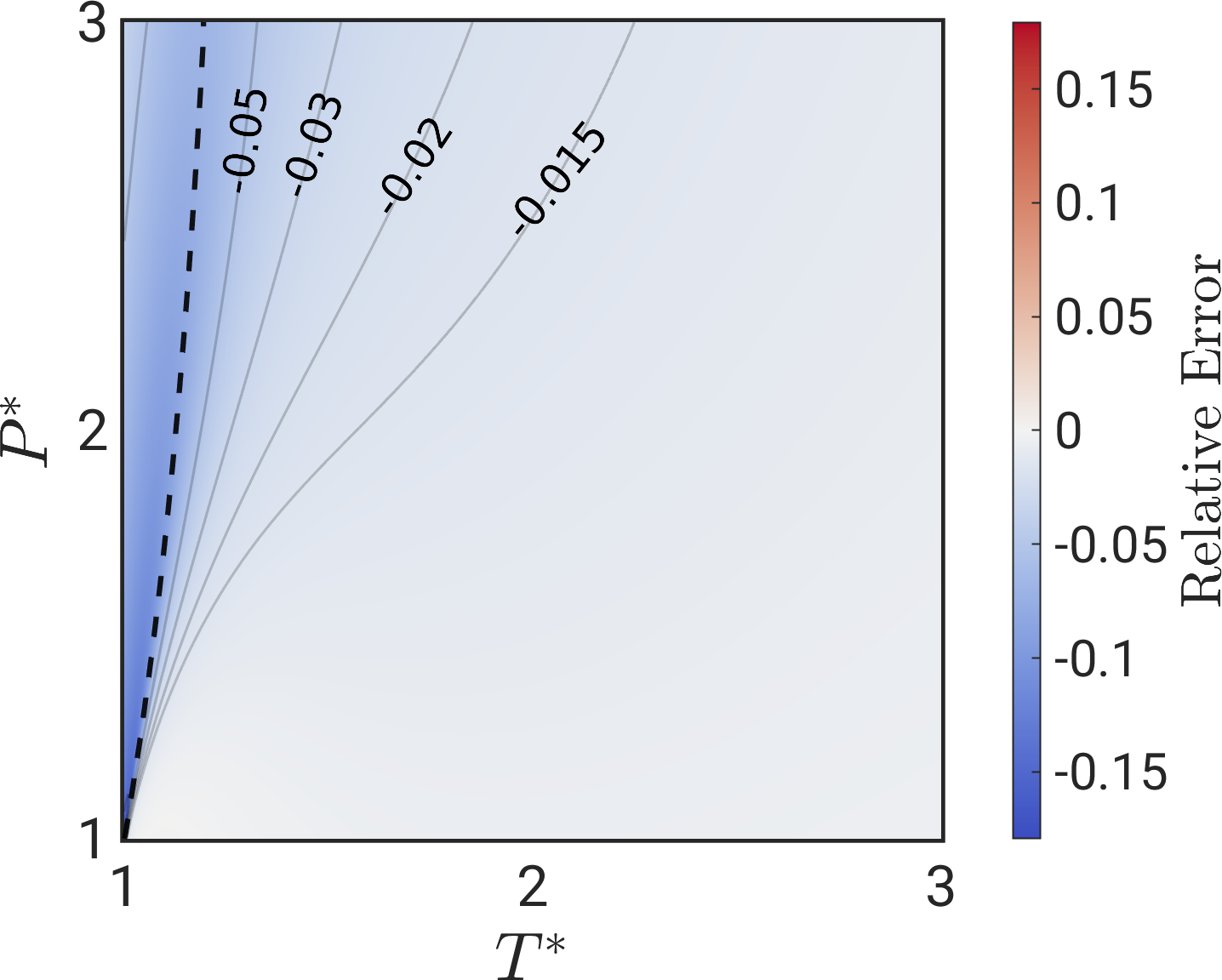}{(b) \ntwo{}, $\denserrtot$}
    \hfill
    \panel{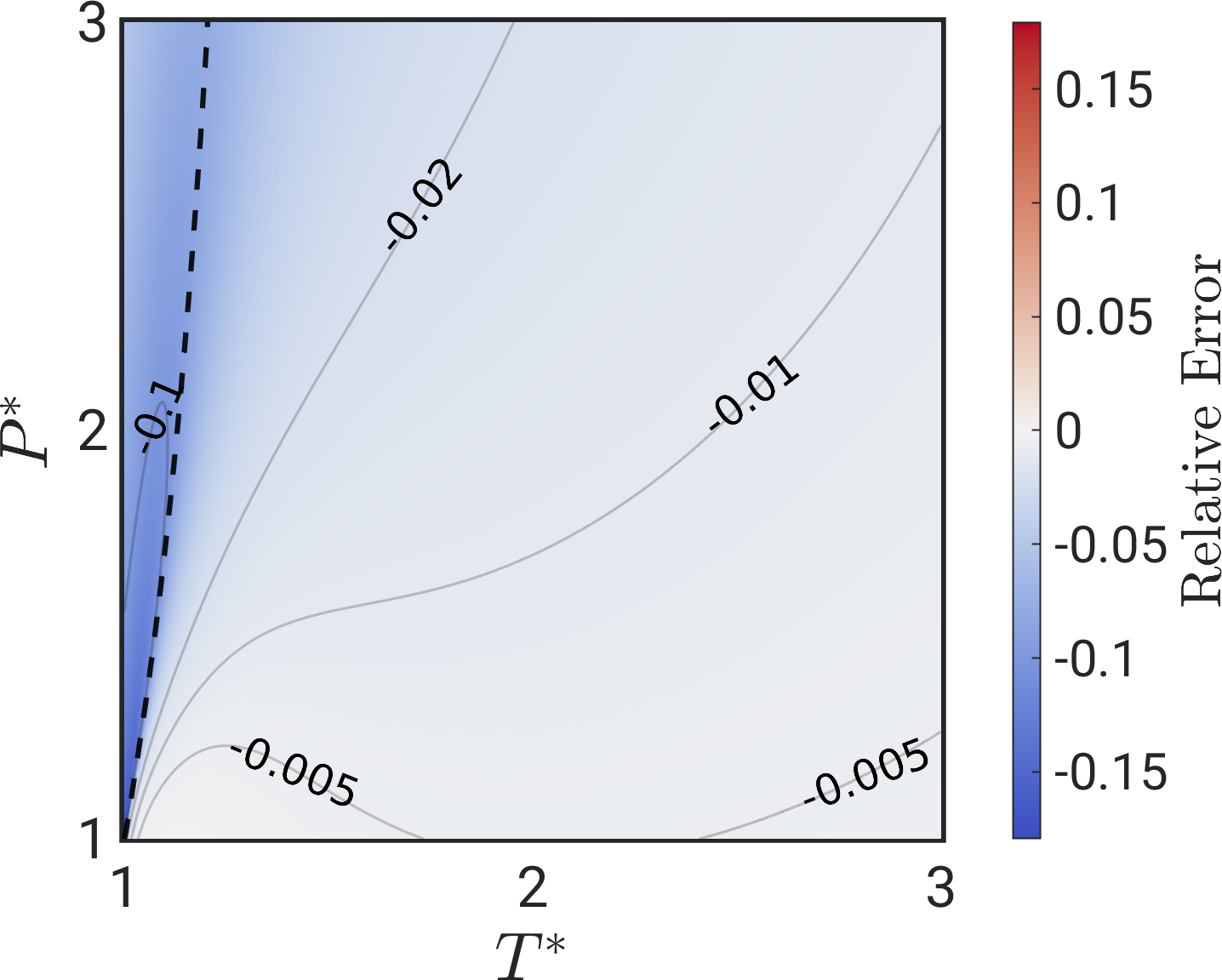}{(c) \chfour{}, $\denserrtot$}
    \caption{Relative density errors $\denserrtot$ for (a) \cotwo{}, (b) \ntwo{}, and (c) \chfour{}.}
    \label{fig:cross_species_density_error}
\end{figure}

Figure~\ref{fig:cross_species_density_error} shows $\denserrtot$ for \ntwo{} and \chfour{}; for ease of comparison, this figure also reproduces $\denserrtot$ for \cotwo{}, which was originally presented in Figure~\ref{fig:CO2_density}. For each species, the SRK EoS underestimates density throughout $\Omega$, with the most severe underestimations of $-22.21\%$, $-17.93\%$, and $-21.65\%$ for \cotwo{}, \ntwo{}, and \chfour{} respectively all occurring in the liquid-like region. 

%

%%%
\subsection{Cross-Species Viscosity Results}
\begin{figure}[!hbt]
    \centering
    \panelThree
        {images/2d_results_NEW/CO2_2d_visc_err_tot.png}
        {(a) \cotwo{},  $\conderrtot$}
    \hfill
    \panelThree
        {images/2d_results_NEW/CO2_2d_visc_err_corr.png}
        {(b) \cotwo{},  $\conderrcorr$}
    \hfill
    \panelThree
        {images/2d_results_NEW/CO2_2d_visc_err_eos.png}
        {(c) \cotwo{},  $\conderrfirst$}
    \hfill
    \panelThree
        {images/2d_results_NEW/CO2_2d_visc_err_secondorder.png}
        {(d) \cotwo{},  $\conderrsecond$}
    \vspace{0.5em} \\
    \hrule
    \vspace{0.5em}
    % 
    % new row
    \panelThree
        {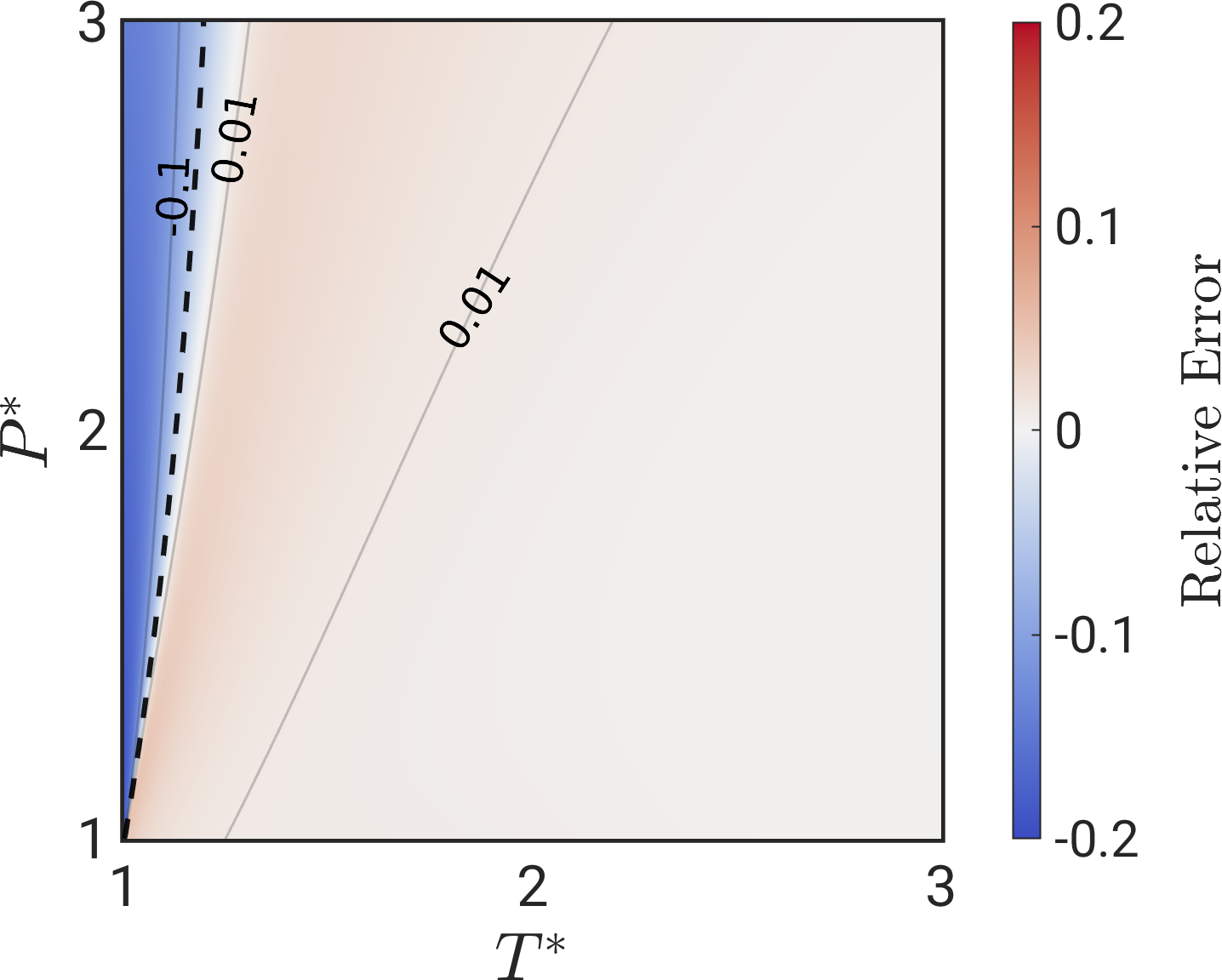}
        {(e) \ntwo{},  $\conderrtot$}
    \hfill
    \panelThree
        {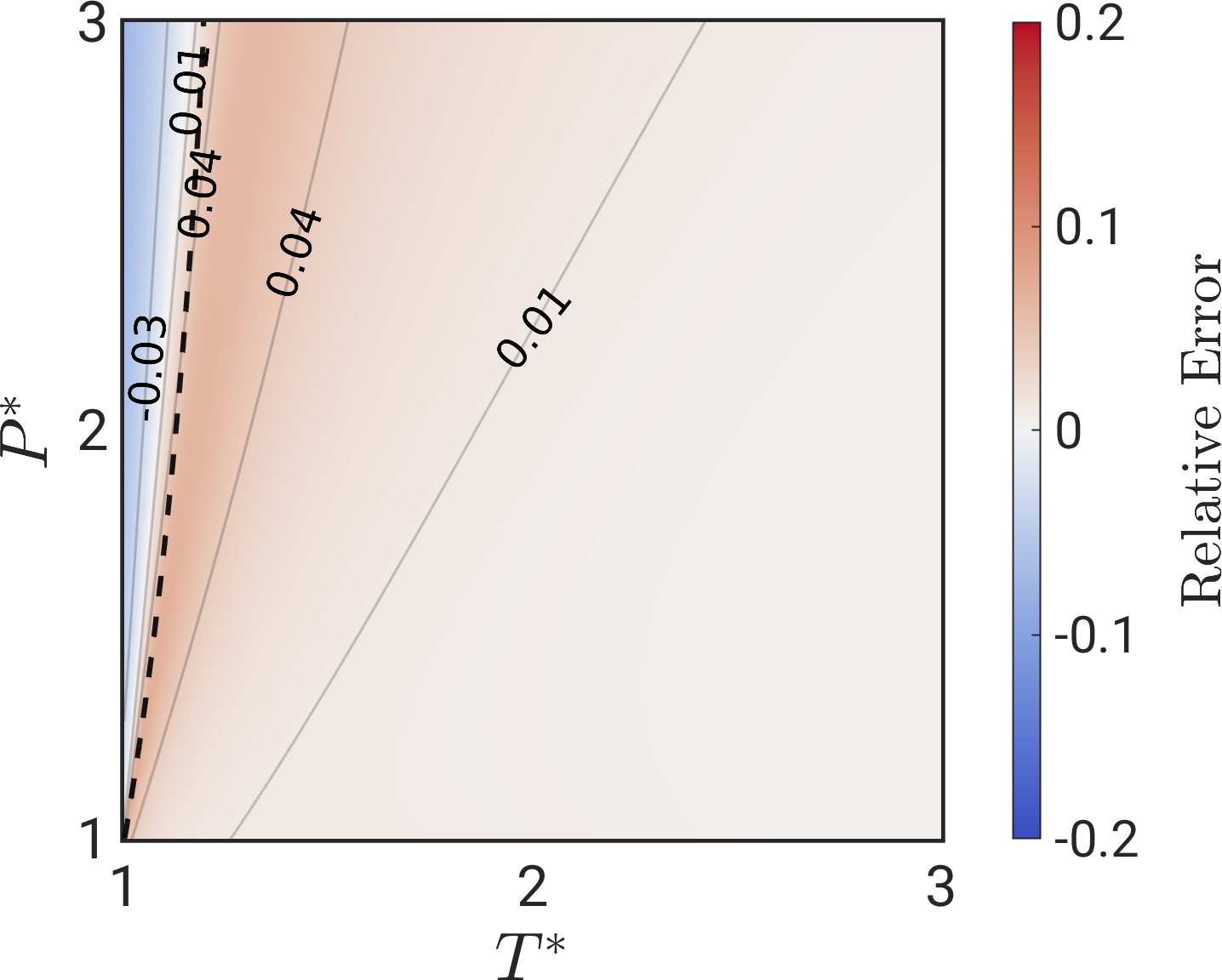}
        {(f) \ntwo{},  $\conderrcorr$}
    \hfill
    \panelThree
        {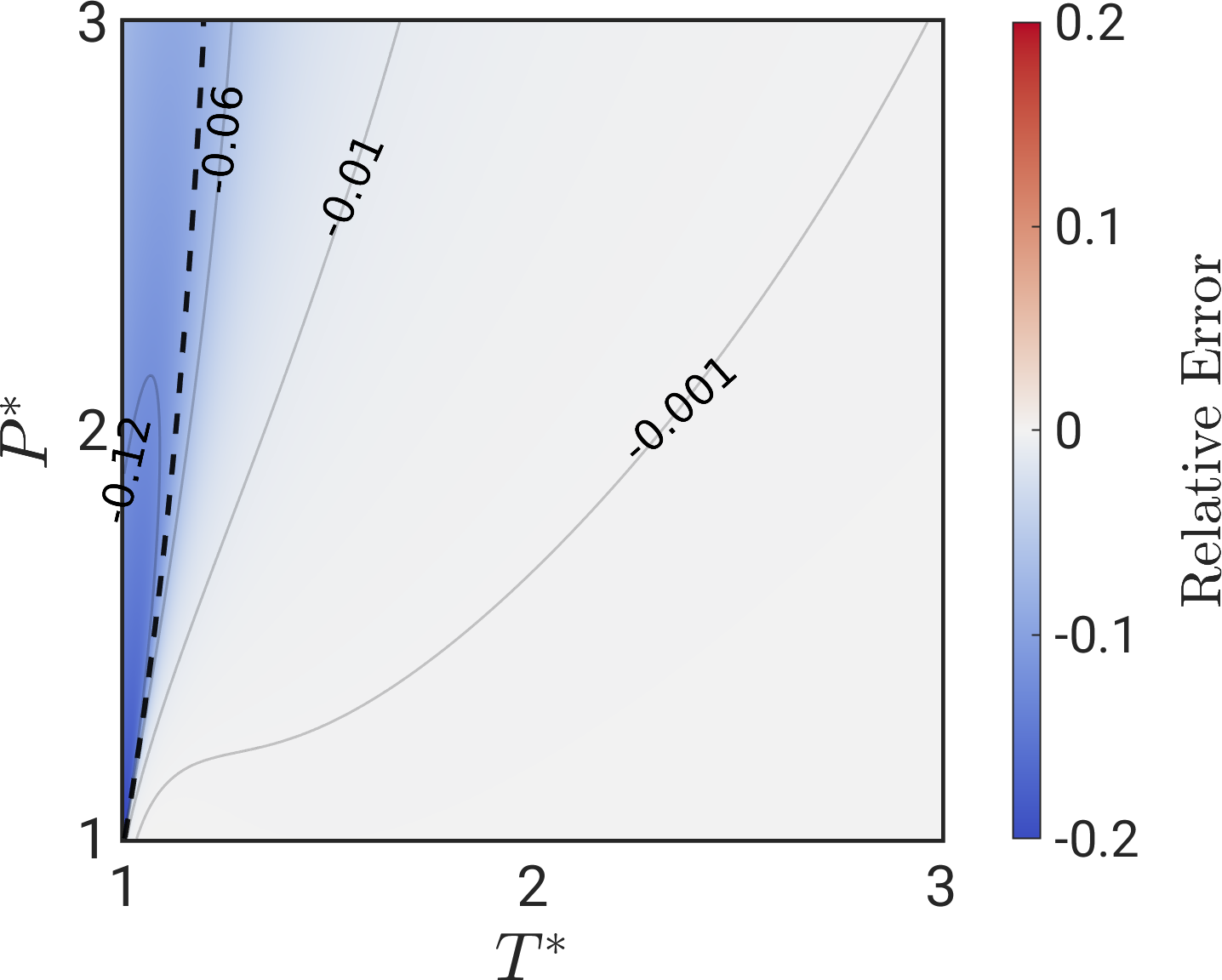}
        {(g) \ntwo{},  $\conderrfirst$}
    \hfill
    \panelThree
        {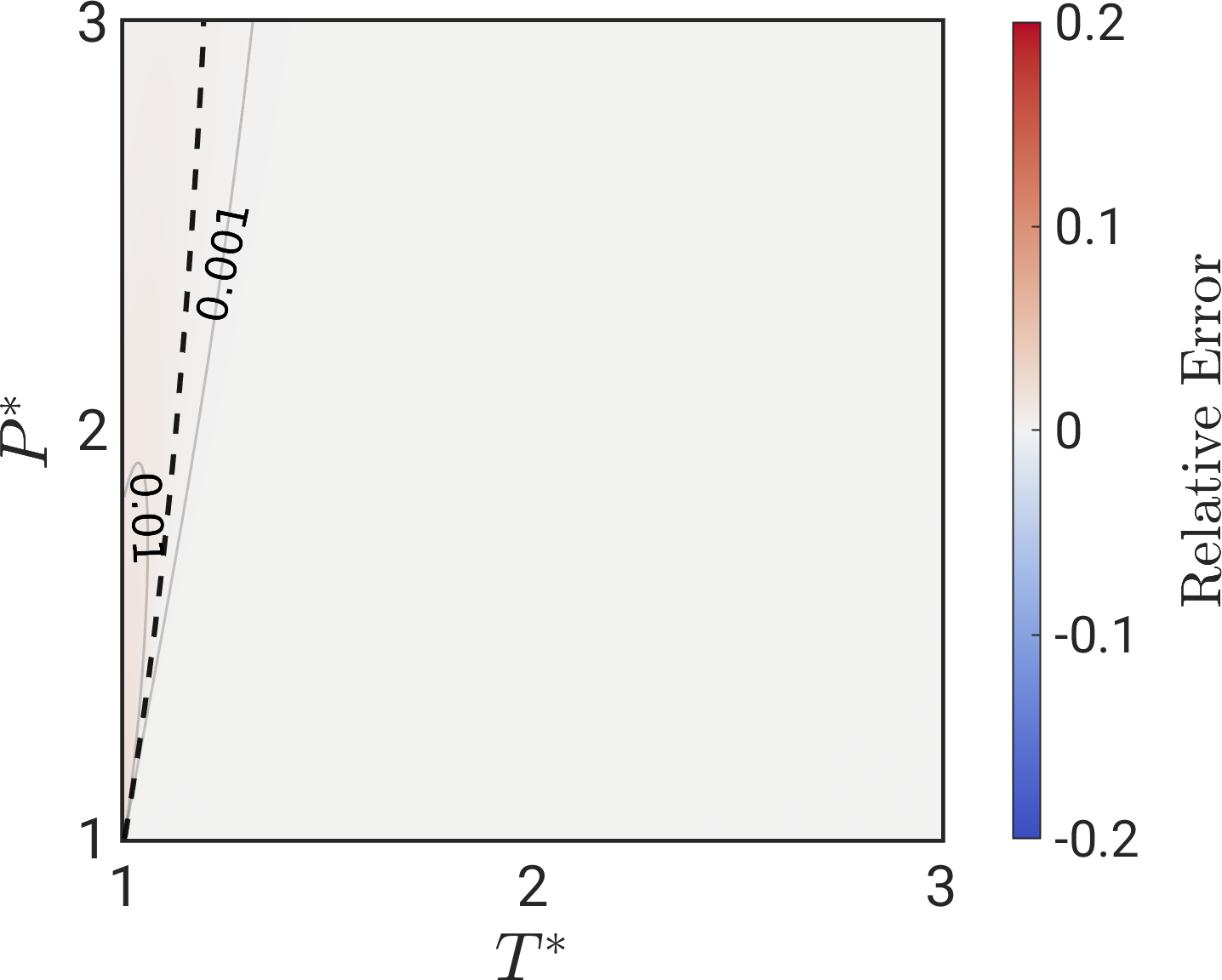}
        {(h) \ntwo{},  $\conderrsecond$}
    \vspace{0.5em} \\
    \hrule
    \vspace{0.5em}
    %
    % new row
    \panelThree
        {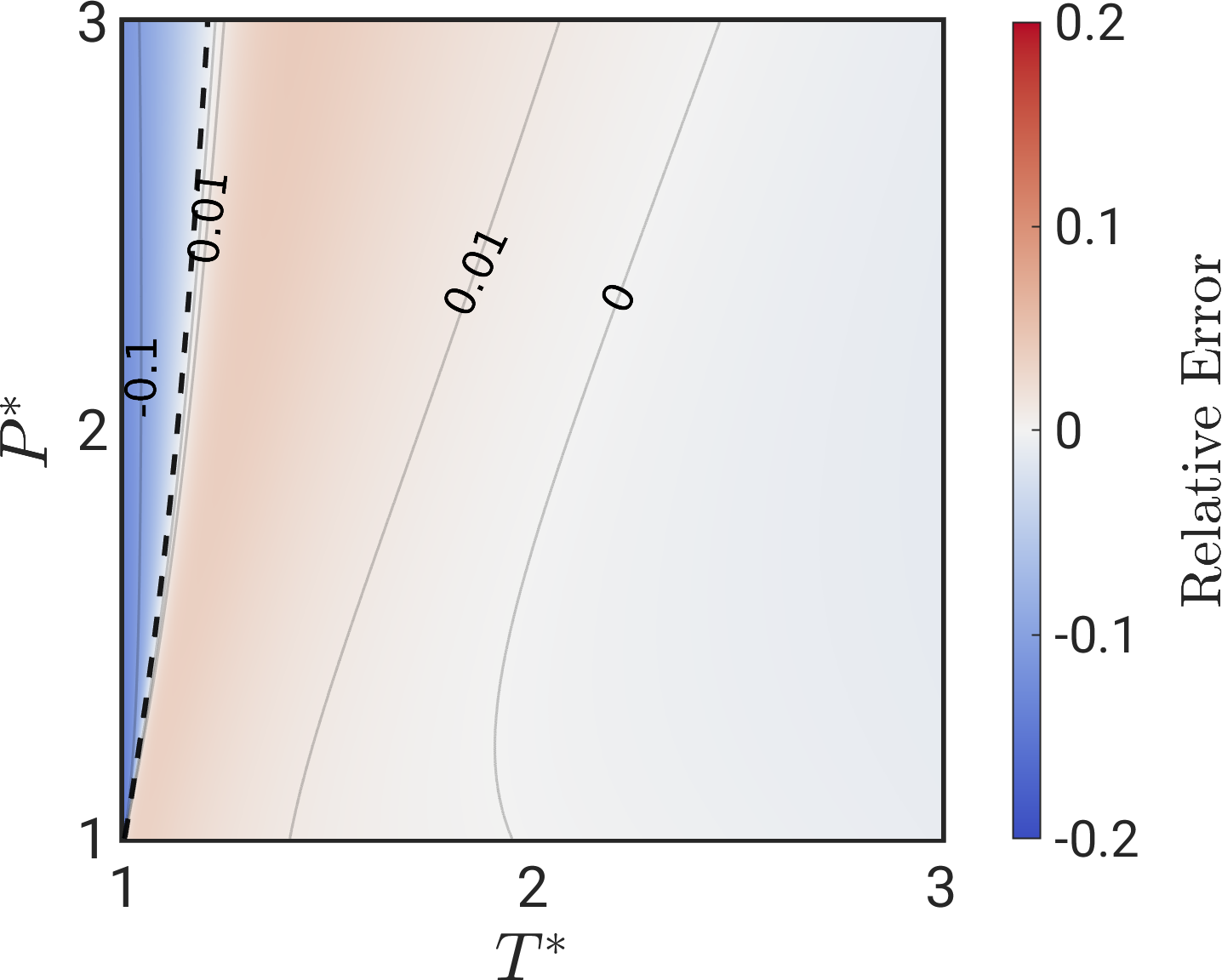}
        {(i) \chfour{},  $\conderrtot$}
    \hfill
    \panelThree
        {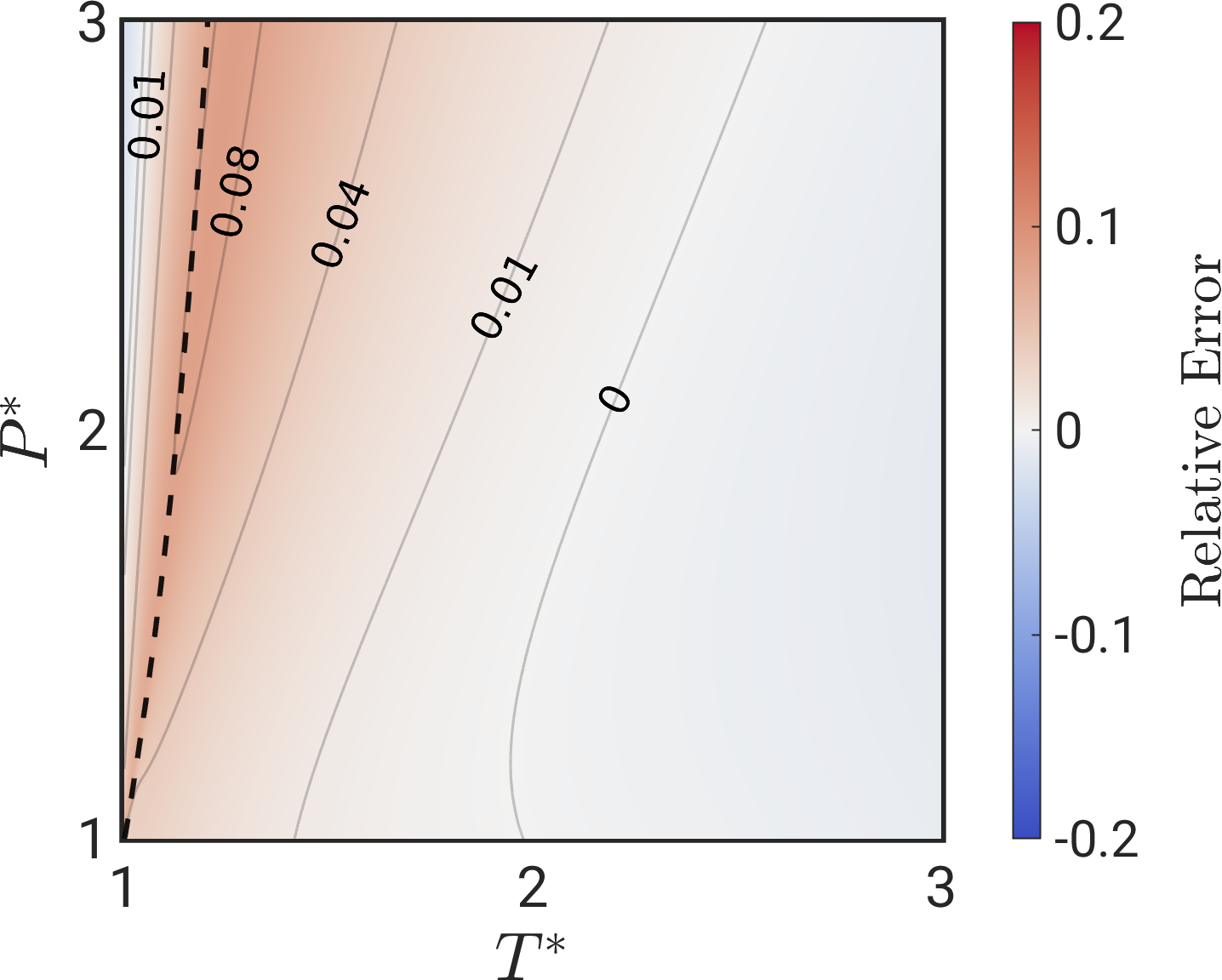}
        {(j) \chfour{},  $\conderrcorr$}
    \hfill
    \panelThree
        {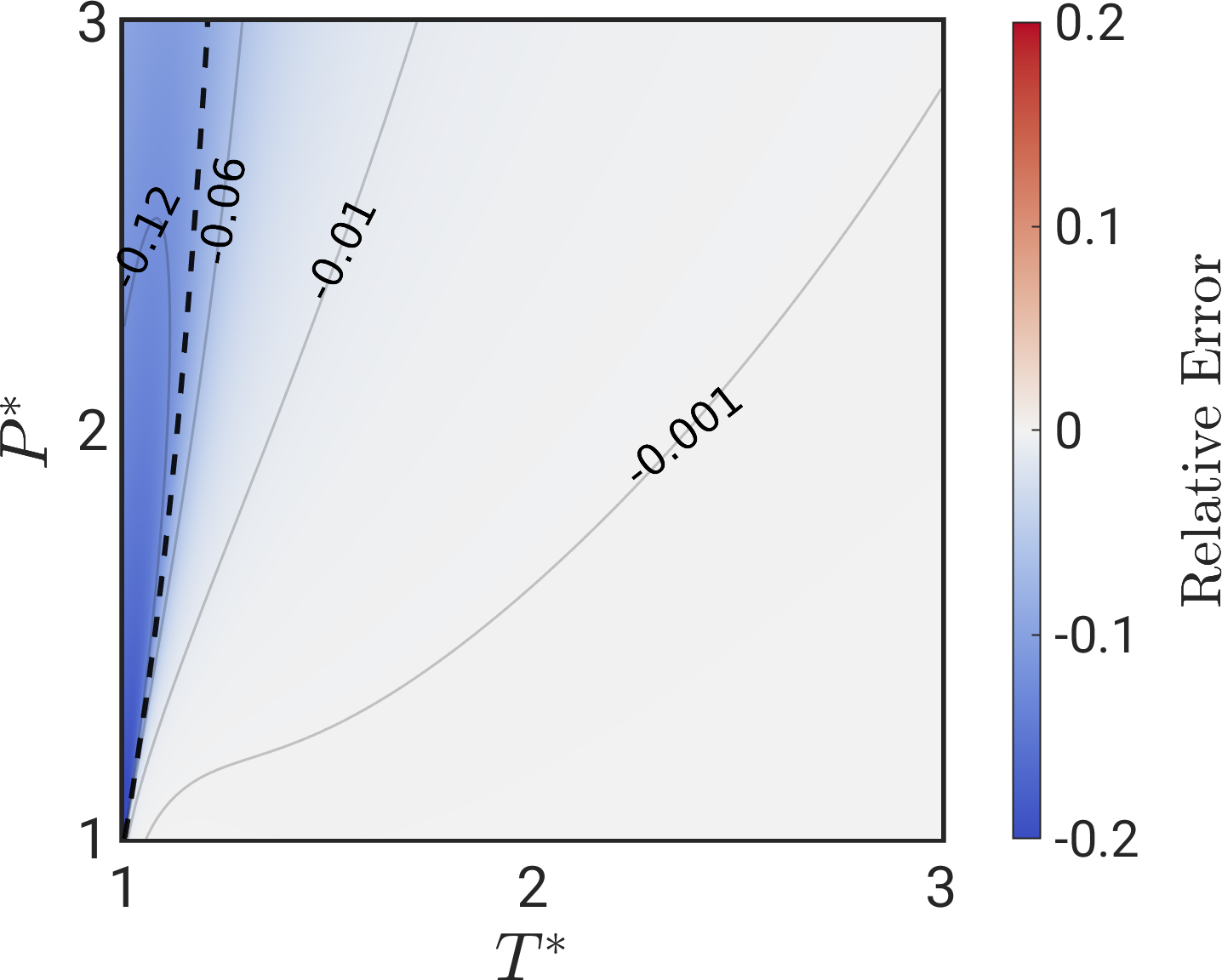}
        {(k) \chfour{},  $\conderrfirst$}
    \hfill
    \panelThree
        {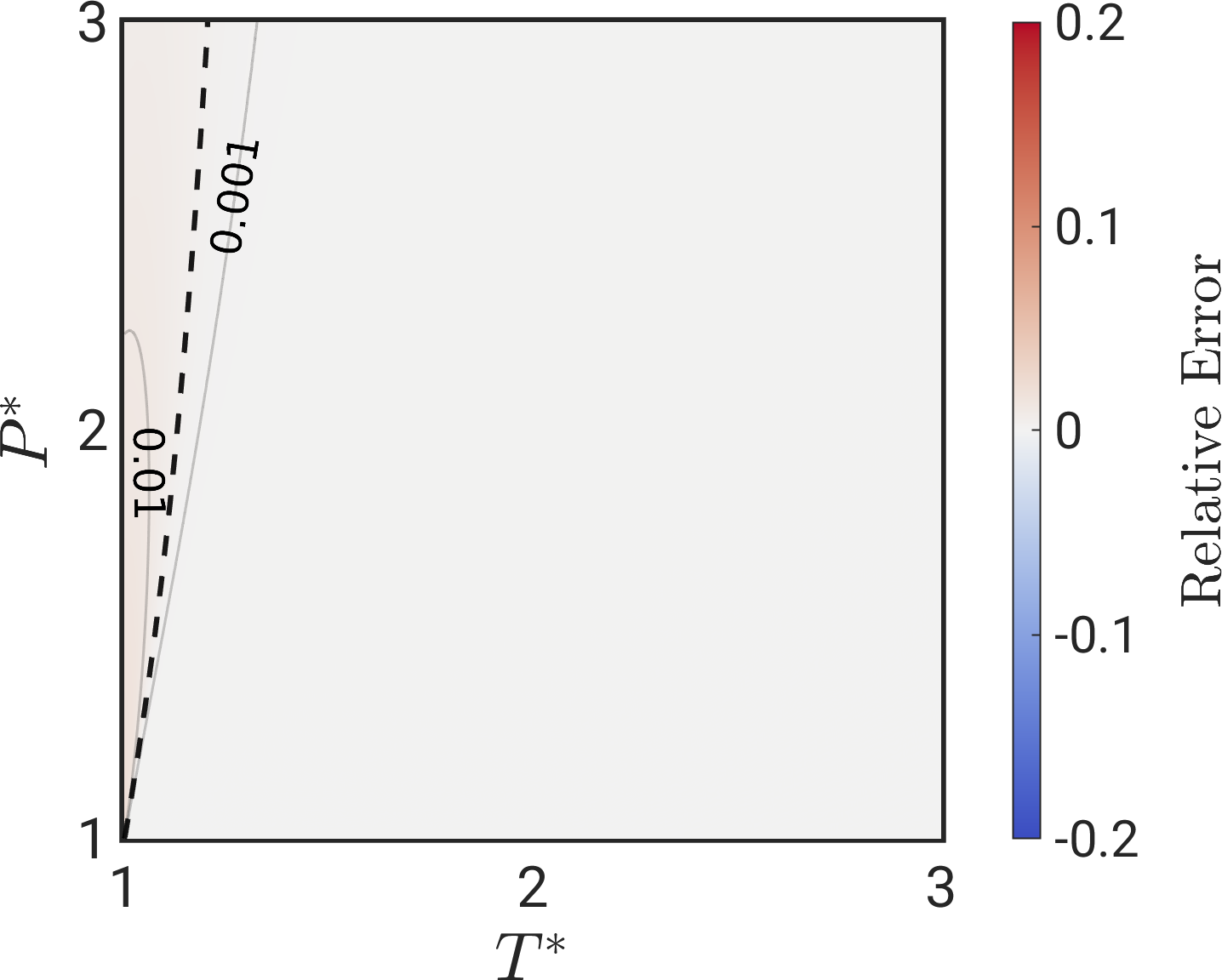}
        {(l) \chfour{},  $\conderrsecond$} 
    \vspace{1em} \\
    \caption{Cross-species viscosity results. Each row shows $\viscerrtot$, $\viscerrcorr$, $\viscerrfirst$, and $\viscerrsecond$ for a given species. The top row shows data for \cotwo{}, the middle row shows data for \ntwo{}, and the bottom row shows data for \chfour{}.}
    \label{fig:cross_species_viscosity}
\end{figure}

Recalling the decomposition
\begin{equation}
    \viscerrtot 
        = \viscerrcorr + \viscerrfirst + \viscerrsecond,
\end{equation}
Figure~\ref{fig:cross_species_viscosity} shows $\viscerrtot$, $\viscerrcorr$, $\viscerrfirst$, and $\viscerrsecond$ for \ntwo{} and \chfour{}. For ease of comparison, this figure also reproduces the similar data for \cotwo{}, which were originally presented in Figures~\ref{fig:CO2_viscosity} and \ref{fig:CO2_viscosity_error_decomposition}. Table~\ref{tab:cross_species_viscosity} gives maximum and minimum values for these errors in $\Omega$. 

For all three species, the total error exhibits the same trend; i.e. viscosity is underestimated in the liquid-like region and overestimated in the gas-like region. For all three species $\viscerrfirst$ is negative throughout $\Omega$. Recalling that $\viscerrfirst = S_1 \denserrtot$, the trend in $\viscerrfirst$ can be explained by the fact that the SRK EoS underestimates density for all three species (see Figure~\ref{fig:cross_species_density_error}), while $S_1$ (not shown) is positive throughout $\Omega$ for all three species, with the largest values of $S_1$ occurring in the liquid-like region. 

On the other hand, $\viscerrcorr$ exhibits a slightly different trend when comparing \cotwo{}, \ntwo{}, and \chfour{}. As observed in Section~\ref{sec:viscosity}, for \cotwo{} the Chung correlation itself overestimates viscosity throughout $\Omega$. However, for \ntwo{} and \chfour{}, the behavior of the Chung correlation is more complicated, with the Chung correlation itself sometimes overestimating and sometimes underestimating viscosity. 

\makeatletter
\def\photowidth{0.24}
\def\gutterwidth{0.5}
\makeatother

\begin{table}[hbt!]
    \centering
      \caption{Minimum and maximum viscosity errors for \cotwo{}, \ntwo{}, and \chfour{}. Data for $\viscerrtot$, $\viscerrcorr$, $\viscerrfirst$, and $\viscerrsecond$ are given.}
      
    \begin{tabular}{c |  c |  c | c }
    \hline
        & \cotwo{}
        & \ntwo{}
        & \chfour{} \\    
    \hline
    \ $\viscerrtot$, min 
        & -0.1850
        & -0.1824
        &  -0.1423\\
    \ $\viscerrtot$, max 
        & 0.1457
        & 0.0478
        &  0.0398 \\
        \hline
   \ $\viscerrcorr$, min 
       & 0.0052
       & -0.0727
       & -0.0238 \\
   \ $\viscerrcorr$, max 
       & 0.1762
       & 0.0714
       & 0.0846 \\
       \hline   
   \ $\viscerrfirst$, min 
       &  -0.3504
       & -0.2148
       & -0.2222 \\
   \ $\viscerrfirst$, max 
       & -2.6283e-04
       & 2.7756e-04
       &  -8.3134e-05 \\
       \hline   
   \ $\viscerrsecond$, min 
       & -1.9715e-06
       & -9.3180e-06
       &  -9.4826e-06 \\
   \ $\viscerrsecond$, max
       & 0.0651
       & 0.0215
       &  0.0199 \\
       \hline
    \end{tabular}
  
    \label{tab:cross_species_viscosity}
\end{table}

%%%
\subsection{Cross-Species Thermal Conductivity Results}
\begin{figure}[!hbt]
    \centering
    \panelThree
        {images/2d_results_NEW/CO2_2d_cond_err_tot.png}
        {(a) \cotwo{},  $\conderrtot$}
    \hfill
    \panelThree
        {images/2d_results_NEW/CO2_2d_cond_err_corr.png}
        {(b) \cotwo{},  $\conderrcorr$}
    \hfill
    \panelThree
        {images/2d_results_NEW/CO2_2d_cond_err_eos.png}
        {(c) \cotwo{},  $\conderrfirst$}
    \hfill
    \panelThree
        {images/2d_results_NEW/CO2_2d_cond_err_secondorder.png}
        {(d) \cotwo{},  $\conderrsecond$}
    \vspace{0.5em} \\
    \hrule
    \vspace{0.5em}
    % 
    % new row
    \panelThree
        {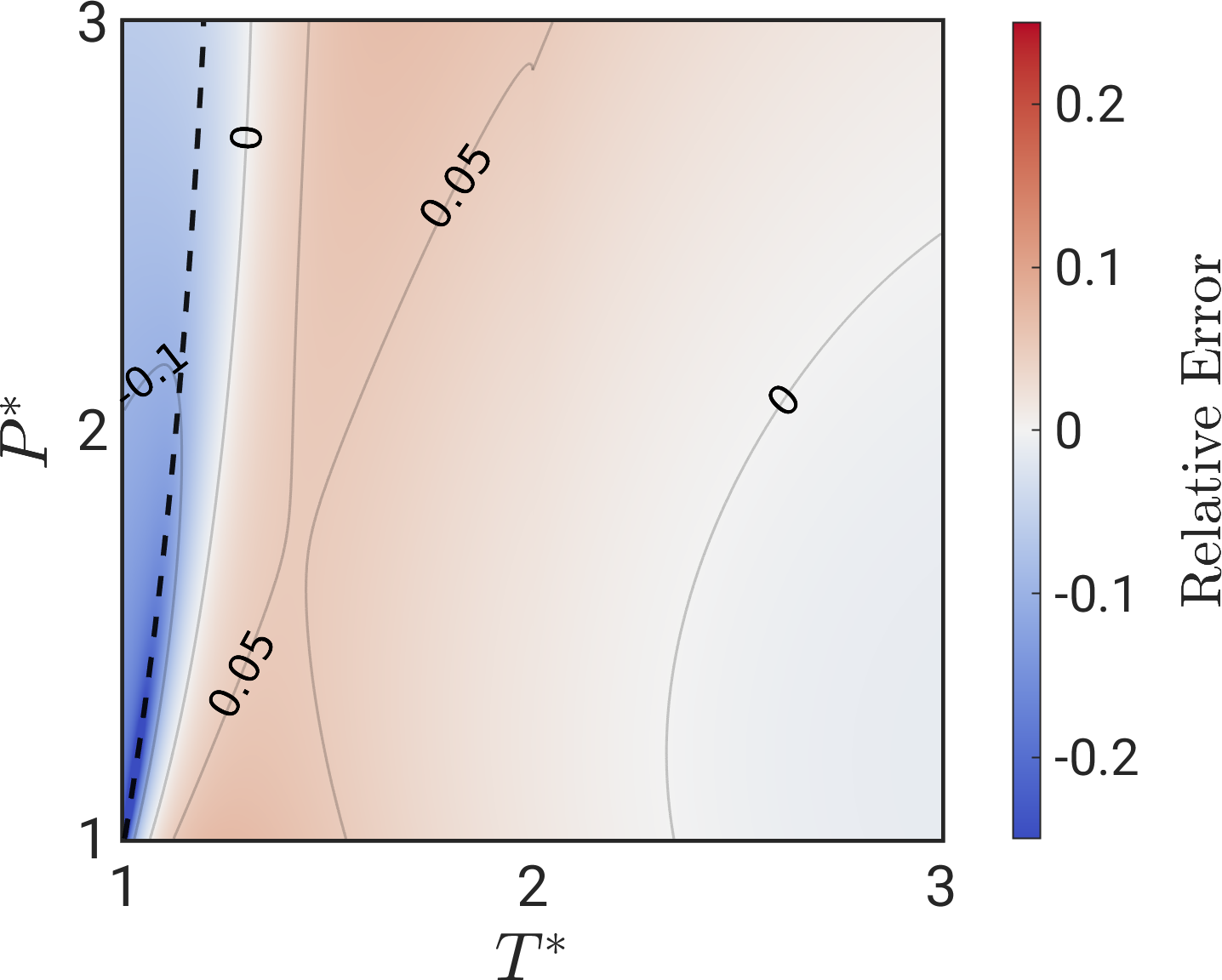}
        {(e) \ntwo{},  $\conderrtot$}
    \hfill
    \panelThree
        {images/2d_results_NEW/N2_2d_cond_err_corr.png}
        {(f) \ntwo{},  $\conderrcorr$}
    \hfill
    \panelThree
        {images/2d_results_NEW/N2_2d_cond_err_eos.png}
        {(g) \ntwo{},  $\conderrfirst$}
    \hfill
    \panelThree
        {images/2d_results_NEW/N2_2d_cond_err_secondorder.png}
        {(h) \ntwo{},  $\conderrsecond$}
    \vspace{0.5em} \\
    \hrule
    \vspace{0.5em}
    %
    % new row
    \panelThree
        {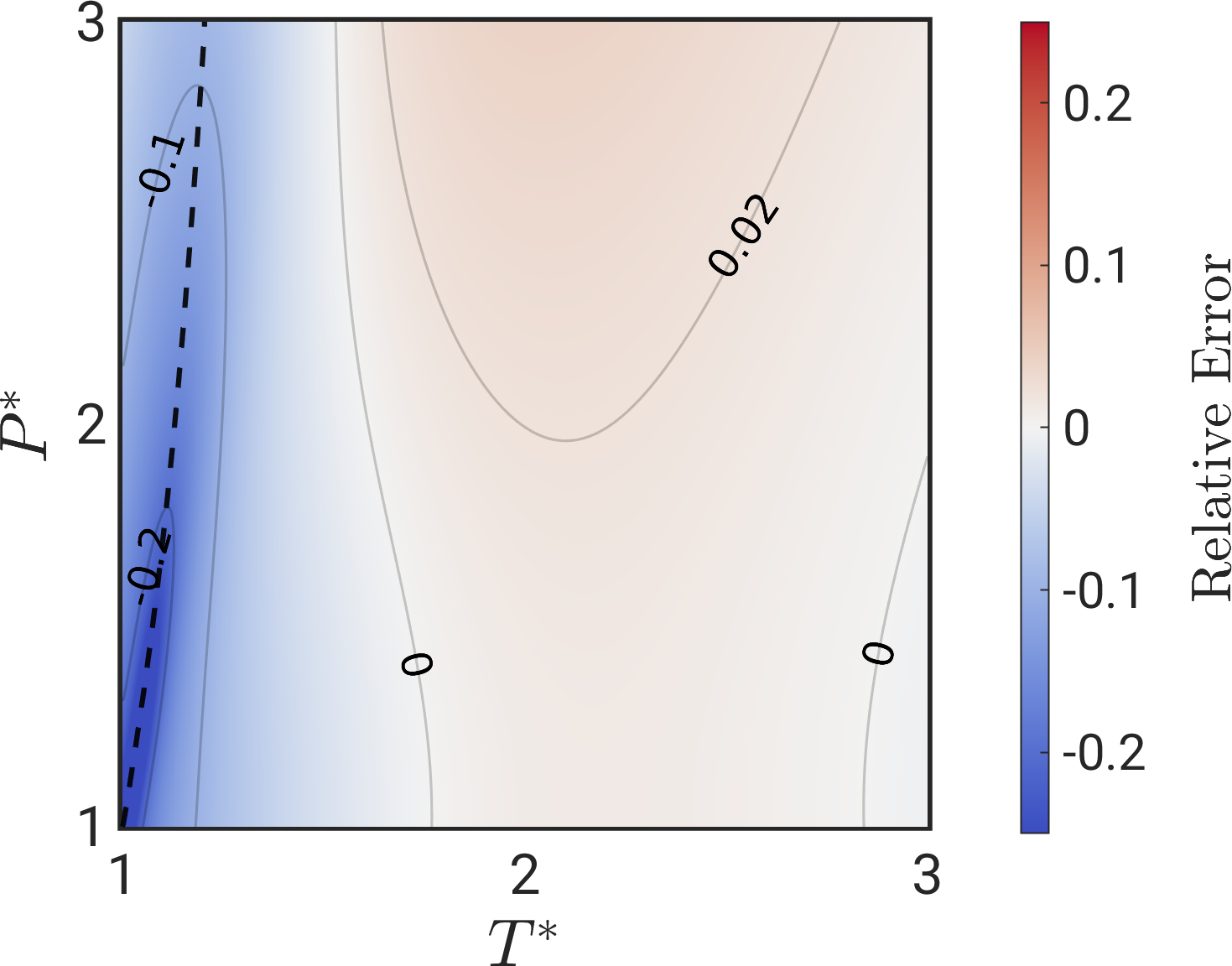}
        {(i) \chfour{},  $\conderrtot$}
    \hfill
    \panelThree
        {images/2d_results_NEW/CH4_2d_cond_err_corr.png}
        {(j) \chfour{},  $\conderrcorr$}
    \hfill
    \panelThree
        {images/2d_results_NEW/CH4_2d_cond_err_eos.png}
        {(k) \chfour{},  $\conderrfirst$}
    \hfill
    \panelThree
        {images/2d_results_NEW/CH4_2d_cond_err_secondorder.png}
        {(l) \chfour{},  $\conderrsecond$} 
    \vspace{1em} \\
    \caption{Cross-species thermal conductivity results. Each row shows $\conderrtot$, $\conderrcorr$, $\conderrfirst$, and $\conderrsecond$ for a given species. The top row shows data for \cotwo{}, the middle row shows data for \ntwo{}, and the bottom row shows data for \chfour{}. Note that the color-scale for \cotwo{} is different than for \ntwo{} and \chfour{}.}
    \label{fig:cross_species_conductivity}
\end{figure}

Figure~\ref{fig:cross_species_conductivity} shows $\conderrtot$, $\conderrcorr$, $\conderrfirst$, and $\conderrsecond$ for \ntwo{} and \chfour{} and also reproduces these data for \cotwo{}, which were originally shown in Figures~\ref{fig:CO2_conductivity} and \ref{fig:CO2_conductivity_error_decomposition}. Table~\ref{tab:cross_species_conductivity} gives maximum and minimum values for these quantities across $\Omega$. 

For all three species, the Chung correlation severely underestimates thermal conductivity at the critical point. However, as Figure~\ref{fig:cross_species_conductivity} shows, just away from the critical point, this underestimation reduces in magnitude to around $-10\%$. 

Unlike the analogous viscosity results, the pattern in total error for thermal conductivity is not consistent across all three species. One similarity between the viscosity and thermal conductivity results is that, for all three species, $\conderrfirst$ is negative throughout $\Omega$, with the most extreme values occurring in the liquid-like region. However, the error in the Chung correlation itself, which is shown by $\conderrcorr$, is quite different for the three species, leading to a different pattern in total error across the three species.

\makeatletter
\def\photowidth{0.24}
\def\gutterwidth{0.5}
\makeatother

\begin{table}[hbt!]
    \centering
   \caption{Minimum and maximum thermal conductivity errors for \cotwo{}, \ntwo{}, and \chfour{}.}
   % Data for $\conderrtot$, $\conderrcorr$, $\conderrfirst$, and $\conderrsecond$ are given.}
      
    \begin{tabular}{c |  c |  c | c }
    \hline
        & \cotwo{}
        & \ntwo{}
        & \chfour{} \\    
    \hline
    \ $\conderrtot$, min 
        & -0.9371
        & -0.9370
        &  -0.9258 \\
    \ $\conderrtot$, max 
        & 0.2061
        &  0.0717
        &  0.0403  \\
        \hline
   \ $\conderrcorr$, min 
       & -0.9308
       & -0.9303
       &  -0.9085\\
   \ $\conderrcorr$, max 
       & 0.4953
       & 0.0825
       & 0.0544 \\
       \hline   
   \ $\conderrfirst$, min 
       &  -0.3467
       & -0.1419
       &  -0.1574\\
   \ $\conderrfirst$, max 
       & -4.0359e-04
       & 2.8631e-04
       &  -6.8765e-05 \\
       \hline   
   \ $\conderrsecond$, min 
       & -0.0106
       & 0.0091
       &  -0.0090 \\
   \ $\conderrsecond$, max
       & 0.0112
       & 3.9471e-05
       & -6.3891e-06  \\
       \hline
    \end{tabular}
    \label{tab:cross_species_conductivity}
\end{table}

%%%
\subsection{Cross-Species Kinematic Viscosity Results}
The ratio $\nu_\text{ref} / \nu_\text{gen}$ is of interest for these gases as well. The minimum and maximum values of $\nu_\text{ref} / \nu_\text{gen}$ on $\Omega$ for \cotwo{}, \ntwo{}, and \chfour{}, as well as the $(T,P)$ values that produce these extrema; i.e. the pair $(T_\text{min}^{(\nu)}, P_\text{min}^{(\nu)})$ is the pair that produces $\min(\nu_\text{ref} / \nu_\text{gen})$ and similarly $(T_\text{max}^{(\nu)}, P_\text{max}^{(\nu)})$ produces $\max(\nu_\text{ref} / \nu_\text{gen})$ define the potential impacts of prediction of Reynolds number. These values are summarized in Table~\ref{tab:cross_species_kinematic_viscosity}.

For each species, the magnitude of $\nu_\text{ref} / \nu_\text{gen}$ follows the same pattern, with $\nu_\text{ref} / \nu_\text{gen} > 1$ throughout most of the liquid-like region and $\nu_\text{ref} / \nu_\text{gen} < 1$ throughout most of the gas-like region, and with the transition $\nu_\text{ref} / \nu_\text{gen} = 1$ roughly corresponding to the Widom line. See, for example, Figure~\ref{fig:CO2_kinematic_viscosity}.

\begin{table}[hbt!]
    \centering
    \caption{Minimum and maximum ratios of kinematic viscosity $\nu_\text{ref} / \nu_\text{gen}$. Shown are the minimum and maximum values of $\nu_\text{ref} / \nu_\text{gen}$ along with the temperature and pressure pairs that produce these extrema. For each temperature and pressure, both the absolute quantity and the reduced quantity are shown.}
      
    \begin{tabular}{c |  c |  c | c }
    \hline
        & \cotwo{}
        & \ntwo{}
        & \chfour{} \\    
    \hline
        \ $T_{\min}^{(\nu)}$, $P_{\min}^{(\nu)}$  
          &  $304.1$,  $74.10$ 
          & $140.7$, $34.18$ 
          &   $501.0$, $57.37$  \\
        \ $T_{\min}^{(\nu)} \, / \, T_{c}$, $P_{\min}^{(\nu)}$ / $P_{c}$  
          &  $1.000$, $1.0178$ 
          & $1.114$, $1.020$
          &  $2.676$, $1.264$ \\    
       \ $\min (\nu_\text{ref} / \nu_\text{gen})$
         & $0.8344$
         & $0.9246$
         & $0.9124$\\
    \hline  
       \ $T_{\max}^{(\nu)}$, $P_{\max}^{(\nu)}$ 
          & $538.9$, $72.81$
          &  $333.7$, $33.51$
          &  $571.7$, $45.39$ \\ 
       \ $T_{\max}^{(\nu)} \, / \, T_{c}$, $P_{\max}^{(\nu)} / P_{c}$   
          & $1.772$,  $1.000$ 
          & $2.644$, $1.000$
          & $3.000$, $1.000$ \\   
       \ $\max (\nu_\text{ref} / \nu_\text{gen})$ 
          & $1.029$
          & $1.122$
          &  $1.070$\\
       \hline
    \end{tabular}

  \label{tab:cross_species_kinematic_viscosity}
\end{table}

In particular, for each species, $\nu_\text{ref} / \nu_\text{gen}$ is maximized in the liquid-like region somewhere along the isotherm corresponding to the critical temperature. On the other hand, it noteworthy that for \cotwo{} and \ntwo{} the minimizer of $\nu_\text{ref} / \nu_\text{gen}$ is near the critical point while for \chfour{} this minimizer is far from the critical point. However, there is pattern here as well: for all three species, include \chfour{}, this minimizer occurs near the Widom line on the gas-like side.

%%%
\subsection{Cross-Species Stability Results}
$\Delta \rho_*^{(\visc)}$ and $\Delta \rho_*^{(\cond)}$ can interpreted as the smallest density errors that could cause the Chung correlation to output a negative transport property. For all three species, $\Delta \rho_*^{(\visc)} < 0$ and $\Delta \rho_*^{(\cond)} < 0$ throughout $\Omega$, indicating that for all considered temperatures and pressures a given EoS would need to underestimate density in order to produce a negative value of viscosity or thermal conductivity. Moreover, recall from the discussion in Section~\ref{sec:viscosity_stability} that a relatively small $|\Delta \rho_*|$ value means that a relatively small density error (e.g. an underestimation of about 50\%) could cause a negative corresponding transport quantity. For each species, $\min |\Delta \rho_*^{(\visc)}|$ and $\min |\Delta \rho_*^{(\cond)}|$, as well as the temperature and pressure pairs that produce these minima; i.e. $(T_{\min}^{(\visc)}, P_{\min}^{(\visc)})$ is the pair that produces $\min |\Delta \rho_*^{(\visc)}|$ and similar for $(T_{\min}^{(\cond)},P_{\min}^{(\cond)})$ are calculated and presented in Table~\ref{tab:cross_species_density_floor}.

For each species, the magnitudes of both $\Delta \rho_*^{(\visc)}$ and $\Delta \rho_*^{(\cond)}$ follow the same pattern, with the least negative values in the top-left of $\Omega$ and the most negative values in the bottom-right of $\Omega$. See, for example, Figures~\ref{fig:CO2_viscosity_density_floor} and \ref{fig:CO2_conductivity_density_floor}. In particular, for all three species, both $\Delta \rho_*^{(\visc)}$ and $\Delta \rho_*^{(\cond)}$ are minimized at $(T_c,3P_c)$, where $3P_c$ was the largest pressure considered. For all species, throughout the liquid-like region, $0 < |\delta \rho_*^{(\visc)}| < 1$ and $0 < |\delta \rho_*^{(\cond)}| < 1$. Recall from the discussion in Section~\ref{sec:viscosity_stability} that these magnitudes indicate the theoretical possibility of obtaining a negative transport property while $\densgen$ is positive. In other words, assuming that any EoS will produce positive values of density, it is nevertheless theoretically possible for all three species that an EoS could underestimate density in the liquid-like region so severely as to produce a negative value of viscosity or thermal conductivity.

\begin{table}[hbt!]
    \centering
    \caption{Minimum magnitudes of $\Delta \rho_*^{(\eta)}$, $\delta \rho_*^{(\eta)}$, $\Delta \rho_*^{(\lambda)}$, and $\delta \rho_*^{(\lambda)}$. Shown are the minimum magnitudes of these quantities, as well as the temperatures $T_\text{min}^{(\eta)}$ and $T_\text{min}^{(\lambda)}$ and pressures $P_\text{min}^{(\eta)}$ and $P_\text{min}^{(\lambda)}$ that produce these minima. Note that for all species, each quantity is minimized at $(T,P) = (T_c, 3P_c)$.} 
      
    \begin{tabular}{c |  c |  c | c }
    \hline
        & \cotwo{}
        & \ntwo{}
        & \chfour{} \\    
    \hline
        \ $T_{\min}^{(\visc)}$, $P_{\min}^{(\visc)}$ 
          &  $304.1$,  $221.4$
          &  $126.2$,  $100.5$
          &  $190.6$,  $136.2$ \\
        \ $T_{\min}^{(\visc)} \, / \, T_{c}$, $P_{\min}^{(\visc)}$ / $P_{c}$  
          & 1.0, 3.0
          &  1.0, 3.0
          &  1.0, 3.0 \\   
       \ $\min |\Delta \rho_*^{(\visc)}|$
          & $298.1$ 
          & $305.4$ 
          & $161.3$  \\   
        \ $\min |\delta \rho_*^{(\visc)}|$
          & $0.3310$
          &  $0.5108$
          &  $0.5832$  \\  
    \hline  
        \ $T_{\min}^{(\cond)}$,  $P_{\min}^{(\cond)}$
          &  $304.1$, $221.4$ 
          &  $126.2$,  $100.5$
          &  $190.6$,  $136.2$  \\
        \ $T_{\min}^{(\cond)} \, / \, T_{c}$, $P_{\min}^{(\cond)}$ / $P_{c}$  
          & 1.0, 3.0
          &  1.0, 3.0
          &  1.0, 3.0 \\
       \ $\min|\Delta \rho_*^{(\cond)}|$
          & $294.9$
          & $348.7$ 
          &  $164.8$   \\   
        \  $\min|\delta \rho_*^{(\cond)}|$
          & $0.3275$
          & $0.5832$
          &  $0.5300$  \\  
       \hline
    \end{tabular}

      \label{tab:cross_species_density_floor}
\end{table}

The cross-species comparison shows that the issue of error propagation in use of correlations is not limited to s\cotwo{}. It is likely to occur for almost any system where transport property correlations are not explicitly fit to the EoS used in the calculation.

%%%%%%%%%%%%%%%%%%%%%%%%%%%%%%%%%%%%%%%%%%%%%%%
% section: end

% section: begin
%%%%%%%%%%%%%%%%%%%%%%%%%%%%%%%%%%%%%%%%%%%%%%%
\section{Conclusion}
These results show the key limitations of using generalized equations of state with generalized correlations for transport properties. Both generalized equations of state and generalized correlations for transport properties contain inherent errors, due to the need to fit over multiple species.  The inherent curve-fit error in the transport equations can then be exacerbated by the mismatch in density created by the use of a generalized equation of state.

These effects lead to substantial challenges in interpretation of computational results for mixtures of gases at supercritical temperatures and pressures at both \textit{global} and \textit{local} scales within the simulation. The combined effect of errors in density and viscosity mean that the simulated system may have a Reynolds number that is ten to fifteen percent different from the Reynolds number of the physical system. These effects are then exacerbated by local errors in key transport properties. While the effect is most likely to be seen in reduced accuracy of simulation results, in extreme cases mismatch can compromise the numeric stability of the simulation either through changing the largest possible time step, as shown in equations \ref{eqn:stability} and \ref{eqn:stability1}, or creating negative values for transport properties.

For a simulation of only one species, the errors created by the use of a generalized equation of state and generalized transport properties, and the propogation of density errors in the transport property correlations, can be quantified by comparison with simulations that use higher-fidelity EoSs and transport property correlations. However, in the absence of high-fidelity models for mixtures, these errors cannot be quantified as effectively for simulations of mixtures, particularly reacting flow in the supercritical regions. However, there are two options that are likely to improve the accuracy of these simulations. 

The first option for improving accuracy of simulations of supercritical mixtures is the \textit{co-development of generalized EoSs and transport property correlations} for density, viscosity, and thermal conductivity. Once a generalized equation of state $\densn(T,P)$ is developed, a correlation for viscosity $\viscn(T,\densn(T,P))$ and thermal conductivity $\condn(T,\densn(T,P))$ that specifically use $\densn(T,P)$ and $T$ as curve-fitting variables, instead of the most accurate possible value of density $\densref(T,P)$ and $T$, will provide the most accurate representation of the physics of the system.

The availability of machine learning tools that allow models to be fit for large data sets relatively rapidly means that accuracy can be further improved by \textit{developing combinations of equations of state and correlations of viscosity and thermal conductivity that are specific for engineering systems.} In this approach, if an engineering application is known to have a set of species over a likely range of temperatures and pressures, the equation of states and transport properties correlations can be refit using \textit{only the species that are known to exist in the system, over the pressures and temperatures that are likely to exist within the system.}  Table 6 shows representative supercritical systems, and the species present within each system.

\begin{table}[hbt!]
\caption{\label{tab:applications} Representative Multi-Species Supercritical Systems}
\centering
\begin{tabular}{ccc } 
\hline
Application & Species with significant & Radicals and Trace \\
 &  mass fractions & Species \\
    \hline
    $\text{H}_{2}$ $\text{-O}_{2}$, 
        & $\text{ O}_{2}$,  
            $\text{ H}_2$, 
            $\text{ H}_{2}\text{O}$  
        & $\text{H}$, 
            $\text{ OH}$, 
            $\text{ O}$, 
            $\text{ H}_{2}\text{O}_2$ \\
        $\text{Liquid Rocket}$ 
        &  
        &  \\
        \hline
    $\text{Methane}$  
        & $\text{CH}_4$, 
            $\text{ O}_2$, 
            $\text{ CO}_2$, 
            $\text{ CO}$    
        & $\text{N}$, 
            $\text{ H}$, 
            $\text{ OH}$, 
            $\text{ O}$, 
            $\text{ CH}_3,$
          \\
   $\text{combustion}$   
        &   $\text{ H}_{2}\text{O}$, 
            $\text{ N}_{2}$, 
            $\text{ NO}$     
        &   $\text{ CH}_3\text{O}_2$, 
            $\text{ CH}_3\text{O}_2\text{H}$, 
            $\text{ CH}_3\text{O}_2$,  \\
    $\text{in air}$   
        &       
        &   $\text{ CH}_3\text{O}$, 
            $\text{ CH}_2\text{O}_2$, 
            $\text{ HCO}$ \\
    \hline
    $\text{Methane}$ 
        & $\text{CH}_4$, 
            $\text{ O}_2$, 
            $\text{ CO}_2$,    
        & $\text{N}$, 
          $\text{ H}$, 
          $\text{ OH}$, 
          $\text{ O}$, 
          $\text{ CH}_3,$
          \\
    $\text{oxycombustion}$ 
            & $\text{ CO}$, 
            $\text{ H}_{2}\text{O}$      
        & $\text{ CH}_3\text{O}_2$,
        $\text{ CH}_3\text{O}$, 
        $\text{ CH}_2\text{O}_2$, \\
     $\text{in sCO}_2$
        &       
        &   $\text{ CH}_3\text{O}$, 
            $\text{ CH}_2\text{O}_2$, 
            $\text{ HCO}$ \\
    \hline
     $\text{Venusian}$ 
        & $\text{CO}_2$, 
            $\text{ N}_2$   
        & $\text{SO}_2$, 
            $\text{ Ar}$, 
            $\text{ H}_2\text{O}$, 
            $\text{ CO}$, \\
         $\text{atmosphere}$ 
        & 
        &   $\text{ He}$, 
            $\text{ Ne}$, 
            $\text{ HCl}$, 
            $\text{ HF}$     \\
    \hline
\end{tabular}
\end{table}

The relative difficulty of creating these systems will vary with the application. For instance, for modeling the Venusian atmosphere, both the major species and the trace species are well-characterized, and an equation of state, viscosity correlation, and thermal conductivity correlation is available for all species. This may not be true for all systems. A bespoke approach to calculating properties for individual simulations or families of simulations does raise issues of direct comparibility of results obtained using other simulations.  This may limit the viability of such an approach.  

Engineering applications with fluids operating at supercritical temperatures and pressures, and the need to accurately simulate these systems, is continuing to increase. Improving the accuracy of these simulations requires re-examining how key properties calculations are implemented, and the impact of errors in these models. Ensuring consistency in EoSs and transport property calculations is a possible route towards achieving this goal.  

%%%%%%%%%%%%%%%%%%%%%%%%%%%%%%%%%%%%%%%%%%%%%%%
% section: end

% section: begin
\section*{Funding Sources}
%%%%%%%%%%%%%%%%%%%%%%%%%%%%%%%%%%%%%%%%%%%%%%%

This work was authored by NLR for the U.S. Department of Energy (DOE), operated under Contract No. DE-AC36-08GO28308. Funding was provided by the Exascale Computing Project (17-SC-20-SC), a collaborative effort of two U.S. Department of Energy (DOE) organizations, the Office of Science and the National Nuclear Security Administration, and the High Performance Computing for Energy Innovation (HPC4EI) program. Mr. Fara's participation was made possible by a fellowship from the National Science Foundation Mathematical Sciences Graduate Student Internship Program (NSF-MSGI). The research was performed using computational resources sponsored by the Department of Energy located at the National Laboratory of the Rockies. The views expressed in the article do not necessarily represent the views of the DOE or the U.S. Government. The U.S. Government retains and the publisher, by accepting the article for publication, acknowledges that the U.S. Government retains a nonexclusive, paid-up, irrevocable, worldwide license to publish or reproduce the published form of this work, or allow others to do so, for U.S. Government purposes.

%%%%%%%%%%%%%%%%%%%%%%%%%%%%%%%%%%%%%%%%%%%%%%%
% section: end

% section: begin
\section*{Appendix}
%%%%%%%%%%%%%%%%%%%%%%%%%%%%%%%%%%%%%%%%%%%%%%%

The sensitivity terms $S_1$ and $S_2$ used in this paper are the partial derivatives of $\visc$ or $\cond$ as computed by the Chung correlation with respect to $\rho$. For convenience, in this Appendix, these sensitivity terms are provided. 

In the Chung correlation, viscosity (in units of Poise, P) is computed as the sum 
\begin{subequations}
    
\begin{equation}
    \eta = \eta_k + \eta_p
\end{equation}
where the density-corrected kinetic term $\eta_k$ and the potential term $\eta_p$ are given by
\begin{equation}
    \eta_k = \eta_0 \,\left[  \frac{1}{G_2} + A_6 Y \right], 
\end{equation}
\begin{equation}
    \eta_p = K^{(\eta)} \, Y^2 \, G_2. 
\end{equation}
\end{subequations}
The definition of the dilute gas viscosity $\eta_0$ (in Poise, P) can be found in \cite{chung}. The reduced density parameter $Y = Y(\rho)$, the Enskog hard-sphere correction factor $G_1 = G_1(Y)$, and the empirical correction factor $G_2 = G_2(Y,G_1)$ are given by
\begin{subequations}
\begin{equation}
    \label{eq:ydef}
    Y = \frac{V_c \rho}{6}, 
\end{equation}
\begin{equation}
   \label{eq:g1def}
    G_1 = \frac{1.0 - 0.5 \,Y}{(1.0 - Y)^3} 
\end{equation}
\begin{equation}
    G_2 = C^{(\eta)} \left[ \, \frac{A_1}{Y}[1.0 - \exp(-A_4 Y)] \, \, 
        + \, A_2 \,G_1 \exp(A_5 Y) 
        + \, A_3 \, G_1 \right],
\end{equation}
\end{subequations}
where the symbols $K^{(\eta)}$ and $C^{(\eta)}$ have been introduced as shorthand for the expressions
\begin{subequations}
\begin{equation}
    K^{(\eta)} 
        = 36.344 \times 10^{-6} \frac{(M T_c)^{1/2}}{V_c^{2/3}} \, A_7 
        \exp \left( A_8 + \frac{A_9}{T^*_\text{LJ}} + \frac{A_{10}}{(T^*_\text{LJ})^2} \right) 
\end{equation}
\begin{equation}
    C^{(\eta)} 
        = 1 / (A_1 A_4 + A_2 + A_3).
\end{equation}
\end{subequations}

Definitions of the constants $A_i$, $1 \leq i \leq 10$ are found in \cite{chung}. The quantity $V_c$ is the critical volume in cm$^3$/mol, $T_c$ is critical temperature in K, $\rho$ is molar density in units mol/cm$^3$, and $T^*_\text{LJ} = k_\text{B} T / \varepsilon$ is the Lennard--Jones reduced temperature with $k_\text{B}$ the Boltzmann constant and $\varepsilon$ the Lennard--Jones characteristic energy parameter.

Using the chain rule for partial derivatives, $S_1 := \partial \eta/ \partial \rho$ can be expanded as
\begin{subequations}
\begin{equation}
    S_1 
        = \left[ \frac{\partial \eta}{\partial Y}
        + \frac{\partial \eta}{\partial G_2} 
            \, \frac{\partial G_2}{\partial Y}
        + \frac{\partial \eta }{\partial G_2} 
            \, \frac{\partial G_2}{\partial G_1}
            \, \frac{\partial G_1}{\partial Y} \right]
        \, \frac{\partial Y}{\partial \rho}
\end{equation}
where
\begin{equation}
    \label{eq:derY}
    \frac{\partial Y}{\partial \rho} 
        = \frac{V_c}{6},
\end{equation}
\begin{equation}
\label{eq:derG1}
    \frac{\partial G_1}{\partial Y} 
        = \frac{2.5 - Y}{(1.0 - Y)^4}, \\
\end{equation}
\begin{equation}
       \frac{\partial G_2}{\partial Y} 
        = C^{(\eta)} \left[ \frac{A_1 A_4}{Y} \exp(-A_4 Y) 
        - \frac{A_1}{Y^2} (1.0 - \exp(-A_4 Y)) 
         + A_2 A_5 G_1 \exp (A_5 Y) \right], 
\end{equation}
\begin{equation}
    \frac{\partial \eta}{\partial Y} 
        = \eta_0 A_6 + 2.0 K^{(\eta)} Y G_2, 
\end{equation}
\begin{equation}
    \frac{\partial \eta}{\partial G_2} 
        = - \frac{\eta_0}{G_2^2} + K^{(\eta)} Y^2. 
\end{equation}
\end{subequations}

Similarly, the Chung correlation computes the thermal conductivity in units of cal/cm-s-K as
\begin{subequations}
\begin{equation}
    \lambda = \lambda_k + \lambda_p,
\end{equation}
where
\begin{equation}
    \lambda_k = \lambda_0 \left[ \frac{1}{H_2} + B_6 Y \right], 
\end{equation}

\begin{equation}
    \lambda_p = K^{(\lambda)} \, Y^2 \, H_2,
\end{equation}
\end{subequations}

where $Y$ and $G_1$ are as in \eqref{eq:ydef} and \eqref{eq:g1def} and the quantity $H_2(Y,G_1)$ is given by
\begin{subequations}
\begin{equation}
    H_2 = C^{(\lambda)} \left[ B_1 [1 - \exp(-B_4 Y)] \, / \, Y 
        + B_2 \, G_1 \exp(B_5 Y) + B_3 \, G_1 \right],
\end{equation}
and where the symbols $K^{(\lambda)}$ and $C^{(\lambda)}$ are shorthand for
%
%\begin{eqnarray}
\begin{equation}
    K^{(\lambda)} 
        = 3.039 \times 10^{-4} \frac{(T_c / M)^{1/2}}{V_c^{2/3}} 
        B_7 \, \left(\frac{T}{T_c}\right)^{1/2}, 
\end{equation}
\begin{equation}
     C^{(\lambda)} = \frac{1}{B_1 B_4 + B_2 + B_3}.
\end{equation}
%\end{eqnarray}
\end{subequations}
The definitions of $\lambda_0$ and $B_i$, $1 \leq i \leq 7$ are found in \cite{chung}.

Therefore the sensitivity factor $S_2 := \partial \lambda / \partial \rho$ is: 
\begin{subequations}
\begin{equation}
    S_2 
        = \left[ \frac{\partial \lambda}{\partial Y}
        + \frac{\partial \lambda}{\partial H_2} 
            \, \frac{\partial H_2}{\partial Y}
        + \frac{\partial \lambda }{\partial H_2} 
            \, \frac{\partial H_2}{\partial G_1}
            \, \frac{\partial G_1}{\partial Y} \right]
        \, \frac{\partial Y}{\partial \rho}
\end{equation}
where
%

%\begin{eqnarray}
\begin{equation}
    \frac{\partial H_2}{\partial Y} 
        = C^{(\eta)} \left[ \frac{B_1 B_4\exp(-B_4 Y)}{Y}\,  
        - \frac{B_1(1.0 - \exp(-B_4 Y))}{Y^2}  \, 
        + B_2 B_5 G_1 \exp (B_5 Y) \right], \\
        \end{equation}
        \begin{equation}
    \frac{\partial \lambda}{\partial Y} 
        = \lambda_0 B_6 + 2.0 K^{(\lambda)} Y H_2, \\
                \end{equation}
        \begin{equation}
    \frac{\partial \eta}{\partial H_2} 
        = - \frac{\lambda_0}{ H_2^2} + K^{(\lambda)} Y^2, \\
                \end{equation}
       
%\end{eqnarray}
\end{subequations}
where $\partial$Y/$\partial\rho$ and $\partial$Y/$\partial$$G_{1}$   are as in \eqref{eq:derY} and \eqref{eq:derG1}. 
%%%%%%%%%%%%%%%%%%%%%%%%%%%%%%%%%%%%%%%%%%%%%%%
% section: end

\bibliographystyle{elsarticle-num} 

\bibliography{bibliography}

\end{document}